\documentclass[12pt,aps,preprint,nofootinbib,superscriptaddress,nobalancelastpage]{revtex4}
\pdfoutput=1

\usepackage{mathrsfs}
\usepackage{amsmath,amsthm,amssymb,slashed}
\usepackage{color}
\usepackage{graphicx}
\usepackage{verbatim}
\usepackage{hyperref}
\usepackage{multirow}
\hypersetup{colorlinks,bookmarksopen,bookmarksnumbered,citecolor=blue,
linkcolor=blue,pdfstartview=FitH,urlcolor=blue}

\newcommand{\be}{\begin{equation}}
\newcommand{\ee}{\end{equation}}
\newcommand{\bea}{\begin{eqnarray}}
\newcommand{\eea}{\end{eqnarray}}

\newcommand{\tab}{Tab.}
\newcommand{\hc}{\mathrm{h.c.}}

\begin{document}

\preprint{FTUAM-16-18, IFT-UAM/CSIC-16-047}

\title{Global constraints on heavy neutrino mixing}

\author{Enrique Fernandez-Martinez}
\email{enrique.fernandez-martinez@uam.es}
\affiliation{Departamento de F\'isica Te\'orica, Universidad Aut\'onoma de Madrid, Cantoblanco E-28049 Madrid, Spain}
\affiliation{Instituto de F\'isica Te\'orica UAM/CSIC,
 Calle Nicol\'as Cabrera 13-15, Cantoblanco E-28049 Madrid, Spain}
\author{Josu Hernandez-Garcia}
\email{josu.hernandez@uam.es}
\affiliation{Departamento de F\'isica Te\'orica, Universidad Aut\'onoma de Madrid, Cantoblanco E-28049 Madrid, Spain}
\affiliation{Instituto de F\'isica Te\'orica UAM/CSIC,
 Calle Nicol\'as Cabrera 13-15, Cantoblanco E-28049 Madrid, Spain}
\author{Jacobo Lopez-Pavon}
\email{jlpavon@ge.infn.it}
\affiliation{INFN, Sezione di Genova, via Dodecaneso 33, 16146 Genova, Italy}

\begin{abstract}
We derive general constraints on the mixing of heavy Seesaw neutrinos with the SM fields from a global fit to present flavour and electroweak precision data. We explore and compare both a 
completely general scenario, where the heavy neutrinos are integrated out without any further assumption, and 
the more constrained case were only 3 additional heavy states are considered. The latter assumption implies non-trivial correlations 
in order to reproduce the correct neutrino masses and mixings as observed by oscillation data and thus some qualitative differences can be found with the more general scenario.
The relevant processes analyzed in the global fit include searches for Lepton Flavour Violating (LFV) decays, 
probes of the universality of weak interactions, CKM unitarity bounds and electroweak precision data. In particular, a comparative and detailed study of the 
present and future sensitivity of the different LFV experiments is performed. We find a mild $1-2\sigma$ preference for non-zero heavy neutrino mixing of order 0.03-0.04 in the electron and tau sectors. At the $2\sigma$ level we derive bounds on all mixings ranging from 0.1 to 0.01 with the notable exception of the $e-\mu$ sector with a more stringent bound of 0.005 from the $\mu \to e \gamma$ process.
\end{abstract}

\maketitle

\section{Introduction}
\label{sec:intro}

The present evidence for neutrino masses and mixings demands an extension of the Standard Model (SM) to account for them and 
represents one of our best windows to new physics. The simplest and one of the most appealing possibilities, 
given its symmetry with the quark sector, is the addition of fermion singlets, right-handed neutrinos, 
to the SM particle content. However, even this simplest extension points towards the existence of a new type 
of term in the SM Lagrangian: a Majorana mass term for the right-handed neutrinos, allowed by the SM gauge symmetry. 
Such a term would imply direct breaking of the otherwise accidental Lepton number symmetry and the introduction 
of a mass scale not directly related to the Higgs mechanism of electroweak symmetry breaking. Depending on the actual scale of 
this mass term, interesting phenomenological consequences follow, such as the possible explanation of the baryon 
asymmetry of the Universe via the Leptogenesis mechanism~\cite{Fukugita:1986hr} or of the mysterious dark matter 
component via these sterile neutrinos~\cite{Dodelson:1993je,Shi:1998km,Abazajian:2001nj,Asaka:2005an}.  

A popular assumption for this mass scale is that it is above that of electroweak symmetry breaking. 
This choice indeed leads to the well-known Seesaw mechanism~\cite{Minkowski:1977sc,Mohapatra:1979ia,Yanagida:1979as,GellMann:1980vs} 
that nicely accommodates the strikingly tiny neutrino masses, as compared with the rest of the SM fermion content. 
In particular, for neutrino Yukawa couplings ranging in value from the electron to the top quark, a Majorana mass 
scale between the electroweak (EW) and grand unification scales can correctly reproduce our present constraints. 
Unfortunately, this huge hierarchy of scales also suppresses any other observable consequence of the model, 
beyond the leading order Weinberg $d=5$ operator~\cite{Weinberg:1979sa} that explains neutrino masses, rendering 
its experimental verification extremely challenging.

An interesting alternative is that of explaining the smallness of neutrino masses, not through a large hierarchy of scales, 
but rather via an approximate symmetry~\cite{Mohapatra:1986bd,Bernabeu:1987gr,Branco:1988ex,Buchmuller:1990du,Pilaftsis:1991ug,Dev:2012sg}. In particular, 
there are choices for the Majorana mass and Yukawa matrices such as the inverse~\cite{Mohapatra:1986bd,Bernabeu:1987gr} 
or linear~\cite{Malinsky:2005bi} Seesaw mechanisms that, for a given assignment of the charges among the extra states, 
approximately conserve $B-L$. Therefore, the Majorana masses obtained via the Weinberg operator by the light neutrinos 
are necessarily suppressed by the small $B-L$-violating parameters. Interestingly, higher order operators that would, a priori, 
be more strongly suppressed than neutrino masses, are not necessarily protected by this symmetry and can thus lead to 
sizable signals. In particular, apart from the Majorana nature of neutrino masses, the most characteristic signals of 
the Seesaw mechanism with right-handed neutrinos are deviations from unitarity of the lepton PMNS mixing matrix generated
by the only $d=6$ operator present at tree level. This 
in turn would lead to signals in lepton flavour violating processes (LFV), non-universality of weak interactions and/or affect 
electroweak precision observables~\cite{Lee:1977tib,Shrock:1980vy,Schechter:1980gr,Shrock:1980ct,Shrock:1981wq,Langacker:1988ur,Bilenky:1992wv,Nardi:1994iv,Tommasini:1995ii,Bergmann:1998rg,Loinaz:2002ep,Loinaz:2003gc,Loinaz:2004qc,Antusch:2006vwa,Antusch:2008tz,Biggio:2008in,Alonso:2012ji,Abada:2012mc,Akhmedov:2013hec,Basso:2013jka,Abada:2013aba,Antusch:2014woa,Antusch:2015mia,Abada:2015oba,Abada:2015trh,Abada:2016awd}. 

In this work we will combine results from all these probes to derive updated constraints on the presently allowed 
mixing among the extra massive neutrinos and the SM flavour eigenstates. We will present our results both, for a 
completely model-independent parametrization without any further assumption about the extra massive states and for the 
more restricted assumption of only three massive neutrinos, in analogy to the three generations for 
all other fermions. In the latter case, since the three extra neutrinos must also reproduce the correct pattern of masses and mixings 
as observed in neutrino oscillations, correlations among the potentially observable effects are predicted and constraints 
are qualitatively different from the general case. 

This paper is organized as follows. In Section~\ref{sec:param} we introduce the parametrizations adopted for our studies for 
the general and three-heavy-neutrino cases. In Section~\ref{sec:obs} we describe the set of observables used to probe for 
the heavy extra neutrinos. In Section~\ref{sec:res} we present and describe our results and finally we conclude in 
Section~\ref{sec:sum}.   

\section{Parametrizations}
\label{sec:param}

Starting from the usual type-I Seesaw Lagrangian:
\begin{eqnarray}\label{eq:The3FormsOfNuMassOp}
\mathscr{L} &=& \mathscr{L}_\mathrm{SM} -\frac{1}{2} \overline{N_\mathrm{R}^i} (M_N)_{ij} N^{c j}_\mathrm{R} -(Y_{N})_{i\alpha}\overline{N_\mathrm{R}^i}  \phi^\dagger
\ell^\alpha_\mathrm{L} +\hc\; ,
\end{eqnarray}
where $\phi$ denotes the SM Higgs field, $M_N$ the Majorana mass allowed for the right-handed neutrinos $N_\mathrm{R}^i$ and $Y_{N}$ the Yukawa couplings between the neutrinos and the 
Higgs field. The vev of the Higgs $v_{\mathrm{EW}}$ will, in addition, induce Dirac masses $m_D = v_\text{EW} Y_N/\sqrt{2}$. In the usual Seesaw limit, for $M_N \gg m_D $, the three light and mostly-active neutrinos observed in the neutrino oscillation phenomenon will be clearly separated from the heavy and mostly-sterile new states. Upon integrating out these heavy states, their low energy phenomenology will be encoded in a series of effective operators. The first such operator is the well-known $d=5$ Weinberg operator~\cite{Weinberg:1979sa} that, upon electroweak symmetry breaking, induces the Majorana masses for the light neutrinos:  
\begin{equation}
\hat{m} \equiv  m_D^t M_N^{-1} m_D = -U_{\rm PMNS} ^* m U_{\rm PMNS} ^\dagger,
\label{eq:mass}
\end{equation}
where $U_{\rm PMNS} = U_{23}(\theta_{23}) U_{13}(\theta_{13},\delta) U_{12}(\theta_{12}) {\rm diag}(e^{-i\alpha_1/2},e^{-i\alpha_2/2},1) $ is the Unitary mixing matrix that diagonalizes the symmetric mass matrix $\hat{m}$ generated from 
the Weinberg operator. At tree level, the only $d=6$ operator obtained upon integrating out the heavy neutrinos induces 
non-canonical neutrino kinetic terms for the three SM active neutrinos when the Higgs develops its vev~\cite{Broncano:2002rw}.
After diagonalizing and normalizing the kinetic terms, the mixing matrix appearing in charged current interactions will 
thus contain, not only the two Unitary rotations to diagonalize the $d=5$ and $d=6$ operators respectively, but also 
the necessary rescaling to bring the neutrino kinetic term to its canonical form. Thus, in all generality, the 
matrix describing the mixing between the light neutrino mass eigenstates and the SM charged leptons via $W$ interactions will 
not be Unitary and to stress this feature we will dub it $N$. Since any general matrix can be parametrized as the product 
of an Hermitian and a Unitary matrix, these deviations from unitarity have been often parametrized as~\cite{FernandezMartinez:2007ms}:
\begin{equation} 
N = (I - \eta) U_{\rm PMNS} ,
\label{eq:Neq}
\end{equation}  
where the small Hermitian matrix $\eta$ (also called $\epsilon$ in other works) encodes the deviations from unitarity in neutrino mixing. This parametrization is very convenient from a phenomenological point of view. Indeed, since the particular neutrino mass eigenstate is never identified in physical observables, its index is always summed upon, while the flavour index labeling the charged leptons participating in the process is normally fixed. Thus, most observables depend on the combination:
\begin{equation} 
\sum_i N_{\alpha i} N^\dagger_{i \beta} = \delta_{\alpha \beta} - 2\eta_{\alpha \beta} +\mathcal{O}\left(\eta_{\alpha \beta}^2\right)
\end{equation}  
and can thus be expressed only though the parameters contained in the Hermitian matrix $\eta$. Moreover, the physical 
interpretation of $\eta$ is also very transparent in terms of the mixing between the extra heavy neutrinos and the SM flavours. Indeed, if the full mass matrix is diagonalized as: 
\begin{equation}
U^T \left(
\begin{array}{cc}
0 & m_D^T \\ m_D & M_N
\end{array}
\right) U = \left(
\begin{array}{cc}
m & 0 \\ 0 & M
\end{array}
\right), \label{eq:diag}
\end{equation}
where $m$ and $M$ are diagonal matrices containing respectively the masses of the 3 light $\nu_i$ and heavy $N_i$ mass eigenstates. 
The diagonalizing matrix $U$ can be written as~\cite{Blennow:2011vn}:
\begin{equation}
U = \left(
\begin{array}{cc}
\ c & s \\ -s^\dagger & \hat{c}
\end{array}
\right) \left(
\begin{array}{cc}
U_{\rm PMNS} & 0 \\ 0 & I 
\end{array}
\right), \label{eq:block}
\end{equation}
where
\begin{equation}
\left(
\begin{array}{cc}
\ c & s \\ -s^\dagger & \hat{c}
\end{array}
\right) \equiv \left(
\begin{array}{cc}
\displaystyle\sum\limits_{n=0}^\infty \frac{ \left(- \Theta \Theta^\dagger \right)^{n}}{2n!} & 
\displaystyle\sum\limits_{n=0}^\infty \frac{ \left(- \Theta \Theta^\dagger \right)^{n}}{\left(2n+1\right)!} \Theta  \\ 
-\displaystyle\sum\limits_{n=0}^\infty \frac{ \left(- \Theta^\dagger \Theta \right)^{n}}{\left(2n+1\right)!} \Theta^\dagger & 
\displaystyle\sum\limits_{n=0}^\infty \frac{ \left(- \Theta^\dagger \Theta \right)^{n}}{2n!}\end{array}
\right) ,
\label{eq:sincos}
\end{equation}
and $\Theta \sim m^\dagger_D M_N^{-1}$ is the general matrix that describes the mixing between the heavy mass eigenstates 
and the active neutrino flavours. Thus, the non-unitary correction $I-\eta$ can be identified with the first term of the 
cosine expansion $1 - \Theta\Theta^\dagger/2$ such that:
\begin{equation}
\eta = \frac{\Theta \Theta^\dagger}{2}.
\label{eq:eta:theta}
\end{equation}
Furthermore, $\eta$ is also ($1/2$ of) the coefficient of the $d=6$ operator  
obtained upon integrating out the heavy neutrino fields:
\begin{equation}
\eta = \frac{m_D^\dagger M_N^{-2} m_D}{2}.
\label{eq:d6eta}
\end{equation}
In all generality the $d=6$ operator $\eta$ is completely independent from the $d=5$ $\hat{m}$ and thus from the measured neutrino masses and mixings in oscillation experiments~\cite{Broncano:2003fq,Antusch:2009gn}. However, both $\hat{m}$ and $\eta$ are ultimately built from $m_D$ and $M_N$ and thus, in particular cases, may not be fully independent. Apart from the completely general parametrization through $\eta$, here we will also investigate one such case. Namely, we will focus on the particular scenario in which:
\begin{itemize}
\item The SM is only extended through 3 right-handed neutrinos.
\item The three extra neutrino mass eigenstates are heavier than the EW scale.
\item Large, potentially observable, $\eta$ is allowed despite the smallness of neutrino masses.
\item The small neutrino masses are radiatively stable.
\end{itemize}
The only way to simultaneously satisfy these requirements is through an underlying $L$ symmetry~\cite{Kersten:2007vk,Abada:2007ux} (see also Ref.~\cite{Adhikari:2010yt,Antusch:2015mia}) which leads to:
\begin{equation}
m_D = \frac{v_\text{EW}}{\sqrt{2}} \left(
\begin{array}{ccc}
Y_{Ne} & Y_{N\mu} & Y_{N\tau} \\ \epsilon_1 Y'_{Ne} & \epsilon_1 Y'_{N\mu} & \epsilon_1 Y'_{N\tau} \\ \epsilon_2 Y''_{Ne} & \epsilon_2 Y''_{N\mu} & \epsilon_2 Y''_{N\tau}
\end{array}
\right)
\qquad
\textrm{and}
\qquad
M_N = \left(
\begin{array}{ccc}
\mu_1 & \Lambda & \mu_3 \\ \Lambda & \mu_2 & \mu_4 \\ \mu_3 & \mu_4 & \Lambda'
\end{array}
\right), \label{eq:texture}
\end{equation}
with all $\epsilon_i$ and $\mu_j$ small lepton number violating parameters  (see also Ref.~\cite{Dev:2013oxa} for a particular scenario where these small parameters arise naturally). By setting all $\epsilon_i=0$ and $\mu_j=0$, lepton number symmetry is indeed recovered with the following $L$ assignments $L_e = L_\mu = L_\tau = L_1 = -L_2 = 1$ and $L_3 = 0$. Also $\hat{m}=0$ (3 massless neutrinos in the $L$-conserving limit), $M_1=M_2=\Lambda$ (a heavy Dirac pair) and $M_3=\Lambda'$ (a heavy decoupled Majorana singlet), but:
\begin{equation}
\eta=  \frac{1}{2} \left(
\begin{array}{ccc}
|\theta_e|^2 & \theta_e \theta_\mu^* & \theta_e \theta_\tau^* \\ \theta_\mu \theta_e^* & |\theta_\mu|^2 & \theta_\mu \theta_\tau^* \\ \theta_\tau \theta_e^* & \theta_\tau \theta_\mu^* & |\theta_\tau|^2
\end{array}
\right)
\quad
\mathrm{with}
\quad
\theta_\alpha \equiv \frac{Y_{N\alpha} v}{\sqrt{2} \Lambda}.
\label{eq:theta}
\end{equation}
So that large $\eta$ is possible even in the limit of massless neutrinos when $L$ is conserved. Upon switching on 
the $L$-violating parameters in Eq.~(\ref{eq:texture}), neutrino masses and mixings $\hat{m}$ that can reproduce 
the observed neutrino oscillations are generated. However, these are not completely independent from $\eta$ and the 
following relationship between the $\theta_\alpha$ in Eq.~(\ref{eq:theta}) and $\hat{m}$ follows~\cite{Fernandez-Martinez:2015hxa}: 
\begin{equation}
\begin{split}
\theta_\tau& \simeq \frac{1}{\hat{m}_{e \mu}^2 - \hat{m}_{ee} \hat{m}_{\mu \mu}}\left(\theta_e\left(\hat{m}_{e \mu}\hat{m}_{\mu \tau}-\hat{m}_{e \tau}\hat{m}_{\mu \mu}\right)+\right.\\
&\left. \theta_\mu\left(\hat{m}_{e \mu}\hat{m}_{e \tau}-\hat{m}_{ee}\hat{m}_{\mu \tau}\right)\pm\sqrt{\theta_e^2\hat{m}_{\mu \mu}-2\theta_e\theta_\mu \hat{m}_{e \mu}+\theta_\mu^2\hat{m}_{ee}}\times \right.\\
&\left.\times\sqrt{\hat{m}_{e \tau}^2\hat{m}_{\mu \mu}-2\hat{m}_{e \mu}\hat{m}_{e \tau}\hat{m}_{\mu \tau}+\hat{m}_{ee}\hat{m}_{\mu \tau}^2+\hat{m}_{e \mu}^2\hat{m}_{\tau \tau}-\hat{m}_{ee}\hat{m}_{\mu \mu}\hat{m}_{\tau \tau}}\right) .
\end{split}
\label{eq:Yt}
\end{equation}
Thus, this extra constraint will lead to correlations among the heavy-active mixing parameters $\theta_\alpha$ and 
therefore also $\eta_{\alpha \beta}$ through Eq.~(\ref{eq:theta}), not present in the completely general scenario 
with more than 3 heavy neutrinos. From now on we will refer to the unrestricted scenario as {\bf G-SS} 
(\emph{general Seesaw}) and to the particular case with 3 extra heavy neutrinos as {\bf 3N-SS}. The parameters 
characterizing the heavy neutrino mixing and the correlations between them in each case are summarized in 
Table~\ref{tab:params}. In particular, the constraints on $\eta$ for the G-SS come from the fact that $\eta$ is positive definite (see Eq.~(\ref{eq:eta:theta})).Regarding $\theta_\tau$ in the 3N-SS case, its value is fixed by $\theta_e$ and $\theta_\mu$ 
through Eq.~(\ref{eq:Yt}) once the SM neutrino masses and mixings encoded in the $d=5$ operator $\hat{m}$ are specified. 
In our analysis we will thus scan the allowed parameter space of the 3N-SS by leaving $\theta_e$ and $\theta_\mu$ free 
in the fit, together with the remaining unknown values characterizing $\hat{m}$: the Dirac phase $\delta$, the Majorana 
phases $\alpha_1$ and $\alpha_2$, the absolute neutrino mass and the mass hierarchy (normal or inverted). Regarding the absolute neutrino mass scale we will add the constraint from Planck on the sum of the light neutrino masses $\sum m_i < 0.23$ at a $95 \%$ CL~\cite{Ade:2015xua}. The rest of the 
oscillation parameters are fixed to their best fits from Ref.~\cite{Gonzalez-Garcia:2014bfa} since they are well-constrained
by present neutrino oscillation data.  
\begin{table}
\begin{center}
\begin{tabular}{|c|c|c|c|c|c|c|}
\hline
  &  $\eta_{ee}$ & $\eta_{\mu \mu}$ & $\eta_{\tau \tau}$ & $\eta_{e \mu}$ & $\eta_{e \tau}$ & $\eta_{\mu \tau}$ \\
\hline
\multirow{2}{*}{G-SS} &  $\eta_{ee} > 0$ & $\eta_{\mu \mu} > 0$ & $\eta_{\tau \tau} > 0$ & $|\eta_{e \mu}| \leq \sqrt{\eta_{ee} \eta_{\mu \mu}}$ & $|\eta_{e \tau}| \leq \sqrt{\eta_{ee} \eta_{\tau \tau}}$ & $|\eta_{\mu \tau}| \leq \sqrt{\eta_{\mu \mu} \eta_{\tau \tau}}$ \\
    &  free &  free & free & free & free & free   \\
\hline
\multirow{2}{*}{3N-SS} &  $\eta_{ee} = \frac{|\theta_e|^2}{2} $ & $\eta_{\mu \mu} = \frac{|\theta_\mu|^2}{2}$ & $\eta_{\tau \tau} = \frac{|\theta_\tau|^2}{2}$ & $ \eta_{e \mu} = \frac{\theta_e \theta_\mu^*}{2}$ & $ \eta_{e \tau} = \frac{\theta_e \theta_\tau^*}{2}$ & $ \eta_{\mu \tau} = \frac{\theta_\mu \theta_\tau^*}{2}$ \\
    &  free &  free & fixed by Eq.~(\ref{eq:Yt}) & fixed by $\theta_e$, $\theta_\mu$  & fixed by $\theta_e$, $\theta_\tau$  & fixed by $\theta_\mu$, $\theta_\tau$   \\
\hline
\end{tabular}
\caption{Summary of the parameters characterizing the mixing between flavour eigenstates and the extra heavy neutrinos 
for a completely general Seesaw scenario ({G-SS}) and the particular case of 3 extra heavy neutrinos (3N-SS). 
The constraints and correlations between parameters in each model are also summarized in the table. The value of 
$\theta_\tau$ for the 3N-SS case is computed through Eq.~(\ref{eq:Yt}) as a function of 
$\theta_e$, $\theta_\mu$, $\delta$, $\alpha_1$, $\alpha_2$, the absolute neutrino mass scale and the mass hierarchy. 
The rest of the oscillation parameters are fixed to their best fits from Ref.~\cite{Gonzalez-Garcia:2014bfa}.}
\label{tab:params}
\end{center}
\end{table}

When presenting the results of the global fit in Section \ref{sec:res} we will derive constraints on the mixing of the heavy neutrinos with the SM active flavours $\theta_\alpha$ in Eq.~(\ref{eq:theta}) for the 3N-SS. Regarding the G-SS, we do not specify the number of heavy neutrinos with which the SM is extended since all the observable effects are simply encoded in the matrix $\eta$. Thus, each heavy neutrino can have a different mixing $\Theta_{\alpha i}$ and, to ease the comparison with the results from the 3N-SS, we will use the combination $\sqrt{2 \eta_{\alpha \alpha}}$ which represents the total mixing from all the additional heavy neutrinos with the flavour $\alpha$ and an upper bound on the individual mixings $\Theta_{\alpha i}$:
\begin{equation}
\Theta_{\alpha i} = \left( m_D^\dagger M^{-1} \right)_{\alpha i} \quad \mathrm{and} \quad 2 \eta_{\alpha \alpha} = \sum_i |\Theta_{\alpha i}|^2.
\label{eq:sqrteta}
\end{equation}
 
\section{Observables}
\label{sec:obs}

Global constraints on the mixing between the heavy and active neutrinos will be derived through a fit to the following 28 observables:

\begin{itemize}
\item{The $W$ boson mass $M_W$}
\item{The effective weak mixing angle $\theta_\text{W}$: $s_\text{W eff}^{2 \text{ lep}}$ and $s_\text{W eff}^{2 \text{ had}}$}
\item{Four ratios of $Z$ fermionic decays: $R_l$, $R_c$, $R_b$ and $\sigma^0_\text{had}$}
\item{The invisible width of the $Z$ $\Gamma_\text{inv}$}
\item{Ratios of weak decays constraining EW universality: $R^\pi_{\mu e}$, $R^\pi_{\tau \mu}$, $R^W_{\mu e}$, $R^W_{\tau \mu}$, $R^K_{\mu e}$, $R^K_{\tau \mu}$, $R^l_{\mu e}$ and $R^l_{\tau \mu}$}
\item{9 weak decays constraining the CKM unitarity}
\item{3 radiative LFV decays: $\mu\rightarrow e \gamma$, $\tau\rightarrow \mu \gamma$ and $\tau\rightarrow e \gamma$}
\end{itemize}

\begin{table}[htbp]
\centering
\begin{tabular}{|c|c|c|}
\hline
Observable & SM prediction & Experimental value \\
\hline
\hline
$M_{W}\simeq M_{W}^{\text{SM}}\left(1+0.20\left(\eta_{ee}+\eta_{\mu\mu}\right)\right)$ & $\left(80.363\pm 0.006\right)$ GeV & $\left(80.385\pm0.015\right)$ GeV\\
$s_\text{W eff}^{2 \text{ lep}}\simeq s_\text{W eff}^{2 \text{ lep SM}}\left(1-1.30\left(\eta_{ee}+\eta_{\mu\mu}\right)\right)$& $0.23152\pm0.00010$  & $0.23113\pm0.00021$\\
$s_\text{W eff}^{2 \text{ had}}\simeq s_\text{W eff}^{2 \text{ had SM}}\left(1-1.30\left(\eta_{ee}+\eta_{\mu\mu}\right)\right)$&$0.23152\pm0.00010$ & $0.23222\pm 0.00027$\\
\hline
$R_{l}\simeq R_{l}^{\text{SM}}\left(1+0.18\left(\eta_{ee}+\eta_{\mu\mu}\right)\right)$ & $20.740\pm 0.010$ & $20.804\pm0.050$\\
$R_{c}\simeq R_{c}^{\text{SM}}\left(1+0.11\left(\eta_{ee}+\eta_{\mu\mu}\right)\right)$ & $0.17226\pm 0.00003$ & $0.1721\pm0.0030$\\
$R_{b}\simeq R_{b}^{\text{SM}}\left(1-0.06\left(\eta_{ee}+\eta_{\mu\mu}\right)\right)$ & $0.21576\pm0.00003$ & $0.21629\pm0.00066$\\
$\sigma^0_\text{had}\simeq\sigma_\text{had}^{0\text{ SM}}\left(1+0.55\left(\eta_{ee}+\eta_{\mu\mu}\right)+0.53\eta_{\tau\tau}\right)$ & $\left(41.479\pm 0.008\right)$ nb & $\left(41.541\pm0.037\right)$ nb\\
$\Gamma_\text{inv}\simeq \Gamma_\text{inv}^{\text{SM}}\left(1-0.33\left(\eta_{ee}+\eta_{\mu\mu}\right)-1.32\eta_{\tau\tau}\right)$ & $\left(0.50166\pm0.00005\right)$ GeV& $\left(0.4990\pm0.0015\right)$ GeV\\
\hline
$R^\pi_{\mu e}\simeq \left(1-\left(\eta_{\mu\mu}-\eta_{ee}\right)\right)$ & 1 & $1.0042\pm0.0022$ \\
$R^\pi_{\tau \mu}\simeq \left(1-\left(\eta_{\tau\tau}-\eta_{\mu\mu}\right)\right)$ & 1 & $0.9941\pm0.0059$\\
$R^W_{\mu e}\simeq \left(1-\left(\eta_{\mu\mu}-\eta_{ee}\right)\right)$ & 1 & $0.992\pm0.020$\\
$R^W_{\tau \mu}\simeq \left(1-\left(\eta_{\tau\tau}-\eta_{\mu\mu}\right)\right)$ & 1 & $1.071\pm0.025$\\
$R^K_{\mu e}\simeq \left(1-\left(\eta_{\mu\mu}-\eta_{ee}\right)\right)$ & 1 & $0.9956\pm0.0040$\\
$R^K_{\tau \mu}\simeq \left(1-\left(\eta_{\tau\tau}-\eta_{\mu\mu}\right)\right)$ & 1 & $0.978\pm0.014$\\
$R^l_{\mu e}\simeq \left(1-\left(\eta_{\mu\mu}-\eta_{ee}\right)\right)$ & 1 & $1.0040\pm 0.0032$\\
$R^l_{\tau \mu}\simeq \left(1-\left(\eta_{\tau\tau}-\eta_{\mu\mu}\right)\right)$ & 1 & $1.0029\pm0.0029$ \\
\hline
$\left|V_{ud}^\beta\right| \simeq \sqrt{1-|V_{us}|^2}(1+\eta_{\mu\mu})$ & $\sqrt{1-|V_{us}|^2}$ & $0.97417\pm0.00021$\\
$\left|V_{us}^{\tau\rightarrow K\nu_{\tau}}\right|\simeq \left|V_{us}\right|\left(1+\eta_{ee}+\eta_{\mu\mu}-\eta_{\tau\tau}\right)$ & $\left|V_{us}\right|$ & $0.2212\pm0.0020$ \\
$\left|V_{us}^{\tau\rightarrow K,\pi}\right|\simeq\left|V_{us}\right|\left(1+\eta_{\mu\mu}\right)$ & $\left|V_{us}\right|$ & $0.2232\pm0.0019$ \\
$\left|V_{us}^{K_{L}\rightarrow \pi e\overline{\nu}_{e}}\right|\simeq\left|V_{us}\right|\left(1+\eta_{\mu\mu}\right)$ & $\left|V_{us}\right|$ & $0.2237\pm0.0011$ \\
$\left|V_{us}^{K_{L}\rightarrow \pi \mu\overline{\nu}_{\mu}}\right|\simeq\left|V_{us}\right|\left(1+\eta_{ee}\right)$ & $\left|V_{us}\right|$ & $0.2240\pm0.0011$ \\
$\left|V_{us}^{K_{S}\rightarrow \pi e\overline{\nu}_{e}}\right|\simeq\left|V_{us}\right|\left(1+\eta_{\mu\mu}\right)$ & $\left|V_{us}\right|$ & $0.2229\pm0.0016$ \\
$\left|V_{us}^{K^{\pm}\rightarrow \pi e\overline{\nu}_{e}}\right|\simeq\left|V_{us}\right|\left(1+\eta_{\mu\mu}\right)$ & $\left|V_{us}\right|$ & $0.2247\pm0.0012$ \\
$\left|V_{us}^{K{\pm}\rightarrow \pi \mu\overline{\nu}_{\mu}}\right|\simeq\left|V_{us}\right|\left(1+\eta_{ee}\right)$ & $\left|V_{us}\right|$ & $0.2245\pm0.0014$ \\
$\left|V_{us}^{K,\pi \rightarrow \mu\nu}\right| \simeq\left|V_{us}\right|\left(1+\eta_{\mu\mu}\right)$ & $\left|V_{us}\right|$ & $0.2315\pm0.0010$ \\
\hline
\end{tabular}
\caption{List of observables input to the global fit. The first column contains the leading dependence on the non-unitarity parameters $\eta$, the second column contains the loop-corrected SM expectation, 
and the third column the experimental measurement used in the fit.}
\label{tab:obs:num}
\end{table}

The dependence of each observable on the non-unitarity mixing matrix $N_{\alpha i}$ and the parameters $\eta_{\alpha \beta}$  will be presented and discussed in this section.
 In Ref.~\cite{Fernandez-Martinez:2015hxa} it was recently shown that loop level corrections
involving the new degrees of freedom can be safely neglected. However,
many SM-mediated loop corrections are relevant for these precision
observables and will therefore be accounted for~\cite{Agashe:2014kda}.
 Notice that, in principle, these SM loop corrections also contain an
indirect dependence on the non-unitarity parameters, notably through
their dependence on $G_F$ as determined in muon decay. This subleading
dependence of the observables will be neglected and only the
corrections from non-unitarity affecting the tree level relations will
be discussed in the following expressions. The numerical analysis, however, contains all relevant 
SM loop corrections when comparing with the corresponding observables. The loop-corrected SM expectation, 
together with the leading non-unitarity correction and the experimental measurements that will be the inputs 
of our global fit are all summarized in \tab (\ref{tab:obs:num}). 

\subsection{Constraints from $\mu$ decay: $G_{F}$, $M_Z$, $M_{W}$ and $\theta_{\text{W}}$}

As usual, all SM predictions will be made in terms of the very accurate measurements of $\alpha$, $M_Z$ and $G_F$ as measured in $\mu$ decay, $G_\mu$~\cite{Agashe:2014kda}:

\begin{eqnarray}
\alpha&=&\left(7.2973525698\pm0.0000000024\right)\cdot 10^{-3}, \nonumber\\
M_Z&=&\left(91.1876\pm0.0021\right) \text{ GeV}, \\
G_\mu&=&\left(1.1663787\pm0.0000006\right)\cdot 10^{-5} \text{ GeV}^{-2}. \nonumber
\end{eqnarray}

However, a non-unitary mixing matrix $N_{\alpha i}$ would modify the expected decay rate of $\mu \to e \nu \bar{\nu}$. Indeed, since the final state neutrinos are not determined, their index must be summed upon obtaining: 

\begin{equation}
\Gamma_{\mu}=\frac{m_{\mu}^5 G_{F}^{2}}{192 \pi^3}\sum_i{|N_{\mu i}|^2} \sum_j{|N_{e j}|^2} \simeq \frac{m_{\mu}^5 G_{F}^{2}}{192 \pi^3}\left(1-2\eta_{ee}-2\eta_{\mu\mu}\right)\equiv
\frac{m_{\mu}^5 G_{\mu}^{2}}{192 \pi^3}.
\end{equation}
Thus, $G_F$ as determined through muon decay ($G_\mu$) acquires a non-unitary correction that will propagate to most observables:
\begin{equation}
G_{F}=G_{\mu}\left(1+\eta_{ee}+\eta_{\mu\mu}\right).
\label{eq:Gmu}
\end{equation}
In particular, the relation between $G_\mu$ and $M_W$ allows to constrain $\eta_{ee}$ and $\eta_{\mu \mu}$ through kinematic measurements of $M_W$:
\begin{equation}
G_\mu = \frac{\alpha \pi M^2_Z\left(1+\eta_{ee}+\eta_{\mu\mu}\right)}{\sqrt{2}M^2_W\left(M^2_Z - M^2_W \right)}.
\end{equation}
Similarly, the weak mixing angle $s_{\text{W}}^2$ will be modified and independent determinations of $s_{\text{W}}^2$ will be used to further constrain $\eta_{ee}$ and $\eta_{\mu \mu}$:
\begin{equation}
s_{W}^{2}=\frac{1}{2}\left(1-\sqrt{1-\frac{2\sqrt{2}\alpha\pi}{G_{\mu}
M_{Z}^{2}}\left(1-\eta_{ee}-\eta_{\mu\mu}\right)}\right), 
\label{eq:sw}
\end{equation}
Regarding different measurements of $s_{\text{W}}^2$ it is important to note that in some low energy determinations, such as 
from the \emph{weak charge of the proton} or \emph{M{\o}ller scattering}, the dependence on this parameter appears through 
the following combination $-1/2 + 2s_{\text{W}}^2$. Since the value of $s_{\text{W}}^2$ is close to $1/4$, there is a partial 
cancellation in this observables that, in the SM, allows for a very accurate determination of $s_{\text{W}}^2$, since small 
changes in its value significantly affect the degree of the cancellation and hence the size of the observable. For the same 
reason, we find that these observables are also very sensitive to corrections of the order of SM loop corrections times the 
non-unitary parameters $\eta$. Indeed, including some of these corrections we find that the corresponding coefficients in front of the $\eta$ parameters in \tab (\ref{tab:obs:num}) would vary up to a factor 2, indicating that our approximation of neglecting 
these terms is not good enough for these precision observables. Since the inclusion of these corrections is beyond the 
scope of this work, we choose not to include these particular determinations of $s_{\text{W}}^2$ in the list of observables 
for our global fit.  

\subsection{Constraints from $Z$ decays}

\subsubsection{$Z$ decays into charged fermions}

The $Z$ decays into charged fermions are not directly modified in presence of heavy neutrinos or a non-unitary lepton mixing matrix at tree level. However, these measurements depend on $G_F$ and $s_W$ and, as such, an indirect dependence on the non-unitarity parameters appears through its determination via muon decay, as described above. In particular:

\begin{equation}
\Gamma\left(Z\rightarrow f\bar{f}\right)\equiv\Gamma_{f}=
\frac{G_{\mu}M_{Z}^{3}\left(g_{V}^{f 2}+g_{A}^{f 2}\right)}{6\sqrt{2}\pi}\left(1+\eta_{ee}+\eta_{\mu\mu}\right)
\label{eq:Z:widths}
\end{equation}
where the vector and axial-vector form factors are given by:
\begin{eqnarray}
g_{V}^{f}&=&N_{C}\left(T_{f}-2Q_{f}s_{\text{W}}^{2}\right) \nonumber	\\
g_{A}^{f}&=&N_{C}T_{f}
\end{eqnarray}
with $N_{C}$ the color factor, $N_{C}=3\text{ }(1)$ for quarks (leptons) and where $Q_f$ and $T_f$ are the electric charge and third component of the weak isospin of the fermion $f$. Notice that an additional dependence on $\eta_{ee}$ and $\eta_{\mu \mu}$ will be present in $g_{V}$ through $s_{\text{W}}^{2}$ and Eq.~(\ref{eq:sw}).

The usual combinations of decay rates will be used as observables for the global fit:
\begin{equation}
R_{q}=\frac{\Gamma_{q}}{\Gamma_{\text{had}}},\quad 
R_{l}=\frac{\Gamma_{\text{had}}}{\Gamma_{l}} \quad \text{and}\quad
\sigma_{\text{had}}^{0}=\frac{12 \pi\Gamma_{ee}\Gamma_{\text{had}}}{M_{Z}^{2}\Gamma_{Z}^{2}}  \,;
\end{equation} 
where $\Gamma_{\text{had}}\equiv\displaystyle\sum_{q\neq t}\Gamma_{q}$.

\subsubsection{Invisible $Z$ width}

In presence of a non-unitary lepton mixing matrix $N_{\alpha i}$, the $Z$ coupling to neutrinos is directly affected and becomes non diagonal since $(N^\dagger N)_{ij} \neq \delta_{ij}$.
Thus, apart from its indirect dependence through $G_F$, the invisible width of the $Z$, from which the number of active neutrinos can be determined, is directly sensitive to the mixing of heavy neutrinos:

\begin{equation}
\Gamma_{\text{inv}}=\frac{G_{F}M_{Z}^{3} \sum_{ij} |(N^\dagger N)_{ij}|^2 }{12\sqrt{2}\pi}
\simeq \frac{G_{\mu}M_{Z}^{3}}{12\sqrt{2}\pi}\big(3-
\left(4\,\eta_{\tau\tau}+\eta_{ee}+\eta_{\mu\mu}\right)\big) \equiv \frac{G_{\mu}M_{Z}^{3} N_{\nu}}{12\sqrt{2}\pi}
\end{equation}
Notice that, since $\eta_{\alpha \beta}$ is positive definite from Eq.~(\ref{eq:eta:theta}), the number of active neutrinos as measured through the invisible $Z$ width will be smaller than 3 in presence of mixing with heavy neutrinos, to be compared with the present determination of $N_\nu = 2.990 \pm 0.007$ from LEP~\cite{ALEPH:2005ab}. 

\subsection{Constraints from weak interaction universality tests}

The lepton flavour universality of weak interactions is strongly constrained through ratios of lepton and meson decays 
differing in the charged lepton generation involved, such as $\pi \to \mu \nu_i$ vs $\pi \to e \nu_i$. Since the final 
state neutrino cannot be determined, these processes are proportional to $\sum_i |N_{\alpha i}|^2 \approx 1 - 2\eta_{\alpha \alpha}$, where $\alpha$ is the flavour of the charged lepton. Thus, a flavour dependence is induced in presence on non-unitary mixing and the weak interaction universality constraints become powerful probes of heavy neutrino mixing:

\begin{equation}
\frac{\Gamma_\alpha}{\Gamma_\beta} \equiv \frac{\Gamma^{\text{SM}}_\alpha}{\Gamma^{\text{SM}}_\beta} R^2_{\alpha\beta} 
= \frac{\Gamma^{\text{SM}}_\alpha}{\Gamma^{\text{SM}}_\beta} \frac{\sum_i |N_{\alpha i}|^2}{\sum_i |N_{\beta i}|^2} \simeq \frac{\Gamma^{\text{SM}}_\alpha}{\Gamma^{\text{SM}}_\beta} \left( 1-2\eta_{\alpha \alpha}+2\eta_{\beta\beta} \right) ,
\label{eq:univ}
\end{equation}
where the ratio of the SM expectations for the decay widths $\Gamma^{\text{SM}}_\alpha$ will be given by a function of the charged lepton masses involved containing the corresponding phase space and chirality flip factors as well as the different loop corrections. Thus, at tree level and for the particular case of $\pi$ decays:
\begin{equation}
\frac{\Gamma^{\pi \text{SM}}_\alpha}{\Gamma^{\text{SM}}_\beta} =\left(\frac{m_{\alpha}\left(m_{\pi}^{2}-m_{\alpha}^{2}\right)}{m_{\beta}^{2}\left(m_{\pi}^{2}-m_{\beta}^{2}\right)}\right)^{2}.
\end{equation}
Constraints on the values of the ratios of weak coupling constants $R_{\alpha \beta}$ as defined in Eq.~(\ref{eq:univ}) have been derived through ratios of different decays~\cite{Pich:2013lsa} and are summarized in \tab~(\ref{tab:obs:num}). 

\subsection{Unitarity of the CKM matrix}

The presence of extra heavy neutrinos leads to unitarity violations of the lepton PMNS mixing matrix leaving the CKM quark mixing unaffected.
However, the processes through which the elements of the CKM matrix $V$ are determined are affected both directly 
(for processes involving leptons) and indirectly (through the determination of $G_F$ in muon decays). In particular, 
the unitarity relation among the elements of the first row of the CKM matrix is very strongly constrained and reads:
\begin{equation}
\left|V_{ud}\right|^{2}+\left|V_{us}\right|^{2}+\left|V_{ub}\right|^{2}=1
\label{CKM:eq}
\end{equation}
%
For the present accuracy on $V_{us}$, the value of $V_{ub} = \left(4.13\pm0.49\right)\times 10^{-3}$~\cite{Agashe:2014kda} can be safely neglected in Eq.~(\ref{CKM:eq}). This relation, together with the measurements from the different processes used to constrain $V_{ud}$ and $V_{us}$ will thus also present indirect sensitivities to $\eta_{\alpha \beta}$. In particular we will rewrite through Eq.~(\ref{CKM:eq}):  
\begin{equation}
\left|V_{ud}\right|=\sqrt{1-\left|V_{us}\right|^{2}}
\label{eq:udus}
\end{equation}
and use the following experimental constraints to fit for $V_{us}$ and the $\eta_{\alpha \beta}$ parameters on which they depend. In our final constraints on $\eta_{\alpha \beta}$ the dependence on $V_{us}$ has been treated as a nuisance parameter and the $\chi^2$ has been minimized with respect to it.

\subsubsection{Superallowed $\beta$ decay}

Superallowed $\beta$ decays provide the best determination of $\left|V_{ud}\right|$. However, in presence of a non-unitary PMNS matrix it will receive a direct correction with $\left(1-2\eta_{ee}\right)$ from the electron and neutrino coupling, as well as the indirect correction from $G_F$ in Eq.~(\ref{eq:Gmu}). All in all the value of $V_{ud}$ extracted from this process corresponds to:
\begin{equation}
\left|V_{ud}^{\beta}\right|=
\left(1+\eta_{\mu\mu}\right)\left|V_{ud}\right|.
\label{eq:betadec}
\end{equation}
The most recent update on $\left|V_{ud}^{\beta}\right|$ based on 20 different superallowed $\beta$ transitions~\cite{Hardy:2014qxa} is listed in \tab~(\ref{tab:obs:num}) and will be an input for our fit.

\subsubsection{$\left|V_{us}\right|$}

$\left|V_{us}\right|$ can be determined through $\tau$ decays and semileptonic or leptonic $K$ decays. 
The values of $f_{+}(0)$ and $f_{K}/f_{\pi}$ involved in these observables have been taken from~\cite{Aoki:2013ldr}.

\begin{itemize}

\item{$K$ decays}

Kaon decays offer a direct way to determine $\left|V_{us}\right|$. Apart from their sensitivity to this parameter, decays with $\mu$ ($e$) final states also have a direct dependence on $\eta_{\mu \mu}$ ($\eta_{e e}$) which cancels against the indirect dependence through $G_\mu$ leading to:
\begin{eqnarray}
\left|V_{us}^{K\rightarrow \pi e\overline{\nu}_{e}}\right|&=&
\left(1+\eta_{\mu\mu}\right)\left|V_{us}\right|,\\
\left|V_{us}^{K\rightarrow \pi \mu\overline{\nu}_{\mu}}\right|&=&
\left(1+\eta_{ee}\right)\left|V_{us}\right|.
\end{eqnarray}
The present determinations of $\left|V_{us}^{K\rightarrow \pi e\overline{\nu}_{e}}\right|$ and 
$\left|V_{us}^{K\rightarrow \pi \mu\overline{\nu}_{\mu}}\right|$ are listed 
in  \tab (\ref{tab:obs:num}) and have been obtained from \cite{Antonelli:2010yf,Moulson:2014cra} together with  
$f_{+}\left(0\right)$ from~\cite{Aoki:2013ldr}, the correlation matrix among observables from \cite{Antonelli:2010yf} has also been taken into account.


An alternative determination of $\left|V_{us}\right|$ stems from the ratio of the branching fractions $\mathcal{B}\left(K \rightarrow \mu \nu \right)/\mathcal{B}\left(\pi \rightarrow \mu \nu \right)$.  Notice that in this ratio any direct or indirect dependence on leptonic non-unitarity cancels allowing to constrain the ratio $\left|V_{us}\right|/\left|V_{ud}\right|$ as in the SM. Since this measurement is latter combined with $\left|V_{ud}^{\beta}\right|$ from Eq.~(\ref{eq:betadec}) to obtain $\left|V_{us}^{K,\pi \rightarrow \mu\nu }\right|$ the same $\left(1+\eta_{\mu\mu}\right)$ correction as for $\left|V_{ud}^{\beta}\right|$ is finally present: 
\begin{equation}
\left|V_{us}^{K,\pi \rightarrow \mu\nu }\right|=
\left(1+\eta_{\mu\mu}\right)\left|V_{us}\right|.
\label{eq:newK}
\end{equation}

\item{$\tau$ decays}

An alternative constraint on $\left|V_{us}\right|$ can be obtained from the $\tau\rightarrow K\nu_{\tau}$ decay rate. In presence of non-unitary leptonic mixing, a direct correction by $\left(1-2\eta_{\tau \tau}\right)$ will be present from the $\tau$ coupling as well as the indirect correction from $G_F$ leading to the following dependence:
\begin{equation}
\left|V_{us}^{\tau\rightarrow K\nu_{\tau}}\right|=
\left(1+\eta_{ee}+\eta_{\mu\mu}-\eta_{\tau\tau}\right)\left|V_{us}\right|.
\end{equation}
The value of $\left|V_{us}^{\tau\rightarrow K\nu_{\tau}}\right|$ is given in \tab~(\ref{tab:obs:num}) \cite{Amhis:2014hma}.\\

Another possibility is to constrain $\left|V_{us}\right|$ from the ratio $\mathcal{B}\left(\tau\rightarrow K\nu_{\tau}\right)/\mathcal{B}\left(\tau\rightarrow \pi\nu_{\tau}\right)$. In complete analogy to Eq.~(\ref{eq:newK}), the sensitivity to the non-unitarity parameters takes the form: 
\begin{equation}
\left|V_{us}^{\tau\rightarrow K,\pi}\right|=
\left(1+\eta_{\mu\mu}\right)\left|V_{us}\right|.
\end{equation}

All these observables with the values listed in \tab~(\ref{tab:obs:num}) will be used to fit for $\eta_{ee}$, $\eta_{\mu \mu}$ and $\eta_{\tau \tau}$. Regarding $|V_{us}|$, its value will be free to vary in the fit and will be treated as a nuisance parameter, choosing the value of $|V_{us}|$ that minimizes the $\chi^2$ for each value of $\eta_{ee}$, $\eta_{\mu \mu}$ and $\eta_{\tau \tau}$. 

\end{itemize}
%

\subsection{LFV observables}

Flavour transitions $\alpha \to \beta$ in presence of non-unitary mixing such that
$(N^\dagger N)_{\alpha \beta} = -2 \eta_{\alpha \beta} \neq 0$ are no longer protected by 
the GIM~\cite{Glashow:1970gm} mechanism. Thus, the stringent constraints that exist on lepton flavour violating (LFV) 
processes translate into strong probes of the PMNS unitarity, in particular on the off-diagonal 
elements $\eta_{\alpha \beta}$. Notice that from Eq.~(\ref{eq:eta:theta}) $\eta$ is a positive-definite matrix 
and its off diagonal elements subject to the Schwarz inequality:

\begin{equation}
|\eta_{\alpha \beta}| \leq \sqrt{\eta_{\alpha \alpha} \eta_{\beta \beta}},
\label{eq:schwarz}
\end{equation}
as summarized in Table~\ref{tab:params}. Thus, the direct constraints on the diagonal elements of $\eta$ stemming 
from the processes discussed above also constrain indirectly the size of the off-diagonal entries. Moreover, for 
the 3N-SS, Eq.~(\ref{eq:theta}) implies that the Schwarz inequality is saturated to an equality. Therefore, in the G-SS a global fit to constrain the diagonal elements of $\eta$ with the list of 
observables described above will be performed. Then, constraints on the off-diagonal entries will be derived indirectly 
through the Schwarz inequality and compared with the direct bounds from LFV processes. For the 3N-SS, the 
LFV observables will be added directly to the global fit since they also constrain the diagonal elements
through the saturation of the inequality. 

\begin{figure}
\centering
\includegraphics[width=0.6\textwidth]{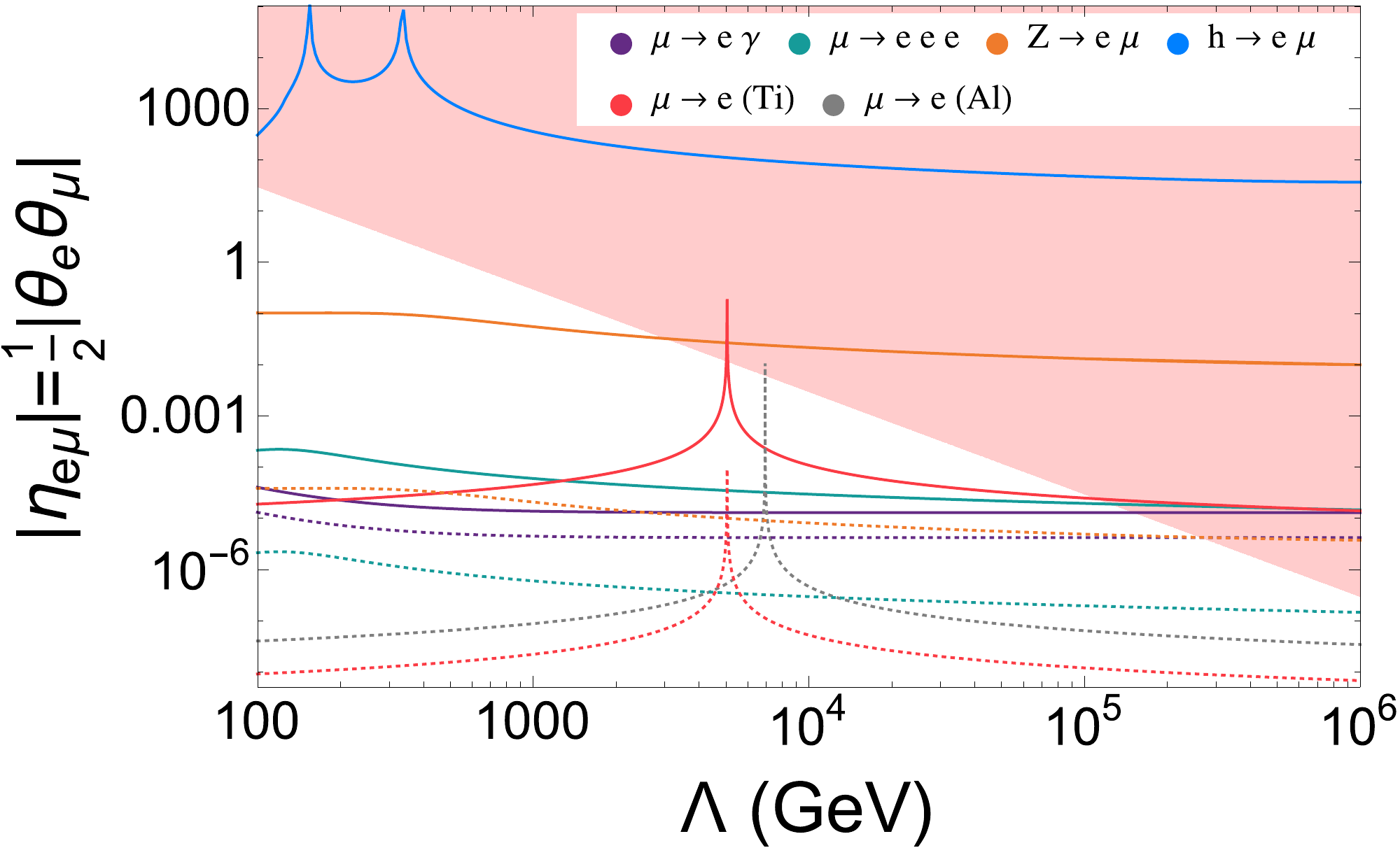}
\includegraphics[width=0.6\textwidth]{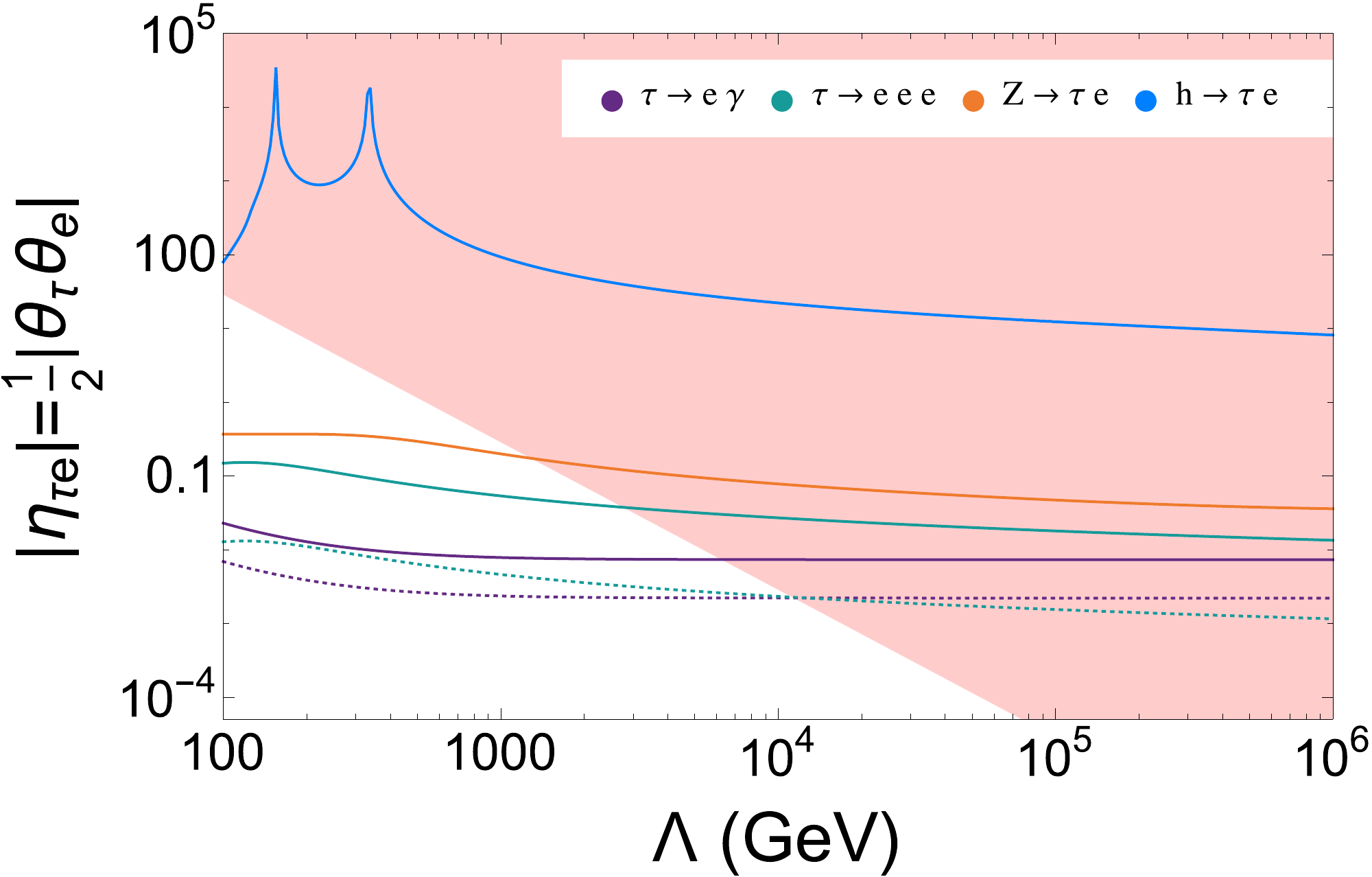}
\includegraphics[width=0.6\textwidth]{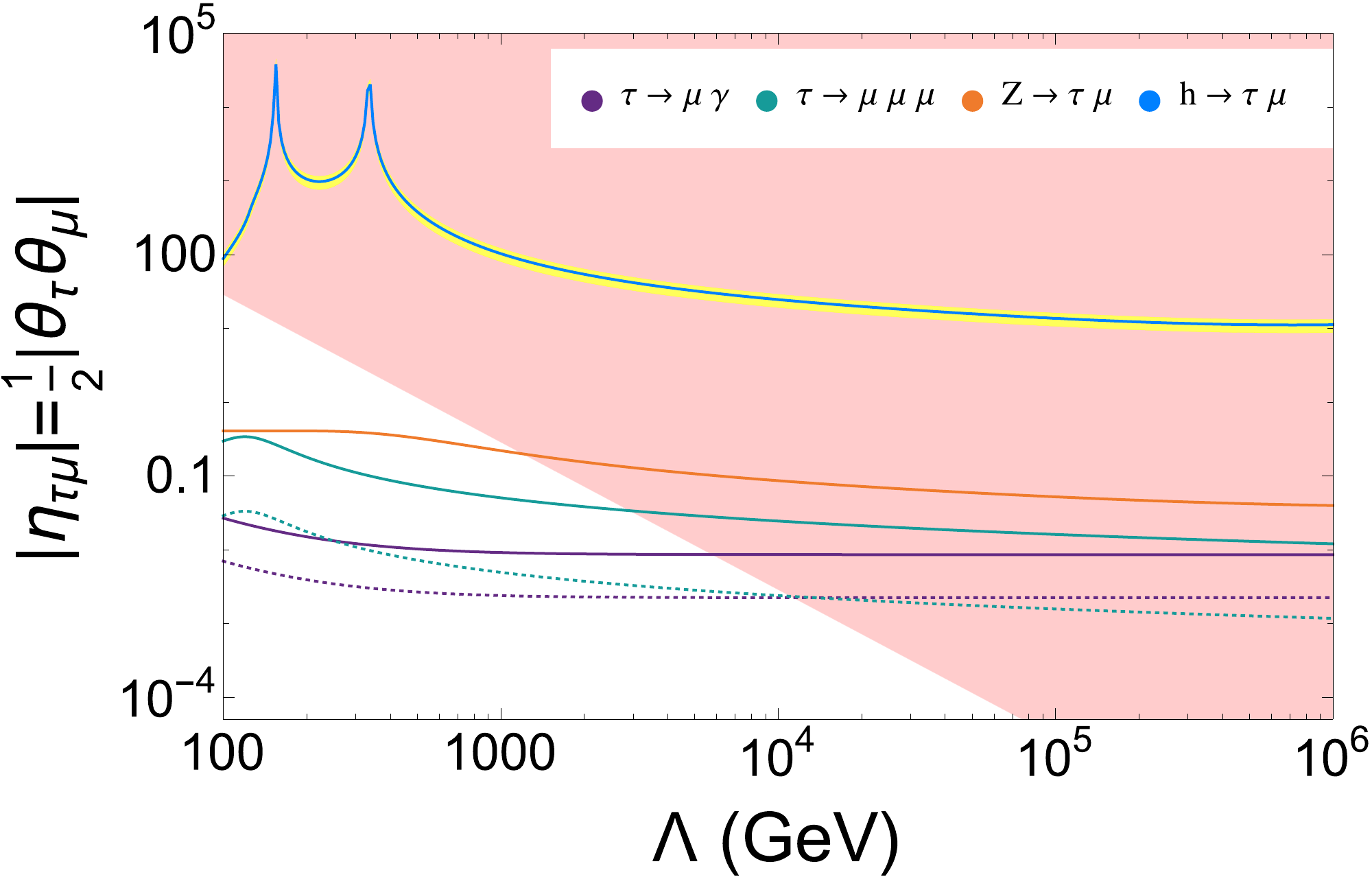}
\caption{$90$\% CL constraints on $\eta_{\alpha\beta}$ from LFV 
observables in the 3N-SS. Solid lines represent current experimental bounds while dotted lines represent future 
sensitivities as listed in Table.~\ref{tab:NDobs:num}.
The red-shadowed region represents the non-perturbative region with $|Y_N|^2>6\pi$. In the bottom panel, given the preference 
for non-zero $h \rightarrow \tau \mu$ \cite{Khachatryan:2015kon,Aad:2015gha} we show the preferred value in blue and 
the the $1 \sigma$ region in yellow.}
\label{fig:bounds}
\end{figure}

Below we list and describe the set of LFV transitions that would take place through non-unitary leptonic mixing. 
The present experimental bounds and future sensitivities are summarized in Table.~\ref{tab:NDobs:num}. A 
comparison summarizing the present relative importance of these observables constraining the off-diagonal elements 
of $\eta$ (solid lines) is presented in Fig.~\ref{fig:bounds}. Since the LFV observables typically depend on the value
of the heavy masses, we have performed the comparison for the 3N-SS, since there is only a common scale that simplifies the comparison. 
As can be seen, radiative decays $\l_\alpha \to l_\beta \gamma$ presently dominate the existing bounds and will thus be added to the global 
fit in the 3N-SS. However, regarding future expectations (dotted lines), 
the constraints on $|\eta_{e \mu}|$ will be dominated by $\mu \to eee$ or $\mu-e$ transitions in nuclei rather 
than by $\mu \to e \gamma$. On the other hand, the present and future sensitivity to $|\eta_{e \tau}|$ and $|\eta_{\mu \tau}|$ 
is completely dominated by the radiative decays $\l_\alpha \to l_\beta \gamma$. In particular, the constraints on $|\eta_{\alpha\beta}|$ from the LFV decays of the $Z$ and Higgs bosons, 
$Z \to \l_\alpha\l_\beta$ and $h\to \l_\alpha\l_\beta$, are at least one or three orders of magnitude weaker than the 
bounds from radiative decays respectively. Unfortunately this precludes the explanation of the present mild preference for non-zero $h\to \mu \tau$~\cite{Khachatryan:2015kon,Aad:2015gha} through heavy neutrino mixing (see yellow band in the lower panel of Fig.~\ref{fig:bounds}). Indeed, the values of the Yukawas required to explain these events are, not only excluded by the other observables depicted in the third panel of Fig.~\ref{fig:bounds}, but also fall into the non-perturbative region, shaded red in the figure.

\begin{table}[htbp]
\centering
\begin{tabular}{|c|c|c|}
\hline
Observable & Experimental bound & Future sensitivity \\
\hline
\hline
$\mu\rightarrow e \gamma$ & $<4.2\cdot 10^{-13}$ \cite{Agashe:2014kda} & $<6\cdot 10^{-14}$ \cite{Baldini:2013ke}\\
$\tau\rightarrow \mu \gamma$ & $<3.3\cdot 10^{-8}$ \cite{Agashe:2014kda}
& $<3\cdot 10^{-9}$ \cite{Bona:2007qt}\\
$\tau\rightarrow e \gamma$ & $<4.4\cdot 10^{-8}$ \cite{Agashe:2014kda} & $<3\cdot 10^{-9}$ \cite{Bona:2007qt}\\
\hline
$Z\rightarrow e \mu$ & $<7.1\cdot 10^{-7}$ \cite{Aad:2014bca} & $<10^{-13}$ \cite{Abada:2014cca} \\
$Z\rightarrow \tau e$ & $<9.3\cdot 10^{-6}$ \cite{Adriani:1993sy, Akers:1995gz}& $-$\\
$Z\rightarrow \tau \mu$ & $<1.1\cdot 10^{-5}$ \cite{Akers:1995gz, Abreu:1996mj}& $-$\\
\hline
$h\rightarrow e \mu$ & $<3.4\cdot 10^{-4}$ \cite{CMS:2015udp} & $-$\\
$h\rightarrow \tau e$ & $<6.6\cdot 10^{-3}$ \cite{CMS:2015udp} & $-$\\
$h\rightarrow \tau \mu$ & $(8.2\pm3.2)\cdot 10^{-3}$ \cite{Khachatryan:2015kon,Aad:2015gha}& $-$\\
\hline
$\mu\rightarrow eee$ & $<10^{-12}$ \cite{Bellgardt:1987du}& $<10^{-16}$ \cite{Blondel:2013ia}\\
$\tau\rightarrow eee$ & $<2.7\cdot 10^{-8}$ \cite{Hayasaka:2010np}& $<2\cdot 10^{-10}$ \cite{Bona:2007qt}\\
$\tau\rightarrow \mu\mu\mu$ & $<2.1\cdot 10^{-8}$ \cite{Hayasaka:2010np}& $<2\cdot 10^{-10}$ \cite{Bona:2007qt}\\
\hline
$\mu\rightarrow e$ (Al) & $-$ & $<10^{-17}$ \cite{Kutschke:2011ux}\\
$\mu\rightarrow e$ (Ti) & $<4.3\cdot 10^{-12}$ \cite{Dohmen:1993mp}& $<10^{-18}$ \cite{Barlow:2011zza}\\
\hline
\end{tabular}
\caption{Summary of the present constraints and expected future sensitivities for the different LFV observables considered.}
\label{tab:NDobs:num}
\end{table}

\subsubsection{LFV $Z$ decays}

For the 3N-SS, the $Z \rightarrow l_\alpha^{\mp}l_\beta^{\pm}$
decay branching ratio is simplified to~\cite{Illana:1999ww}
\be
\mathcal{B}\left(Z \rightarrow l_\alpha^{\mp}l_\beta^{\pm}\right)= \frac{\alpha^2 M_Z^3 G_\mu}{24\sqrt{2}\pi^3 s_w \Gamma_Z}
|\eta_{\alpha\beta}|^2\big|F(\lambda)-F(0)+G(\lambda,0)+G(0,\lambda)-2G(0,0)\big|^2,
\ee
where
\bea
G(\lambda_i,\lambda_j)&=&2C_{24}-1-\lambda_Q\left(C_0+C_{11}+C_{12}+C_{23}\right)-\frac{\lambda_i\lambda_j}{2}C_0,
\nonumber\\
F(\lambda)&=& 2c_w^2\big[\lambda_Q\left(\bar{C}_{11}+\bar{C}_{12}+\bar{C}_{23}\right)-6\bar{C}_{24}+1\big]
-\lambda(1-2s_w^2)\bar{C}_{24}\\
&-&2s_w^2\lambda\bar{C}_{0}+\frac{1-2c_w^2}{2}\big[(1+\lambda)B_1+1\big], \nonumber
\eea
and $\lambda=\Lambda^2/M_W^2$, $\lambda_Q=(p_\alpha-p_\beta)^2/M_W^2=M_Z^2/M_W^2+\mathcal{O}(m_l^2/M_W^2)$
and $C_{\left\{0,11,12,23\right\}}$, $C_{\left\{0,11,12,23,24\right\}}$ and $B_1$ defined in Appendix C
of~\cite{Illana:1999ww}.

As shown in Fig.~\ref{fig:bounds}, at present $\l_\alpha \to l_\beta \gamma$ is able to set bounds much stronger than through this process.

\subsubsection{LFV $h$ decays}

In the case of the LFV Higgs decay the expression at $\mathcal{O}\left(\eta_{\alpha\beta}^2\right)$ for the 
branching ratio is much more involved than in the $Z \rightarrow l_\alpha^{\mp}l_\beta^{\pm}$ case. In Fig.~\ref{fig:bounds} 
we have used the complete computation presented in~\cite{Pilaftsis:1992st,Arganda:2004bz,Arganda:2014dta}. Nevertheless, we instead present here an 
approximate expression which can be useful in order to understand the dependence on the parameters in the 3N-SS.
\be
\mathcal B\left(h\rightarrow l_\alpha^{\mp}l_\beta^{\pm}\right)\approx \frac{\alpha^3}{64\,\pi^2 s_w^6 \Gamma_h}
\left(\frac{\Lambda}{M_W}\right)^4M_h\,|\eta_{\alpha\beta}|^2
\left(\frac{m_\alpha^2}{M_W^2}\,|f_L|^2+\frac{m_\beta^2}{M_W^2}\,|f_R|^2\right),
\ee
where
\bea
f_L &=& 
\frac{M_h^2}{2}\left(C_0+C_{11}-C_{12}\right),
\nonumber\\
f_R &=& 
\frac{M_h^2}{2}\left(C_0+C_{12}\right),
\eea
and $C_{\left\{0,11,12\right\}}= C_{\left\{0,11,12\right\}}(m_\alpha^2,M_h^2,\Lambda^2,M_W^2,M_W^2)$. This approximate 
result is reasonably accurate for scales above few TeV and works very well for $\Lambda \gtrsim 10$ TeV. However, 
since here we are neglecting $\mathcal{O}\left(M_W^2/\Lambda^2\right)$ contributions, it fails for 
$\Lambda \lesssim 1$ TeV. In any case, the full calculation shows that the constraints on $|\eta_{\alpha\beta}|$ are
still very far from the present radiative bounds, falling indeed in the non perturbative region.

\subsubsection{$l_\alpha\rightarrow l_\beta l_\beta l_\beta$ decay}

Another LFV observable that would be induced by heavy neutrino mixing is the $l_\alpha\rightarrow l_\beta l_\beta l_\beta$. Its branching
ratio, for the 3N-SS, is given by~\cite{Ilakovac:1994kj}
\begin{eqnarray}
\mathcal{B}\left(l_\alpha\rightarrow l_\beta l_\beta l_\beta\right)&=&\dfrac{G_\mu^4 M_W^4 m_{\alpha}^5 \left| \eta_{\alpha\beta} \right|^2}
{18432 \pi^7\Gamma_{\alpha}}\left\lbrace 54-1188 s_\text{W}^2 + s_\text{W}^4 \left(1105 + 96 \log{\left(\frac{m_{\alpha}^2}{m_{\beta}^2}\right)} \right) \right. \\ \nonumber
&+& \left. 2 \log^2{\dfrac{\Lambda^2}{M_W^2}} \left(27 -96 s_\text{W}^2 +128 s_\text{W}^4 \right) - 4 \log{\dfrac{\Lambda^2}{M_W^2}} \left(27 - 219 s_\text{W}^2 + 296 s_\text{W}^4 \right) \right\rbrace.
\nonumber
\end{eqnarray}

Notice that, while additional non-unitarity corrections from $G_\mu$ and $s_\text{W}^2$ (also through $\Gamma_\alpha$ when 
$\alpha\neq\mu$) would be present, these are higher order in $\eta$ and therefore subleading since the whole process is already proportional to $|\eta_{\alpha\beta}|^2$.

Fig.~\ref{fig:bounds} shows that the present $\mu\rightarrow eee$ decay bound on 
$|\eta_{e\mu}|$ is quite competitive with the one coming from $\mu \to e \gamma$. The constraint is presently dominated
by $\mu \to e \gamma$, but it is expected to be overcome by $\mu\rightarrow eee$ in the future. On the other hand,
the present and future sensitivity to $|\eta_{e\tau}|$ and $|\eta_{\mu\tau}|$ is dominated by the radiative decays.

\subsubsection{$\mu\rightarrow e$ conversion} 

In the 3N-SS, the ratio between  $\mu\rightarrow e$ conversion rate over the capture 
rate $\Gamma_{capt}$ in light nuclei is given by~\cite{Alonso:2012ji}

\be
\label{Rmue} 
R_{\mu\rightarrow e}\simeq  \dfrac{G_\mu^2 \alpha^5m_\mu^5}{2s_w^4\pi^4\Gamma_\text{capt}}\dfrac{Z_\text{eff}^4}{Z}|\eta_{e\mu}|^2F_p^2   
 \Big[\left(A+Z\right)F_u+\left(2A-Z\right)F_d\Big]^2 \ \ 
     \,. 
\ee
where $A$ corresponds to the mass number, $Z$ ($Z_\text{eff}$) stands for the (effective) atomic number, $F_p$
is a nuclear form factor and
 \begin{eqnarray}
F_{u} &=&\dfrac{2}{3}s_W^2\dfrac{16\log\left(\frac{\Lambda^2}{M_W^2}\right)-31}{12}
-\dfrac{3+3\log\left(\frac{\Lambda^2}{M_W^2}\right)}{8},
\nonumber\\
F_{d}&=&-\dfrac{1}{3}s_W^2\dfrac{16\log\left( \frac{\Lambda^2}{M_W^2}\right)-31}{12}
-\dfrac{3-3\log\left(\frac{\Lambda^2}{M_W^2}\right)}{8},
\nonumber
\end{eqnarray}
The bounds shown in Fig.~\ref{fig:bounds} have been obtained from $\mu\rightarrow e$ conversion transitions in 
$_{13}^{27}\mbox{Al}$ and $_{22}^{48}\mbox{Ti}$. The input values for the nuclear parameters $F_p$, $Z_\text{eff}$ and 
$\Gamma_\text{capt}$  have been extracted from ~\cite{Kitano:2002mt,Suzuki:1987jf} and are summarized in Table 1 
of~\cite{Alonso:2012ji}.

According to the forecasted performances the future sensitivity to $|\eta_{e\mu}|$ will be dominated by this observable. Remarkably, future 
$\mu\rightarrow e$ searches~\cite{Barlow:2011zza} could improve the present bound by three orders of magnitude making 
it a very promising channel to probe for new physics signal in LFV decays.

\subsubsection{Radiative decays}

In the G-SS, the branching ratio for the radiative decays 
$\l_\alpha \to l_\beta \gamma$ is given by:

\begin{equation}
\frac{\Gamma\left(l_{\alpha}\rightarrow \l_{\beta} \gamma \right)}{\Gamma\left(l_{\alpha}
\rightarrow\l_{\beta}\nu_{\alpha}\overline{\nu}_{\beta}\right)}=\frac{3\alpha}{32\pi}
\frac{\left|\displaystyle\sum^{n}_{k=1}U_{\alpha k}U_{k\beta}^{\dagger}F\left(x_{k}\right)\right|^{2}}
{\left(UU^\dagger\right)_{\alpha\alpha}\left(UU^\dagger\right)_{\beta\beta}},
\end{equation}
where $x_k\equiv \frac{M_{k}^2}{M_W^2}$, and 
\begin{equation}
F(x_k)\equiv\frac{10-43x_k+78x_k^2-49x_k^3+4x_k^4+18x_k^3\ln x_k}{3(x_k-1)^4}.
\end{equation}
For $M_k\gg M_W$ the limit can be simplified to:

\begin{equation}
\label{eq:muegamma}
\frac{\Gamma\left(l_{\alpha}\rightarrow \l_{\beta} \gamma \right)}
{\Gamma\left(l_{\alpha}\rightarrow\l_{\beta}\nu_{\alpha}\overline{\nu}_{\beta}\right)}
\simeq \frac{3 \alpha}{8 \pi}\left|\eta_{\alpha\beta}\right|^{2}\big(F\left(\infty\right)-F\left(0\right)\big)^{2}
=\frac{3 \alpha}{2 \pi}\left|\eta_{\alpha\beta}\right|^{2}.
\end{equation}
This expression shows how the non-unitarity induced in the PMNS by the heavy neutrinos and the
separation of the two scales prevents the GIM cancellation. Indeed, the cancellation is recovered
in the limit $x_k\ll 1$. 

These radiative decays are the observables dominating the present constraints on $\eta_{\alpha\beta}$ as shown in Fig.~\ref{fig:bounds} and
will thus be the ones introduced in the fit through Eq.~(\ref{eq:muegamma}) for the 3N-SS. In the G-SS, these constraints will be compared with the bounds stemming from the Schwarz inequality Eq.~(\ref{eq:schwarz}) from the outcome of the global fit to the diagonal entries. 

\section{Results}
\label{sec:res}

With the list of observables described in the previous section and under a Gaussian approximation we construct a $\chi^2$ function to scan the parameter 
spaces of the G-SS and the 3N-SS. For the G-SS the free parameters of the fit are directly $\eta_{ee}$, 
$\eta_{\mu \mu}$ and $\eta_{\tau \tau}$ without further constraints and all the observables listed in 
Section \ref{sec:obs} except for the LFV transitions will be used to constrain them. The LFV radiative decays 
rather constrain the off-diagonal elements of the matrix $\eta$. Therefore, to obtain the global constraints 
on the off-diagonal elements, the LFV radiative decays will be combined and compared with the indirect bounds 
implied by the Schwarz inequality Eq.~(\ref{eq:schwarz}) from the lepton flavour conserving observables.

Regarding the 3N-SS, the free parameters for the fit are $\theta_e$ and $\theta_\mu$ (modulus and phase) 
while $\theta_\tau$ is given by Eq.~(\ref{eq:Yt}) once the light neutrino masses and mixings are specified 
through the $d=5$ operator $\hat{m}$. Thus, we also take as free parameters of the fit the values of 
the unknown phases of the PMNS matrix Dirac ($\delta$) and Majorana ($\alpha_1$ and $\alpha_2$) as well as 
the mass of the lightest neutrino mass eigenstate for both a normal and an inverted neutrino mass ordering. 
The rest of the oscillation parameters are fixed to their best fits from Ref.~\cite{Gonzalez-Garcia:2014bfa} 
since they are well-constrained by present neutrino oscillation data. Notice that, a priori, the number of 
free parameters we fit for in the 3N-SS case is larger than in the G-SS. However, this larger number of 
parameters is only included to take into account the constraints affecting $\theta_\tau$ (and therefore 
$\eta_{\tau \tau}$) via Eq.~(\ref{eq:Yt}) that are absent in the G-SS. Indeed, as we will see from the results 
of the fit, these constraints imply extra correlations between the parameters of the 3N-SS and there is in fact less freedom in the relevant parameters $\eta_{ee}$, $\eta_{\mu \mu}$ and $\eta_{\tau \tau}$ to 
fit for the observables. Since for the 3N-SS the Schwarz inequality Eq.~(\ref{eq:schwarz}) is saturated 
$|\eta_{\alpha \beta}| = \sqrt{\eta_{\alpha \alpha} \eta_{\beta \beta}}$, the LVF radiative decays also imply non-trivial 
constraints on the values of $\theta_\alpha$ and the diagonal elements $\eta_{\alpha \alpha}$ and will hence be included 
in the list of observables of the global fit.  Notice that, under the approximation of Eq.~(\ref{eq:muegamma}), 
the LFV radiative decays do not depend on the Majorana mass scale. Therefore, since none of the observables for the G-SS or 
3N-SS cases depend directly on the Majorana masses, the bounds on the mixing derived apply for any choice of the heavy 
neutrino masses above the electroweak scale. 

\begin{figure}
\centering
\includegraphics[width=0.45\textwidth]{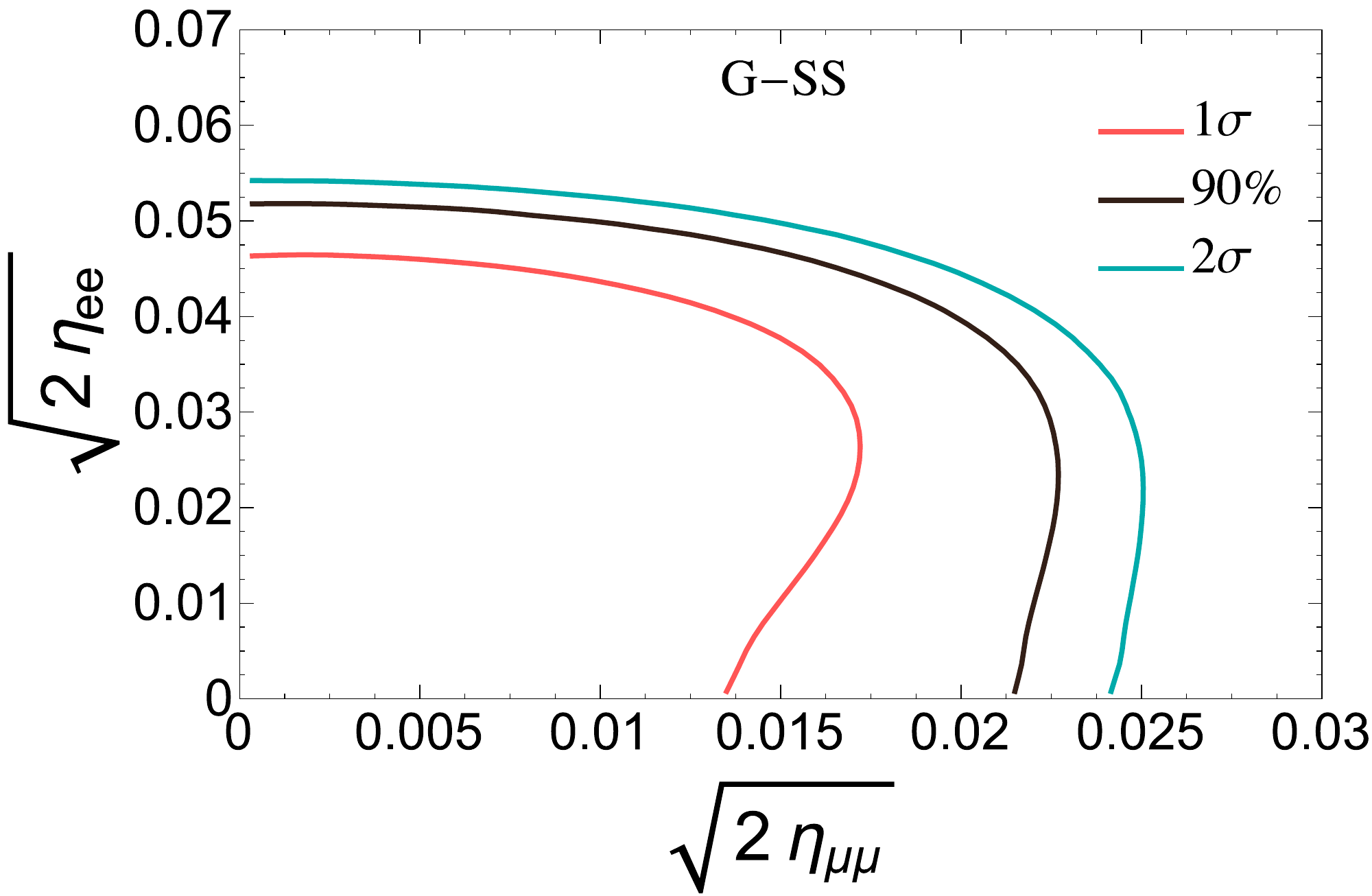}
\includegraphics[width=0.45\textwidth]{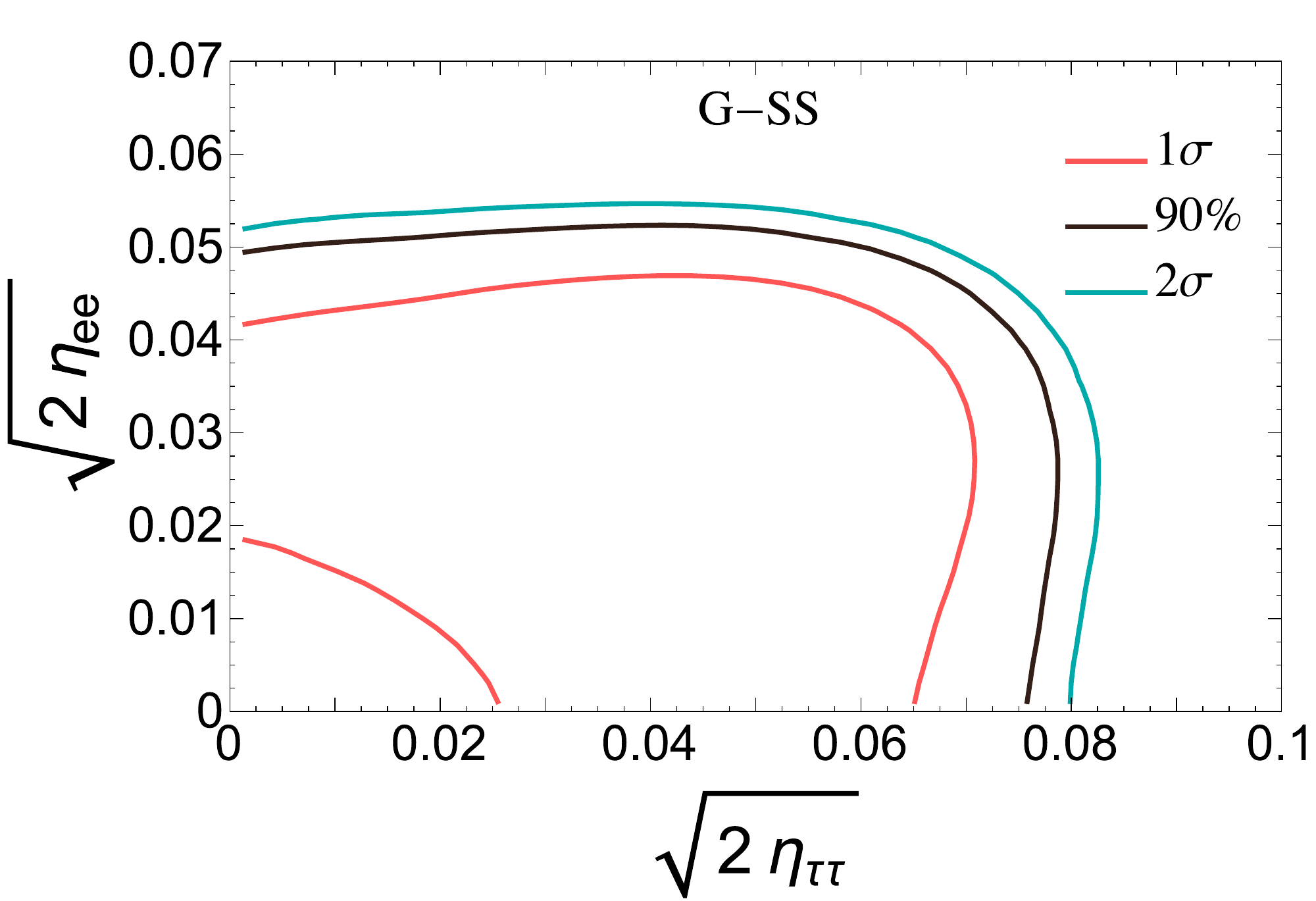}
\includegraphics[width=0.45\textwidth]{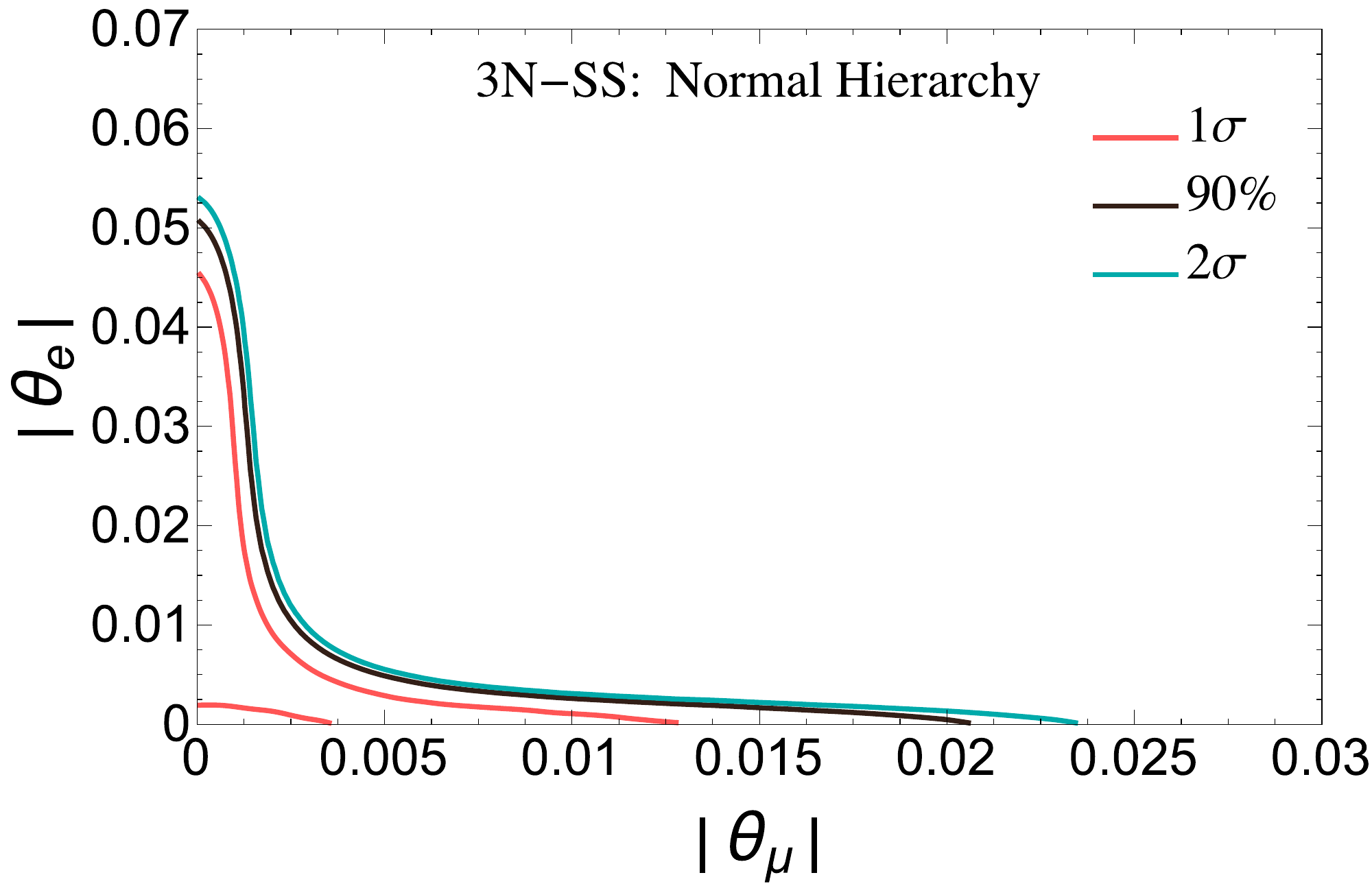}
\includegraphics[width=0.45\textwidth]{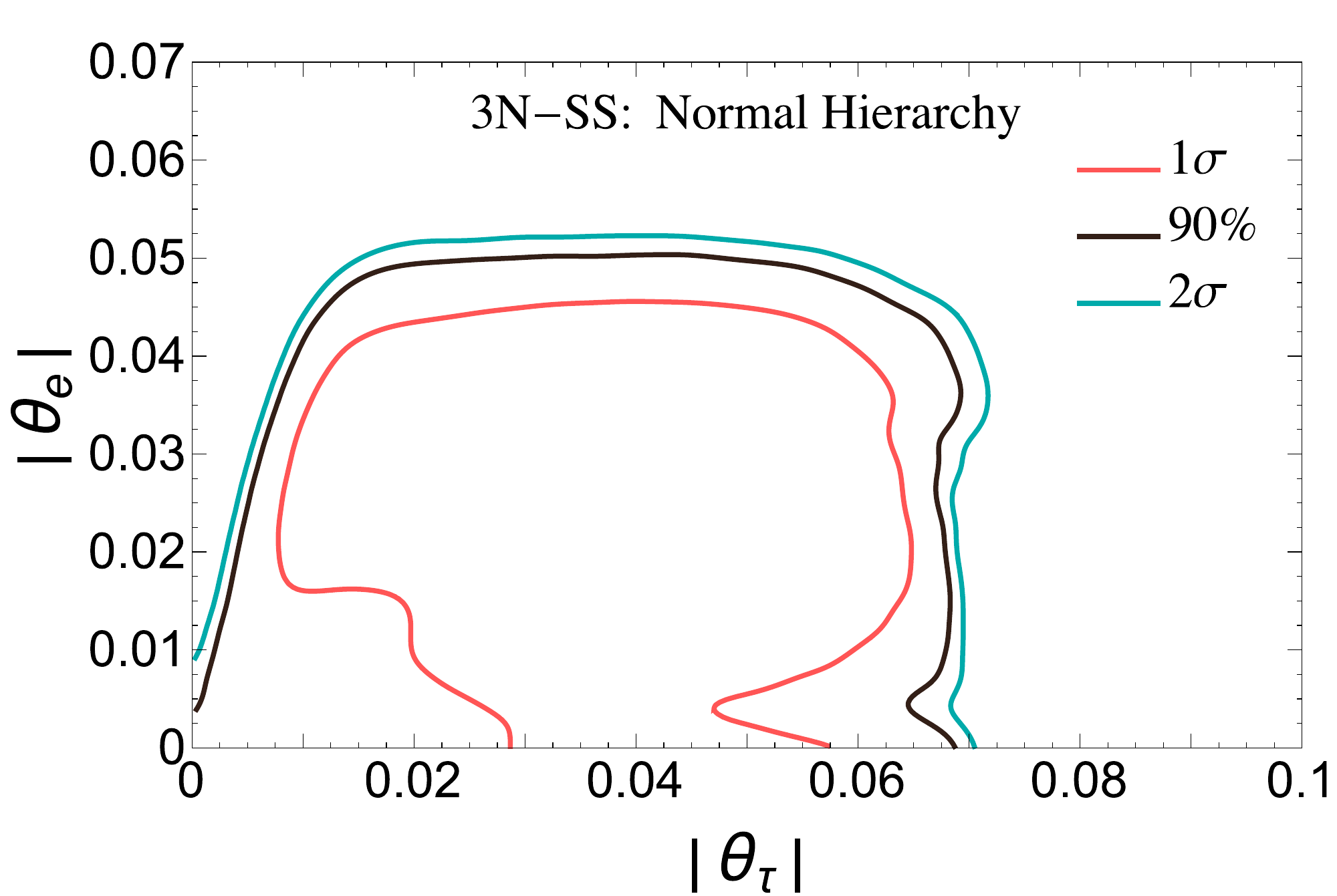}
\includegraphics[width=0.45\textwidth]{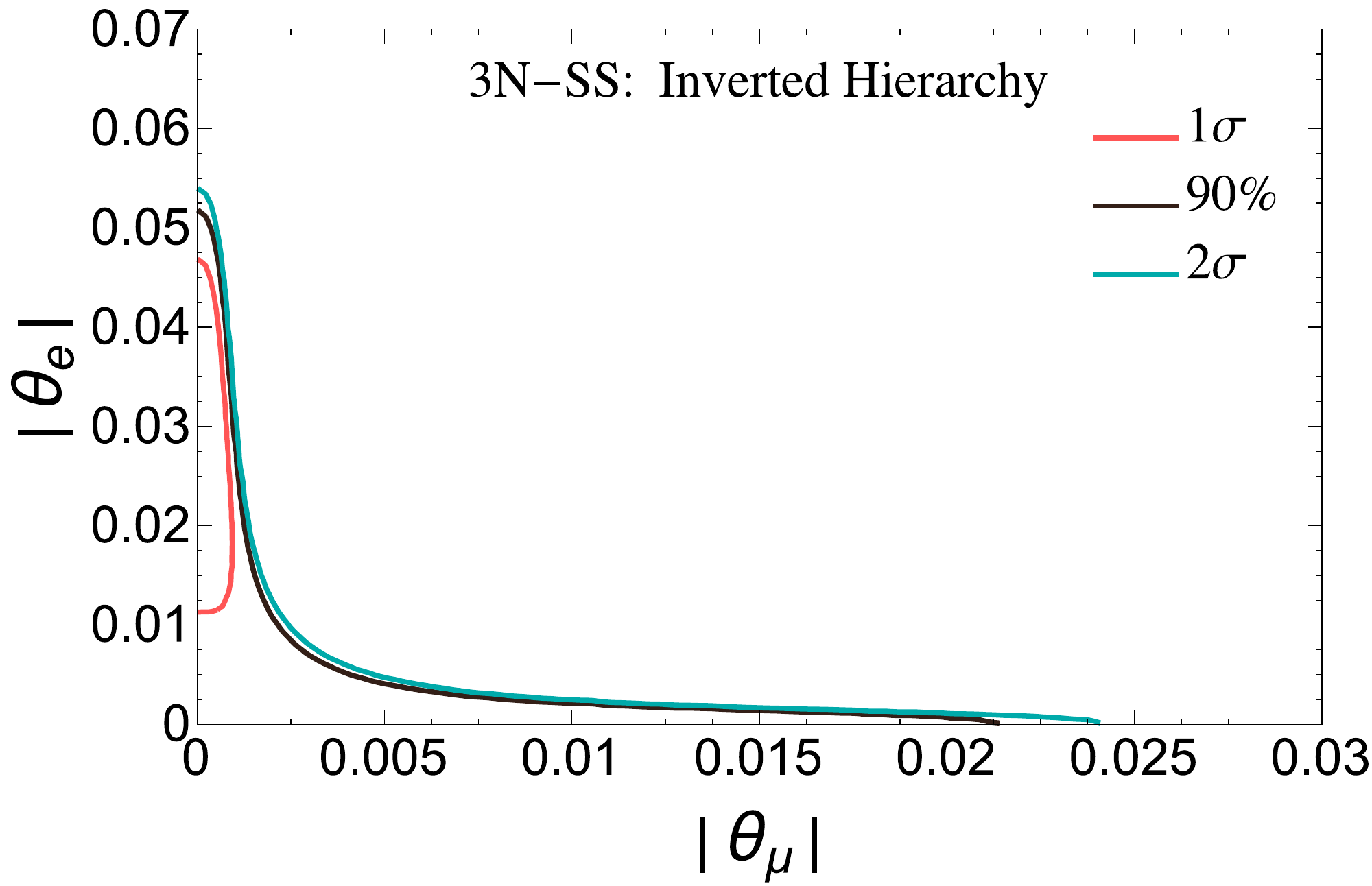}
\includegraphics[width=0.45\textwidth]{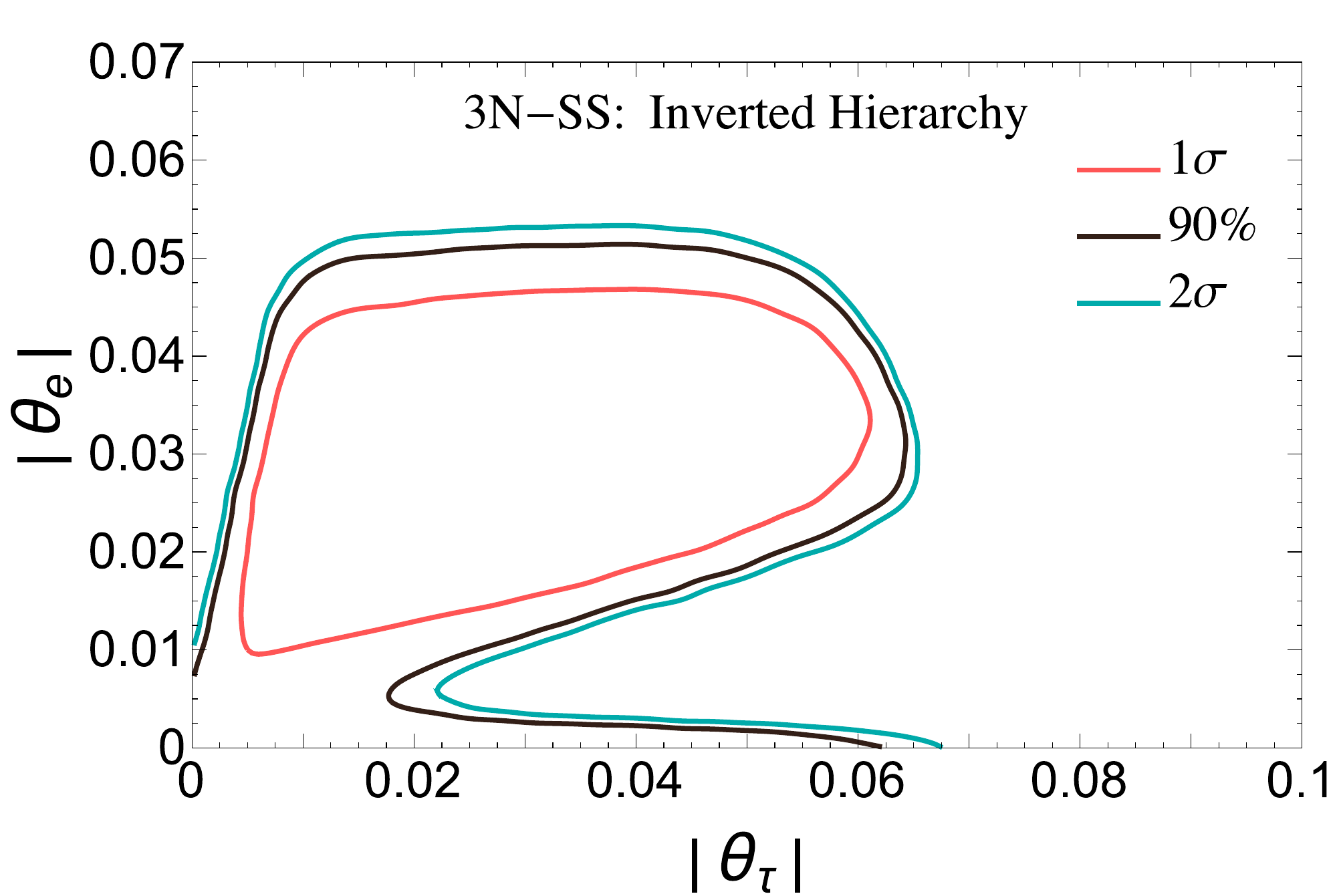}
\caption{Frequentist confidence intervals at $1 \sigma$, $90\%$ and $2 \sigma$ on the parameter space of the G-SS (upper panels) and the 3N-SS for normal hierarchy (middle panels) and inverted hierarchy (bottom panels).}
\label{fig:contours}
\end{figure}

In Fig.~\ref{fig:contours} we present our results from the global fit, performed by scanning the relevant parameter spaces 
through a Markov chain Monte Carlo algorithm. The results presented here correspond to the frequentist confidence intervals for 
$1 \sigma$, $90\%$ and $2 \sigma$ significance. We present the results directly in the heavy-active neutrino mixing 
$\theta_\alpha$ for the 3N-SS under the assumption of a normal neutrino ordering (middle panels) and 
inverted neutrino ordering (lower panels). To ease the comparison of the constraints, we present the results for the G-SS 
(upper panels) in the variable $\sqrt{2 \eta_{\alpha \alpha}}$, which can be identified with the total effective mixing of 
the different heavy mass eigenstates with the flavour $\alpha$, see Eq.~(\ref{eq:sqrteta}), and an upper bound on the individual mixing $\Theta_{\alpha i}$ of any additional heavy neutrino $N_i$. As can be seen, while the bounds on the 
individual parameters are comparable in strength for the two scenarios, the constraints imposed by Eqs.~(\ref{eq:theta}) 
and (\ref{eq:Yt}) for the 3N-SS reflect in strong correlations for their allowed regions. In particular, 
$\mu \to e \gamma$ imposes a very stringent constraint in the product $\theta_e \theta_\mu$ leading to the hyperbolic 
constraints in the middle-left and bottom-left panels of the figure and absent in the upper for the G-SS. On the other hand, 
in the middle and bottom-right panels of the figure non-trivial correlations between $\theta_e$ and $\theta_\tau$, 
absent in the upper-right panel for the G-SS, can be observed. This stems from the fact that $\theta_\tau$ is not free 
to take any value preferred by the observables, but constrained by $\theta_e$, $\theta_\mu$ and the neutrino masses 
and mixings through Eq.~(\ref{eq:Yt}). 

\begin{figure}
\centering
\includegraphics[width=0.32\textwidth]{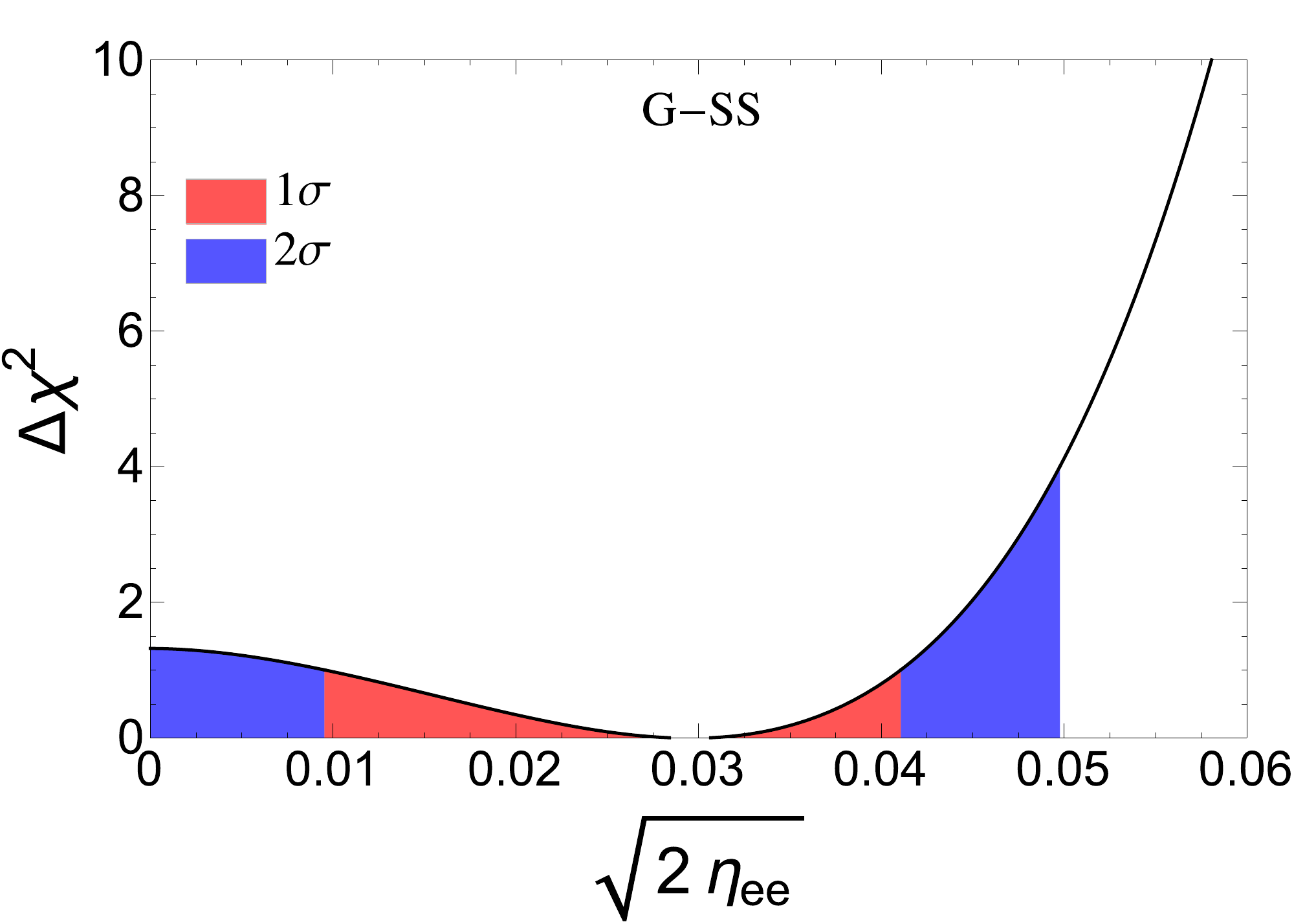}
\includegraphics[width=0.32\textwidth]{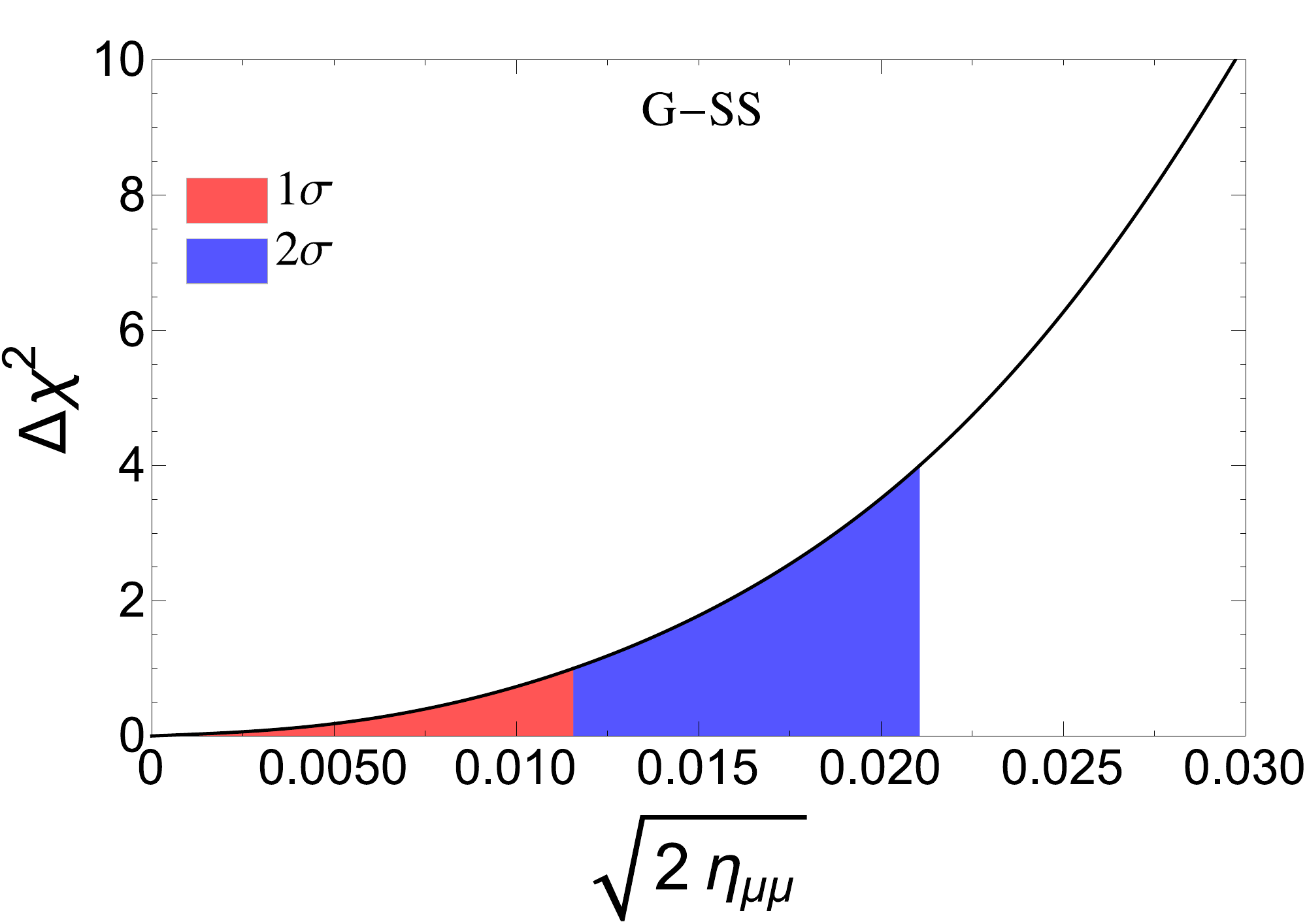}
\includegraphics[width=0.32\textwidth]{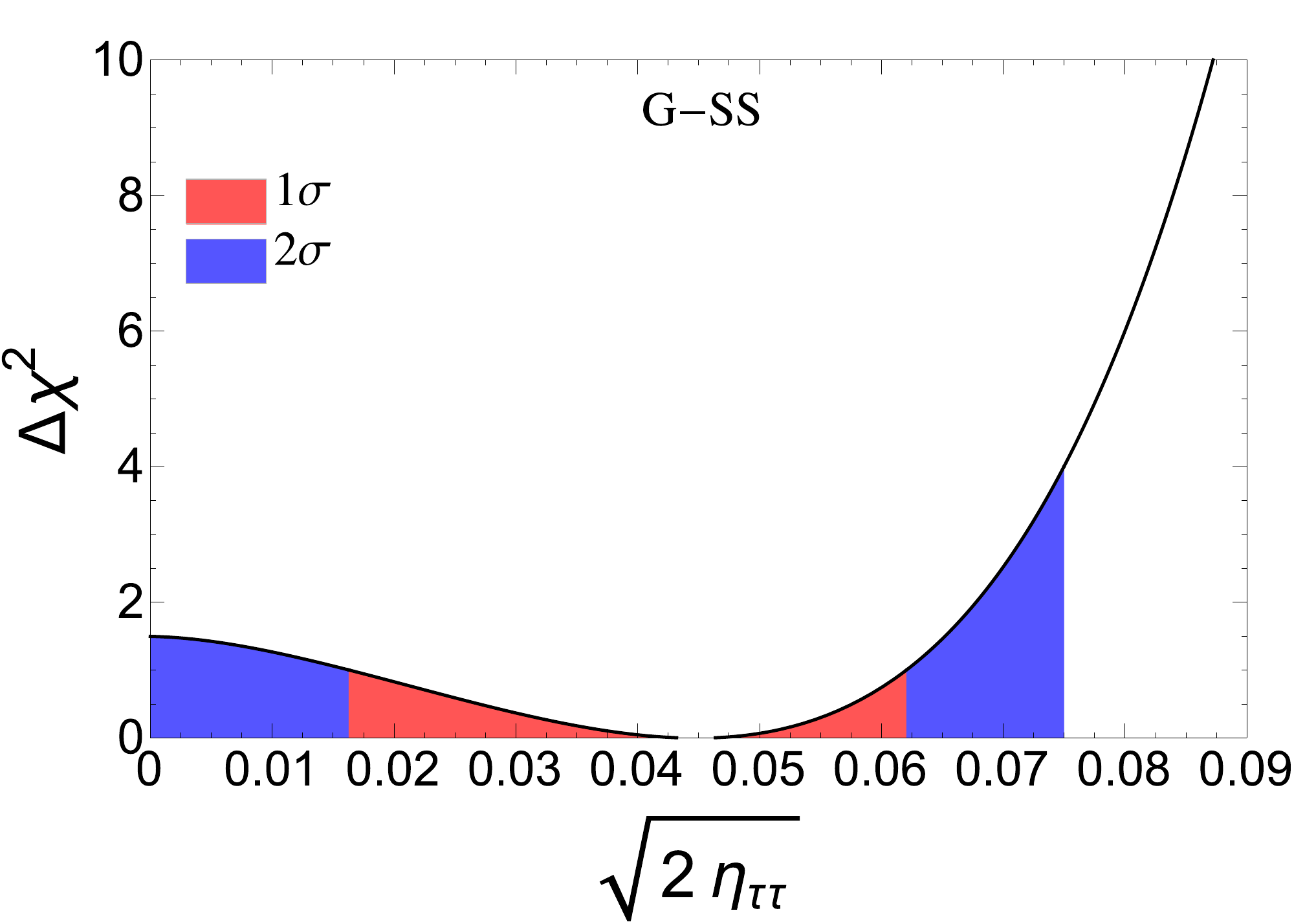} \\
\includegraphics[width=0.32\textwidth]{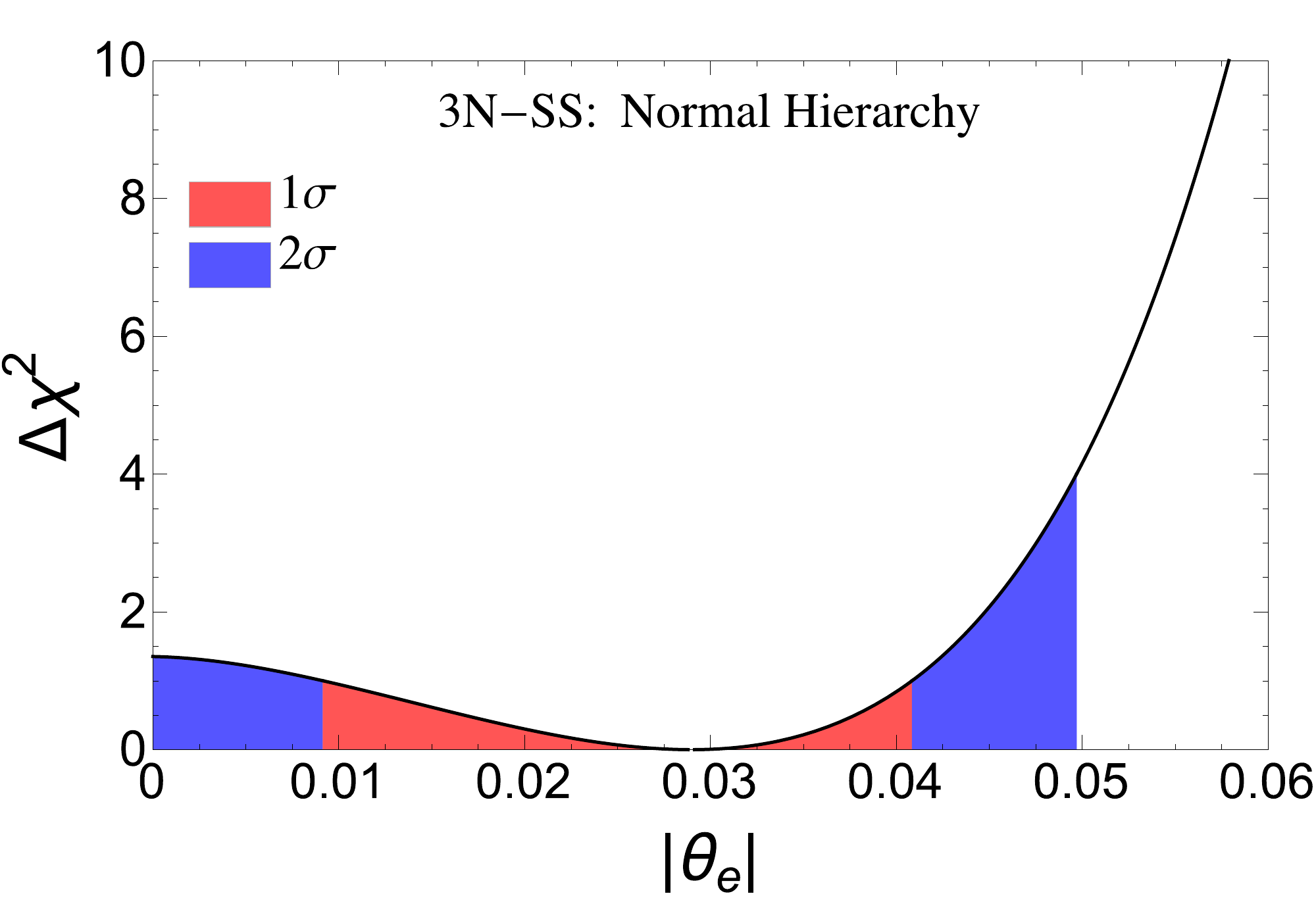}
\includegraphics[width=0.32\textwidth]{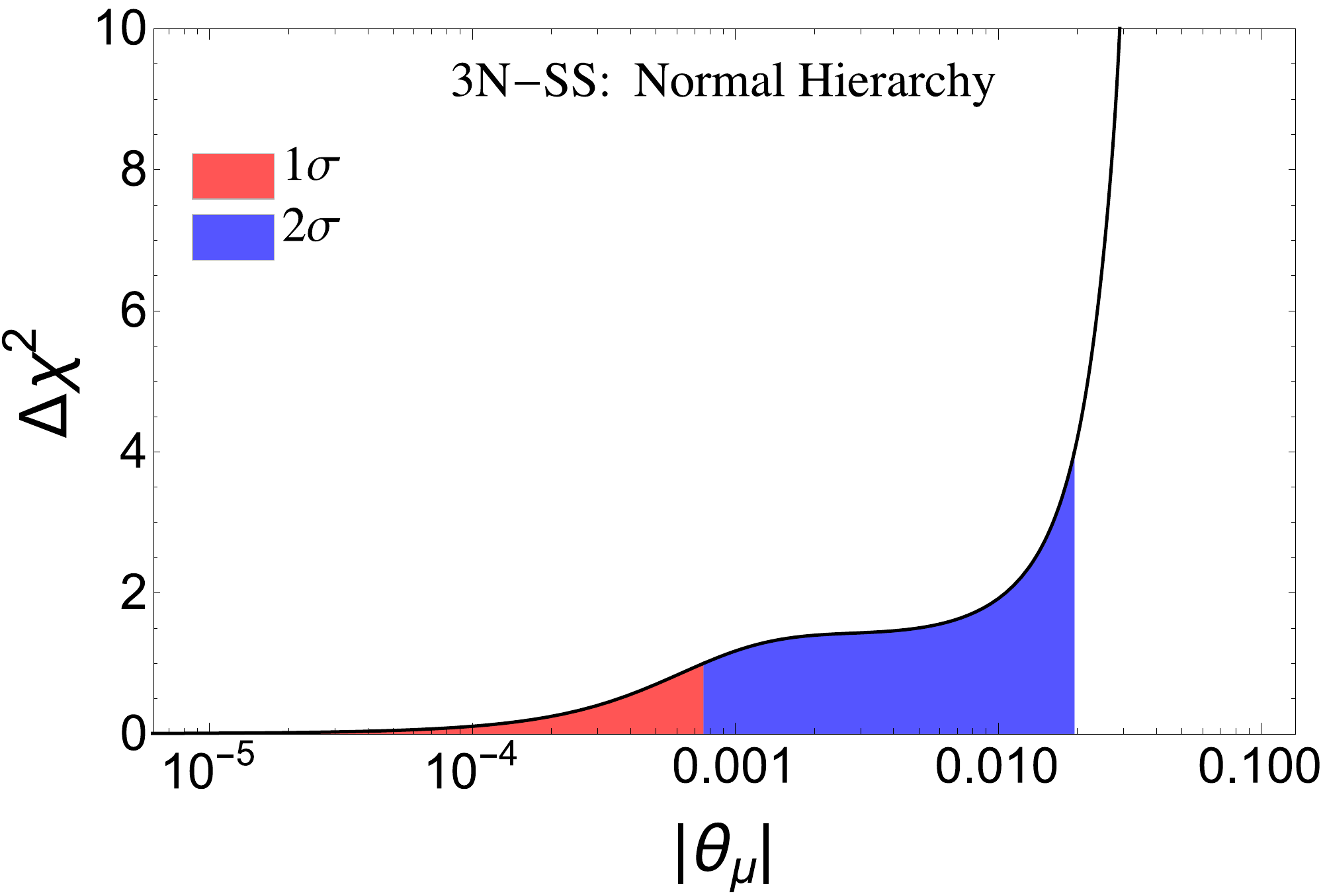}
\includegraphics[width=0.32\textwidth]{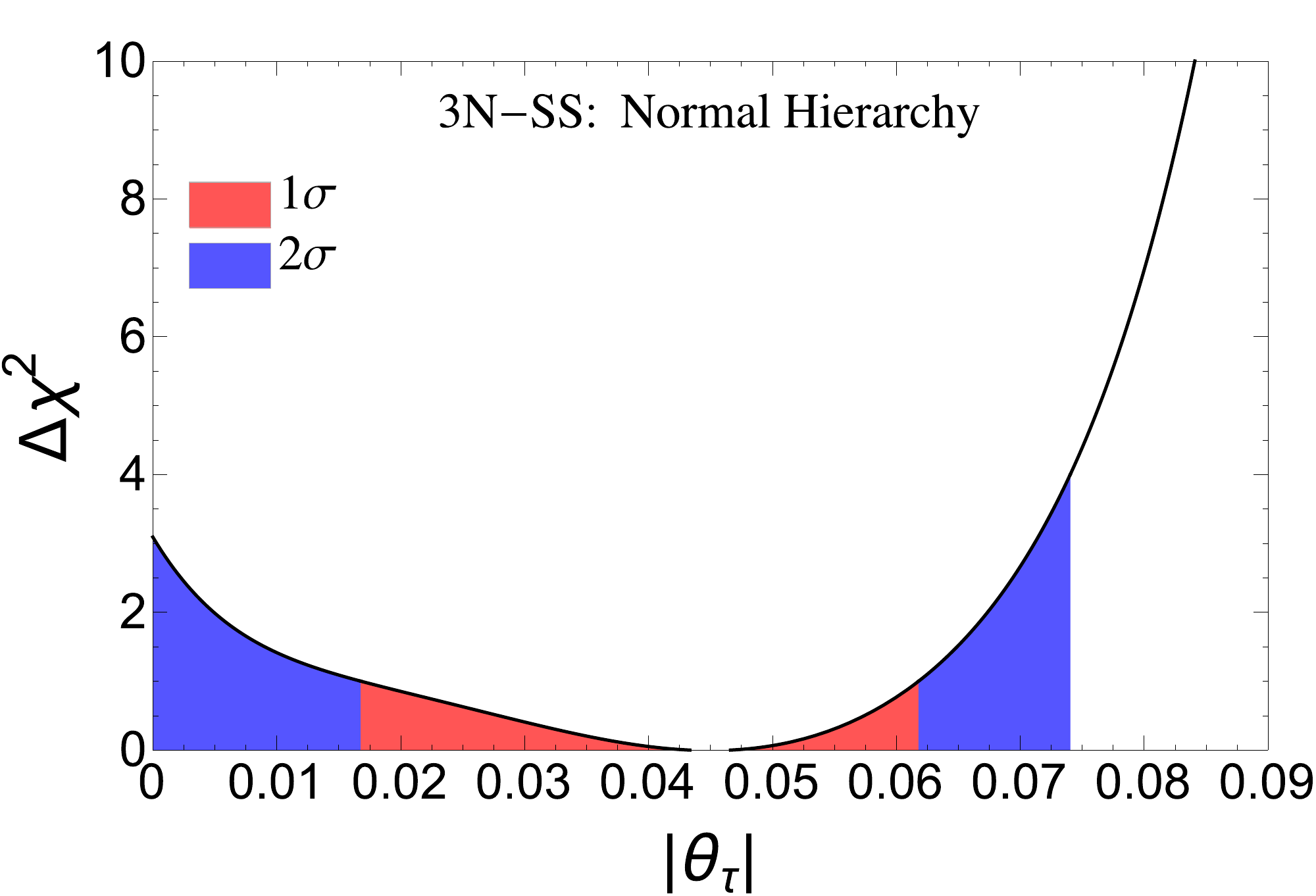} \\
\includegraphics[width=0.32\textwidth]{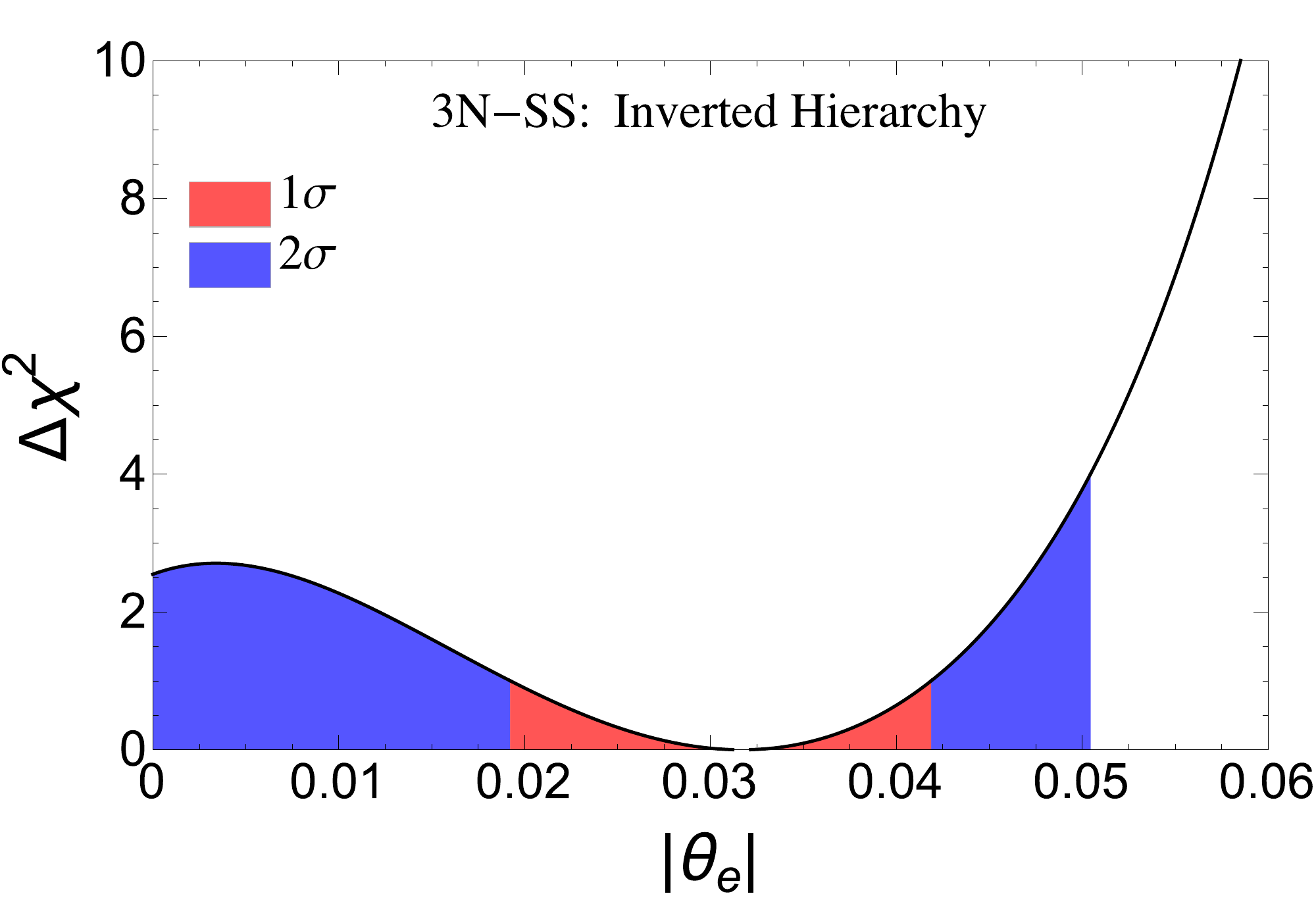} 
\includegraphics[width=0.32\textwidth]{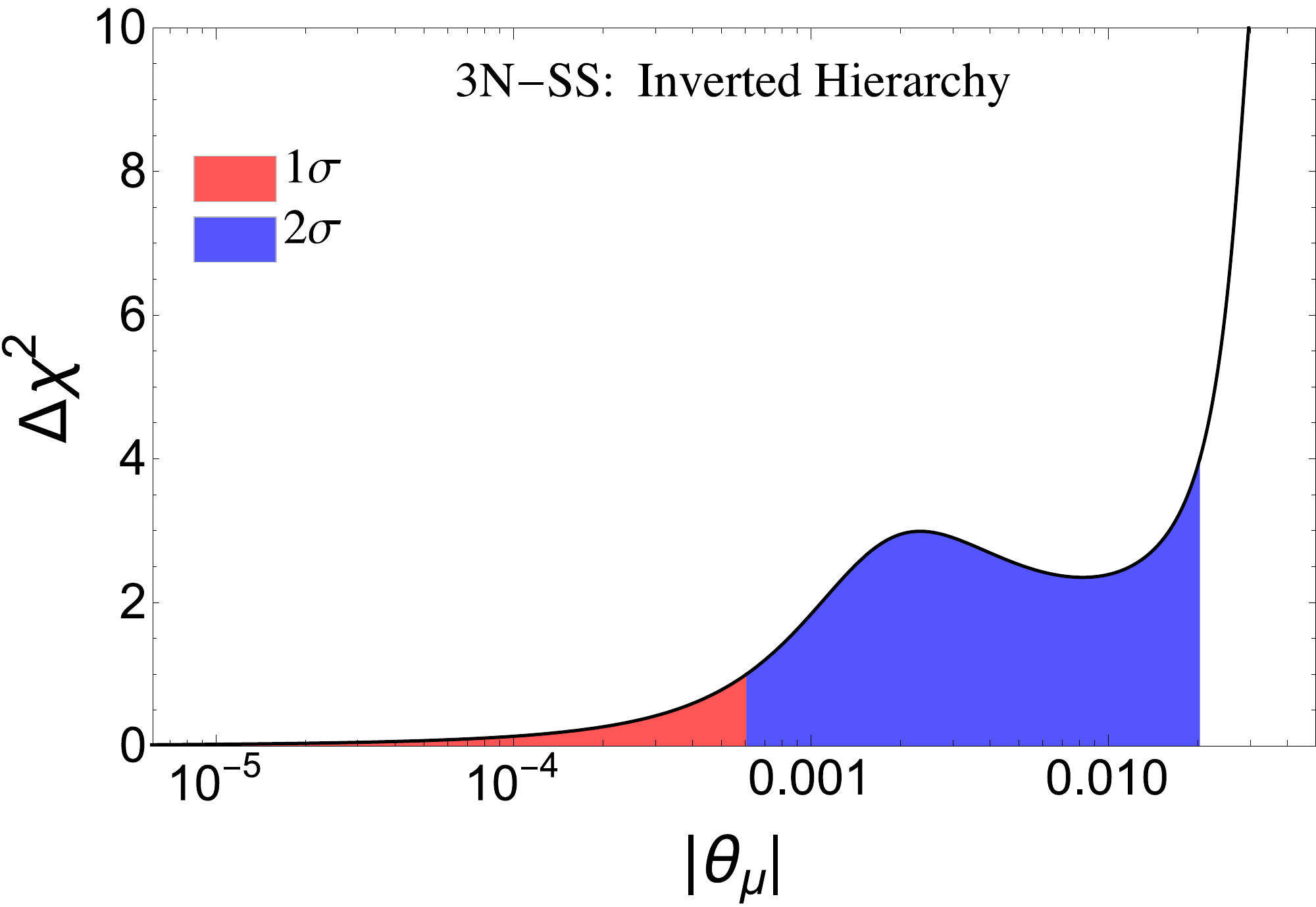} 
\includegraphics[width=0.32\textwidth]{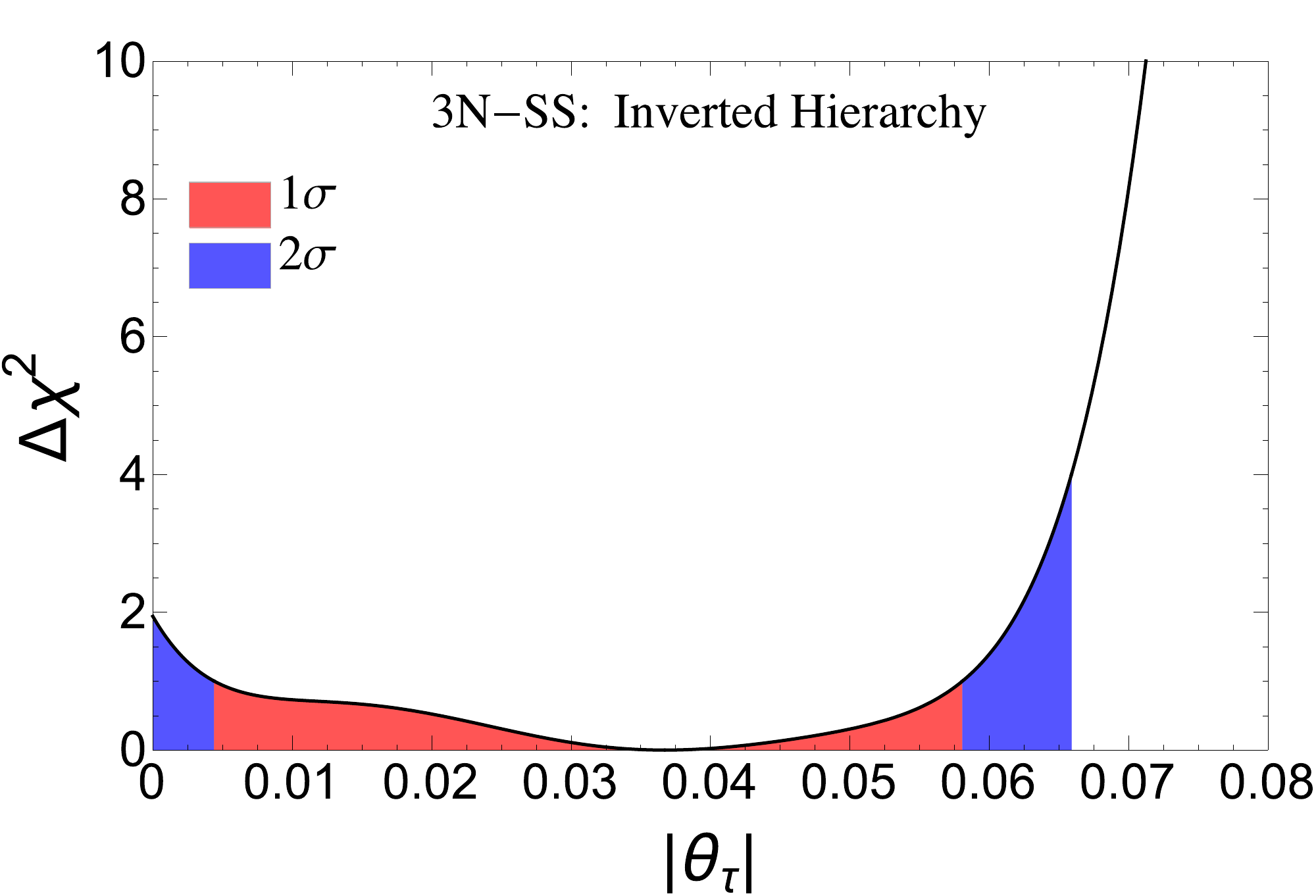}
\caption{$\Delta \chi^2$ profile minimized over all fit variables except for one $\theta_\alpha$ 
(or $\sqrt{2 \eta_{\alpha \alpha}}$) in the case of the G-SS) at a time. The upper panels are for the G-SS, 
and the middle and lower panels for the 3N-SS for a normal and inverted hierarchy respectively.}
\label{fig:chi2}
\end{figure}

\begin{figure}
\centering
\includegraphics[width=0.45\textwidth]{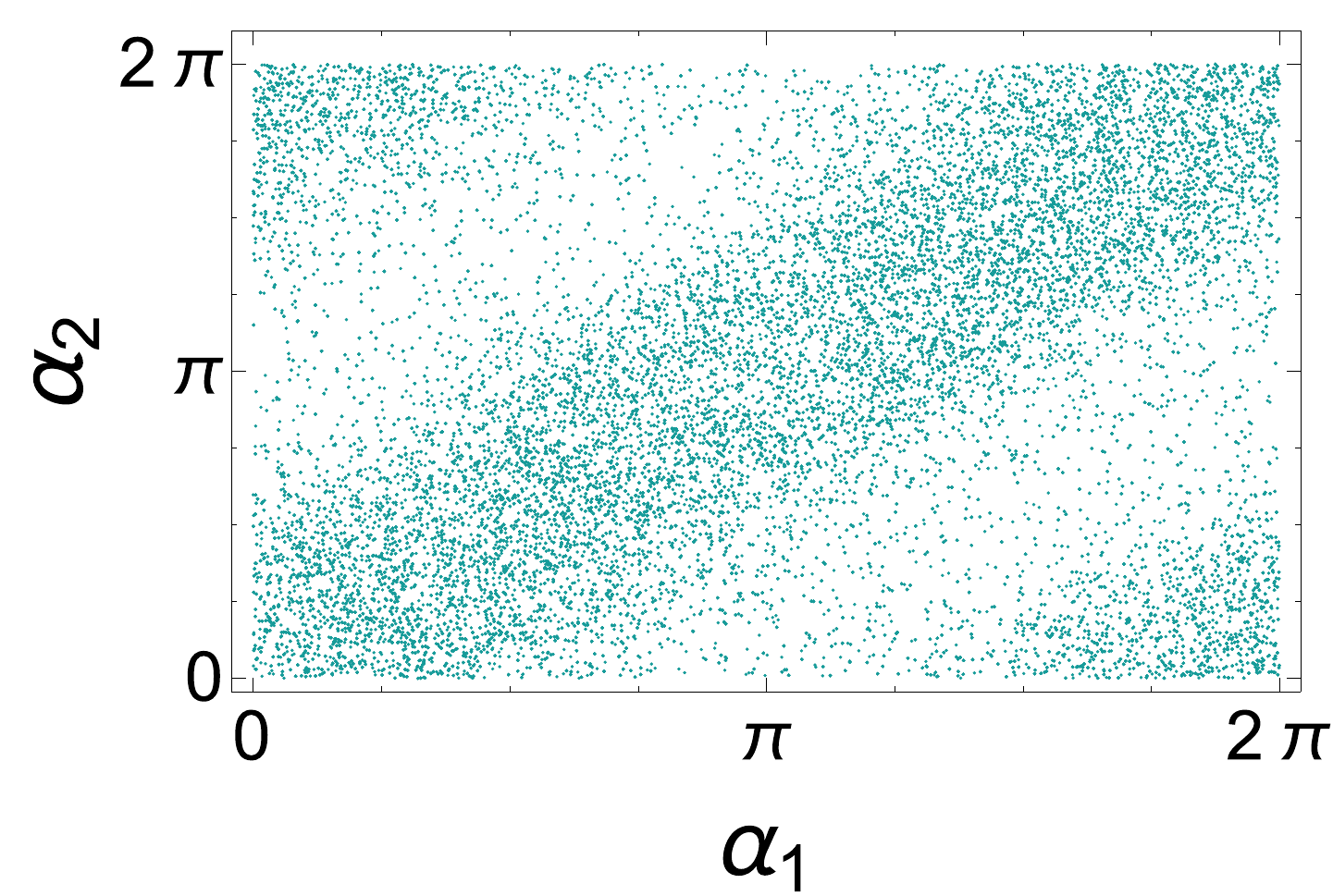}
\includegraphics[width=0.45\textwidth]{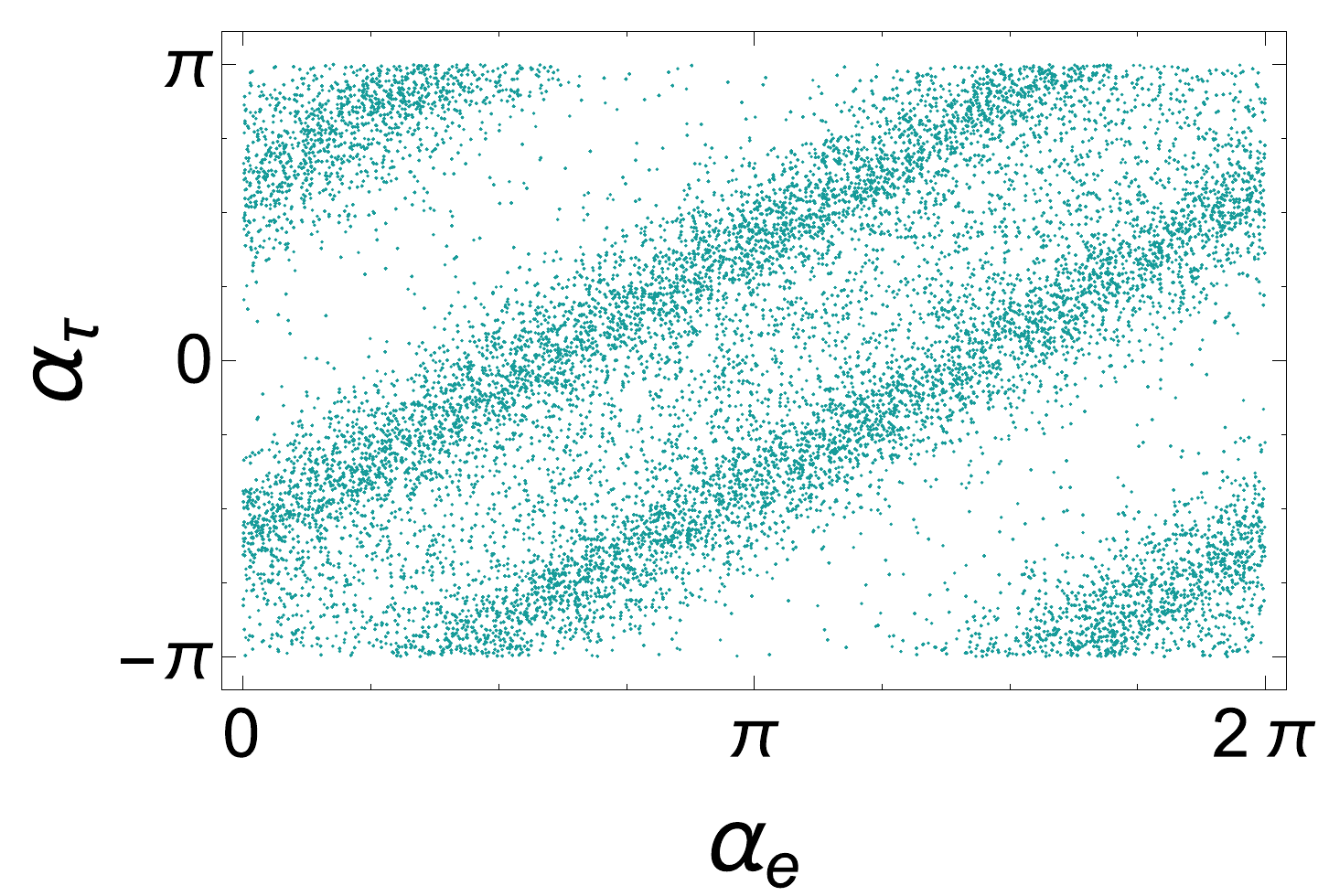}
\caption{Points scanned by the MCMC algorithm with a $\Delta \chi^2 <1$ showing the mild preferred correlation between the two Majorana phases of the PMNS matrix $\alpha_1$ and $\alpha_2$ (left panel) and between the phases of $\theta_e$ and $\theta_\tau$: $\alpha_e$ and $\alpha_\tau$ (right panel) for the 3N-SS and under a normal hierarchy assumption.}
\label{fig:phases}
\end{figure}

To summarize the results of the global fit we present in Fig.~\ref{fig:chi2} the profiles of the $\Delta \chi^2$ 
obtained as a function of the individual $\theta_\alpha$ and minimized over all the other 
parameters. The 1 and $2 \sigma$ regions are colored in red and blue respectively. As can be seen, 
the observables considered (notably the invisible width of the Z and $M_W$) overall show a mild 
(between 1 and $2\sigma$) preference for some degree of non-unitarity $\theta \sim 0.03-0.04$. The constraints 
on the universality of the weak interactions, particularly from ratios of pion and lepton decays, prefer these 
unitarity deviations with non-vanishing mixing with the heavy neutrinos to take place in the electron and tau sectors. 
This preference is clear in the upper panels of Fig.~(\ref{fig:chi2}), which show the constraints for the 
unbounded G-SS. But, even in the more constraint case of a 3N-SS (middle panels for normal hierarchy and lower panels 
for inverted), there is enough freedom to accommodate this general preference shown by the datasets considered. The more characteristic feature that distinguishes the 3N-SS from the G-SS in Fig.~\ref{fig:chi2} is the constraint in $\theta_{\mu}$ which, for the 3N-SS shows a very non-Gaussian behaviour with a very stringent $1\sigma$ limit and a much milder $2 \sigma$ bound comparable to the one found for the G-SS. The reason for the comparatively much stronger $1\sigma$ constraint stems from the very stringent constraint from $\mu \to e \gamma$, which for the 3N-SS imply either a very small $\theta_e$ or $\theta_\mu$. Together with the $1 \sigma$ preference for non-vanishing $\theta_e$, this implies a very strong $1 \sigma$ upper bound for $\theta_\mu$. On the other hand, at the $2 \sigma$ level $\theta_e$ can be arbitrarily small and thus the bound on $\theta_\mu$ from $\mu \to e \gamma$ is evaded. 
Regarding the G-SS, $\mu \to e \gamma$ only constrains the element $\eta_{e \mu}$ and not $\eta_{ee}$ or $\eta_{\mu \mu}$ since, contrary to the 3N-SS, the Schwarz inequality Eq.~(\ref{eq:schwarz}) is not saturated. Regarding $\theta_e$ and $\theta_\tau$, the limits for the 3N-SS and the G-SS are much more similar between them. Indeed, despite the constraint from Eq.~(\ref{eq:Yt}) on $\theta_\tau$, the preferred value for this parameter in the 3N-SS does not show significant deviations with respect to the G-SS. However, non-trivial correlations among the Majorana phases $\alpha_1$ and 
$\alpha_2$ as well as among the phases of $\theta_e$ and $\theta_\tau$: $\alpha_e$ and $\alpha_\tau$ when a normal 
neutrino mass ordering is assumed are required to satisfy Eq.~(\ref{eq:Yt}). These phase correlations are shown in Fig.~\ref{fig:phases}. 
Two interesting features can be observed: (i) 
The values of the PMNS Majorana phases such that $\alpha_1-\alpha_2\sim 2n\pi$ are favoured (left plot); (ii) The
data prefers values for the phases of $\theta_\tau$ and $\theta_e$ which satisfy $\alpha_\tau-\alpha_e\sim \left(2n+1\right)\pi$ (right plot). In the IH case, we have not found any significant correlation among the phases. 

\begin{figure}
\centering
\includegraphics[width=0.32\textwidth]{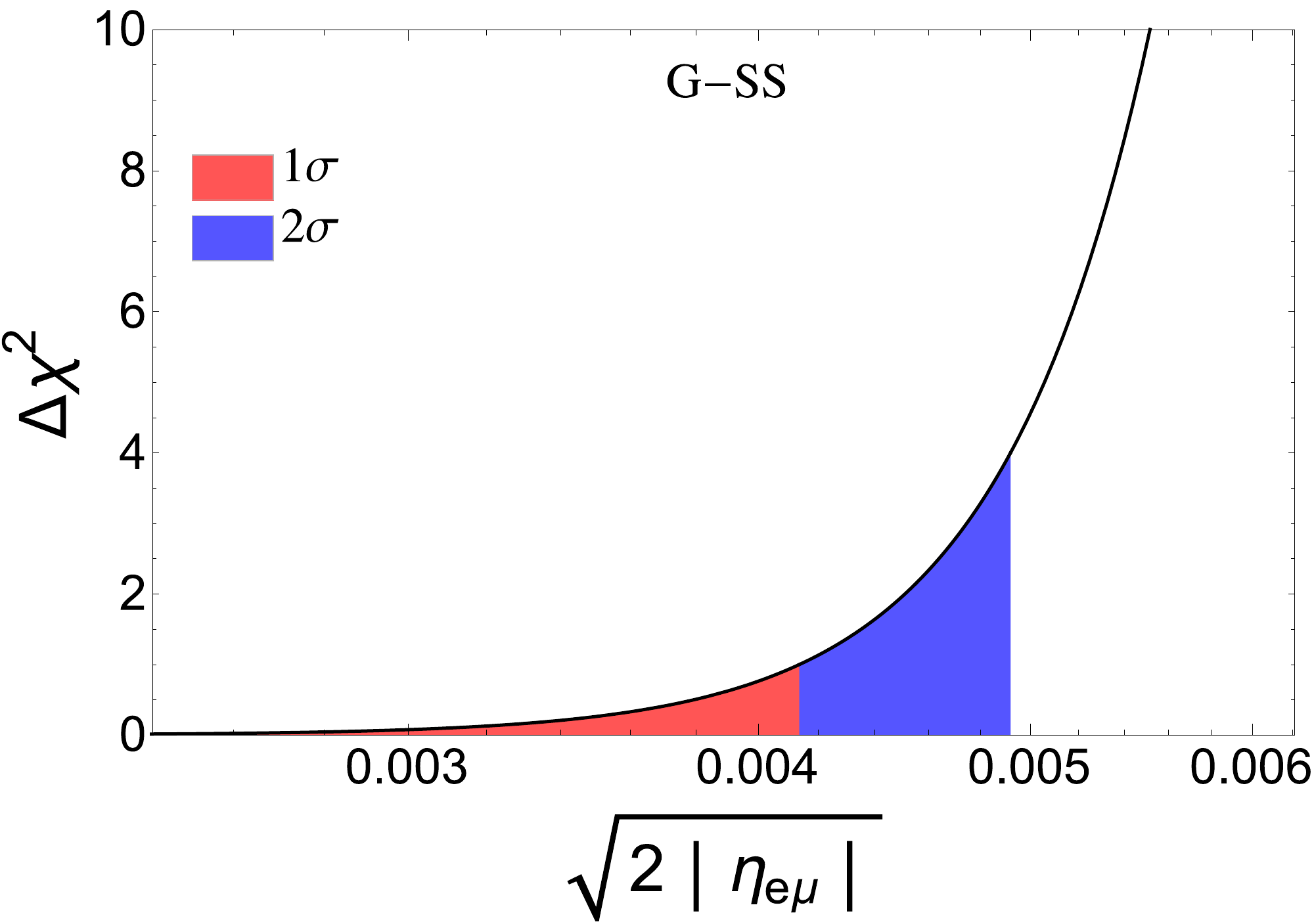}
\includegraphics[width=0.32\textwidth]{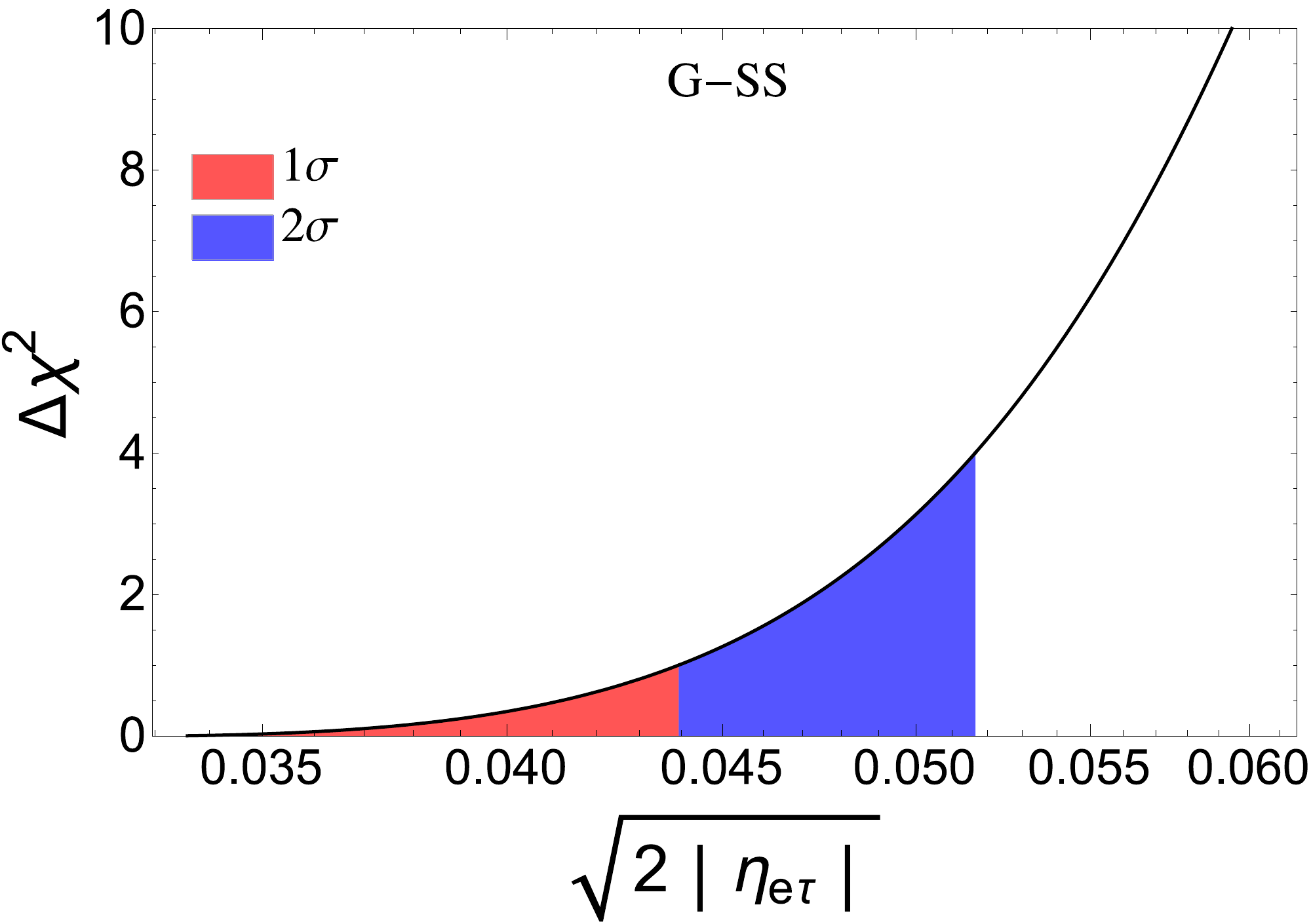}
\includegraphics[width=0.32\textwidth]{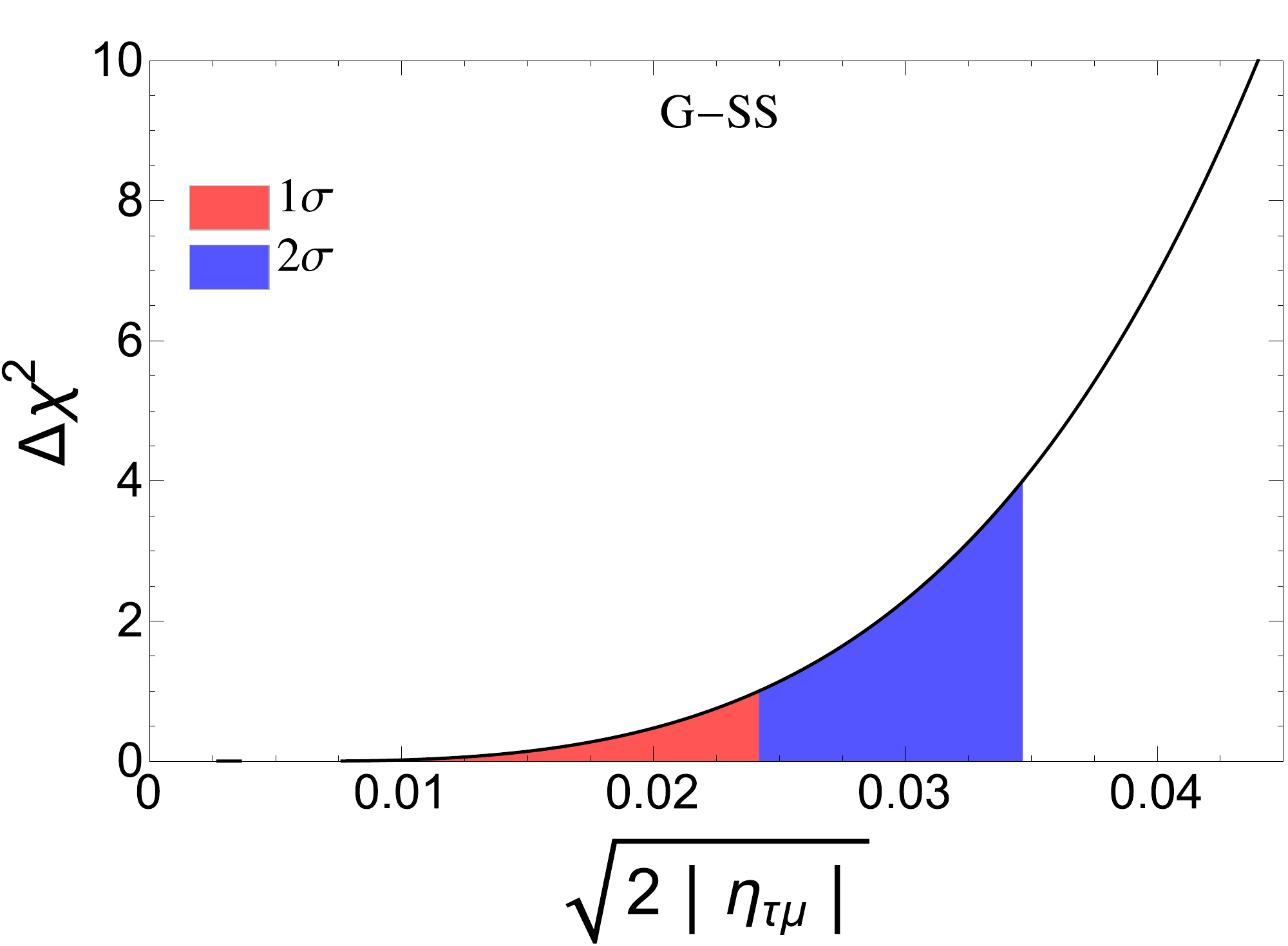} \\
\includegraphics[width=0.32\textwidth]{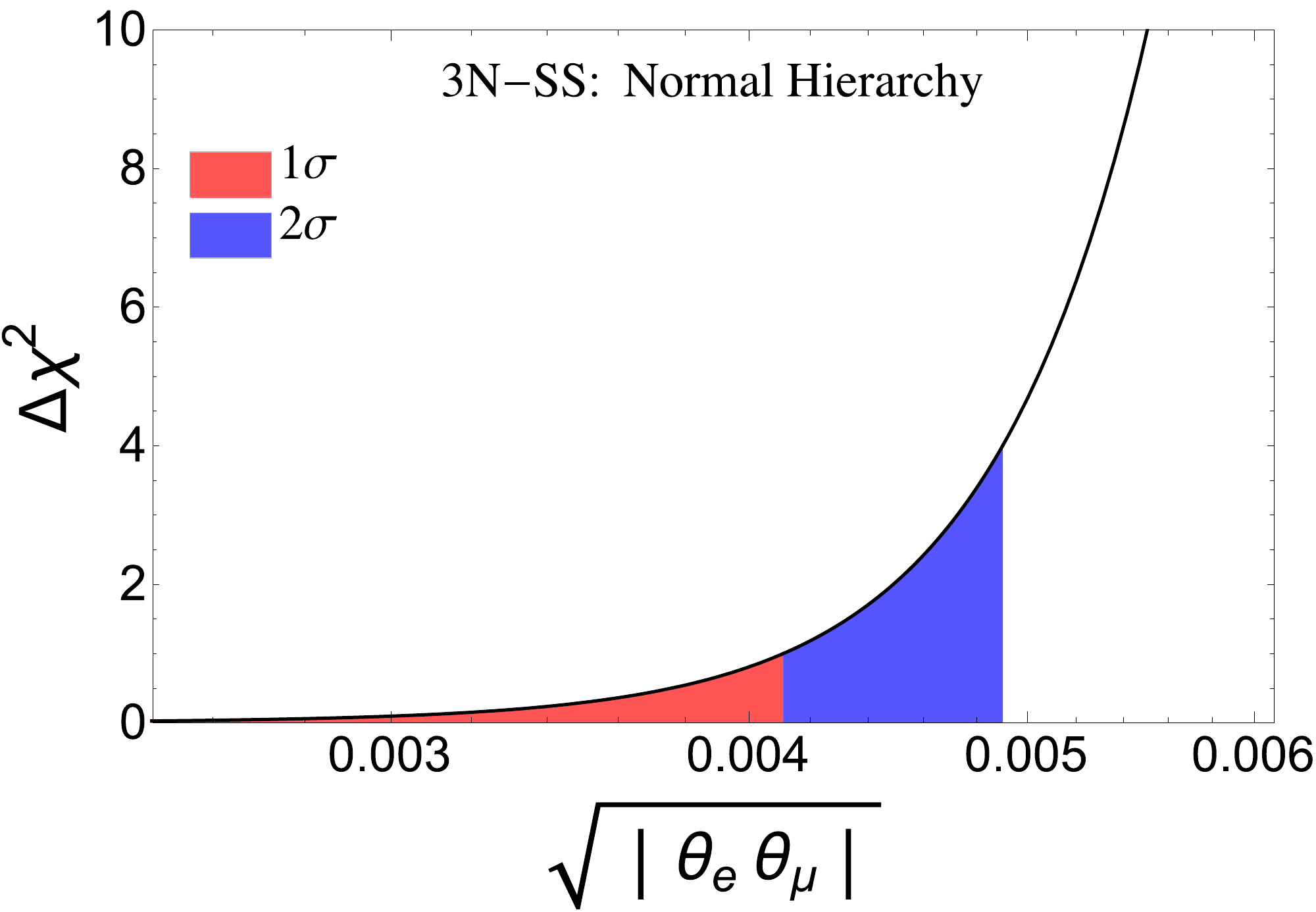}
\includegraphics[width=0.32\textwidth]{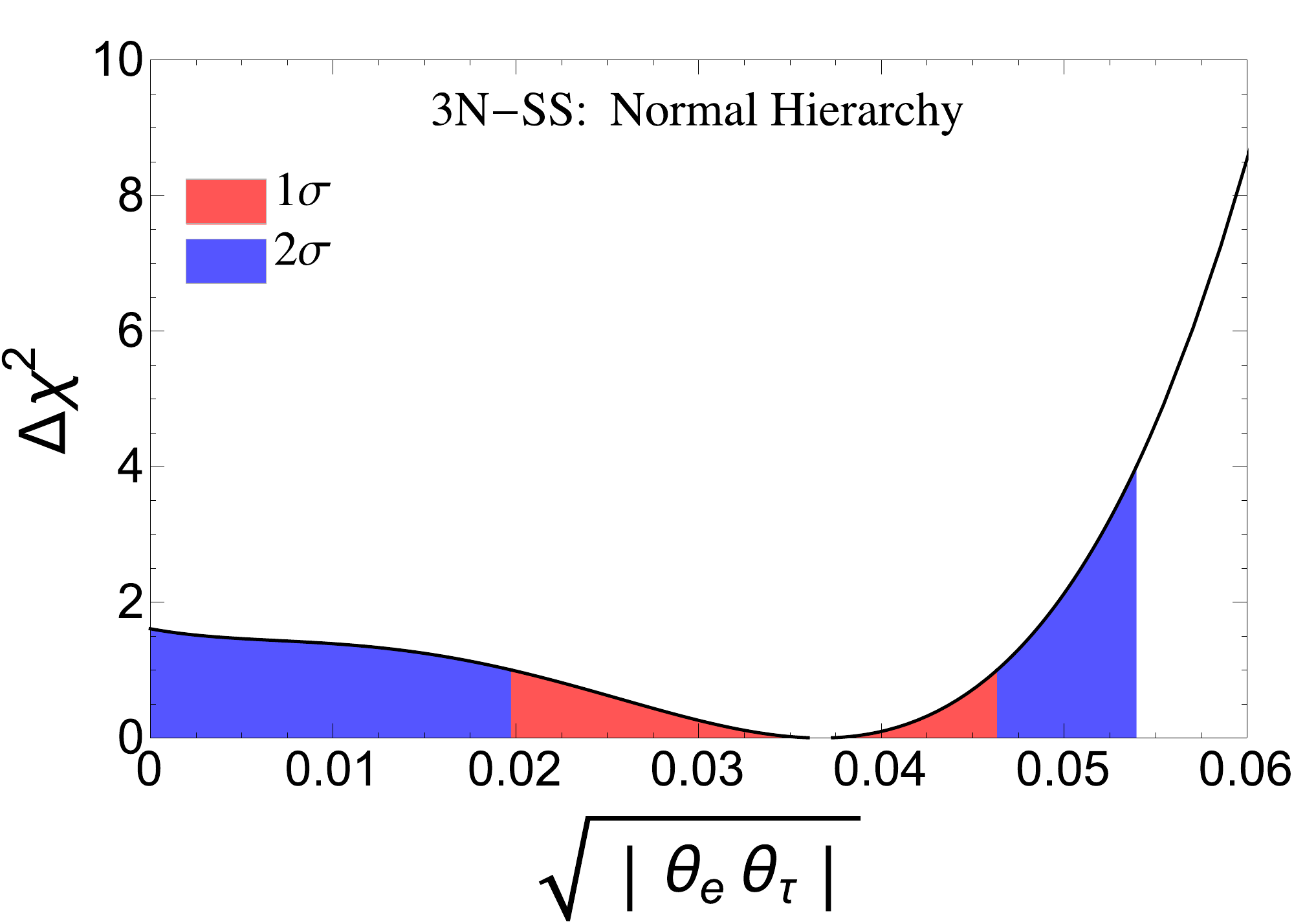}
\includegraphics[width=0.32\textwidth]{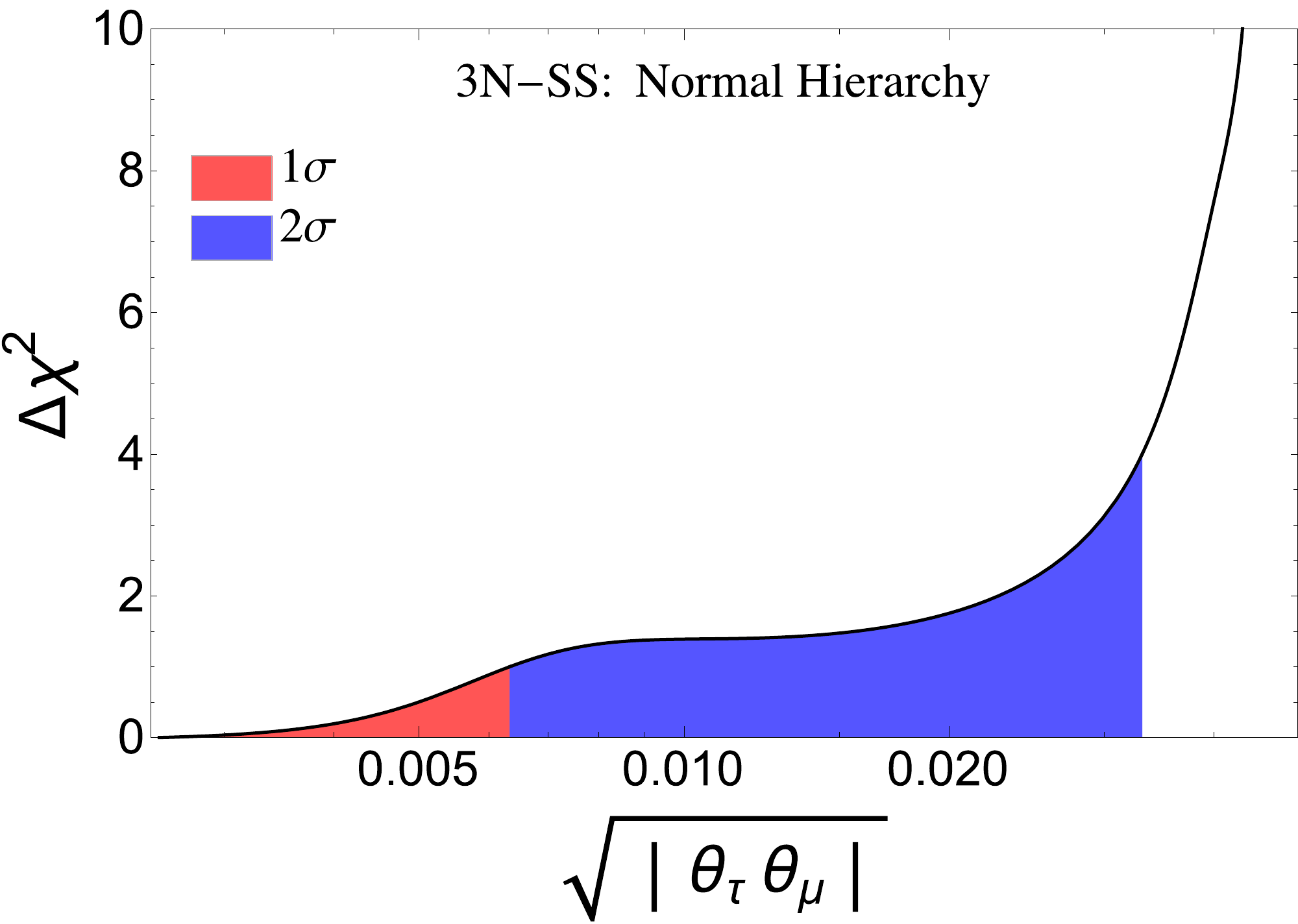} \\
\includegraphics[width=0.32\textwidth]{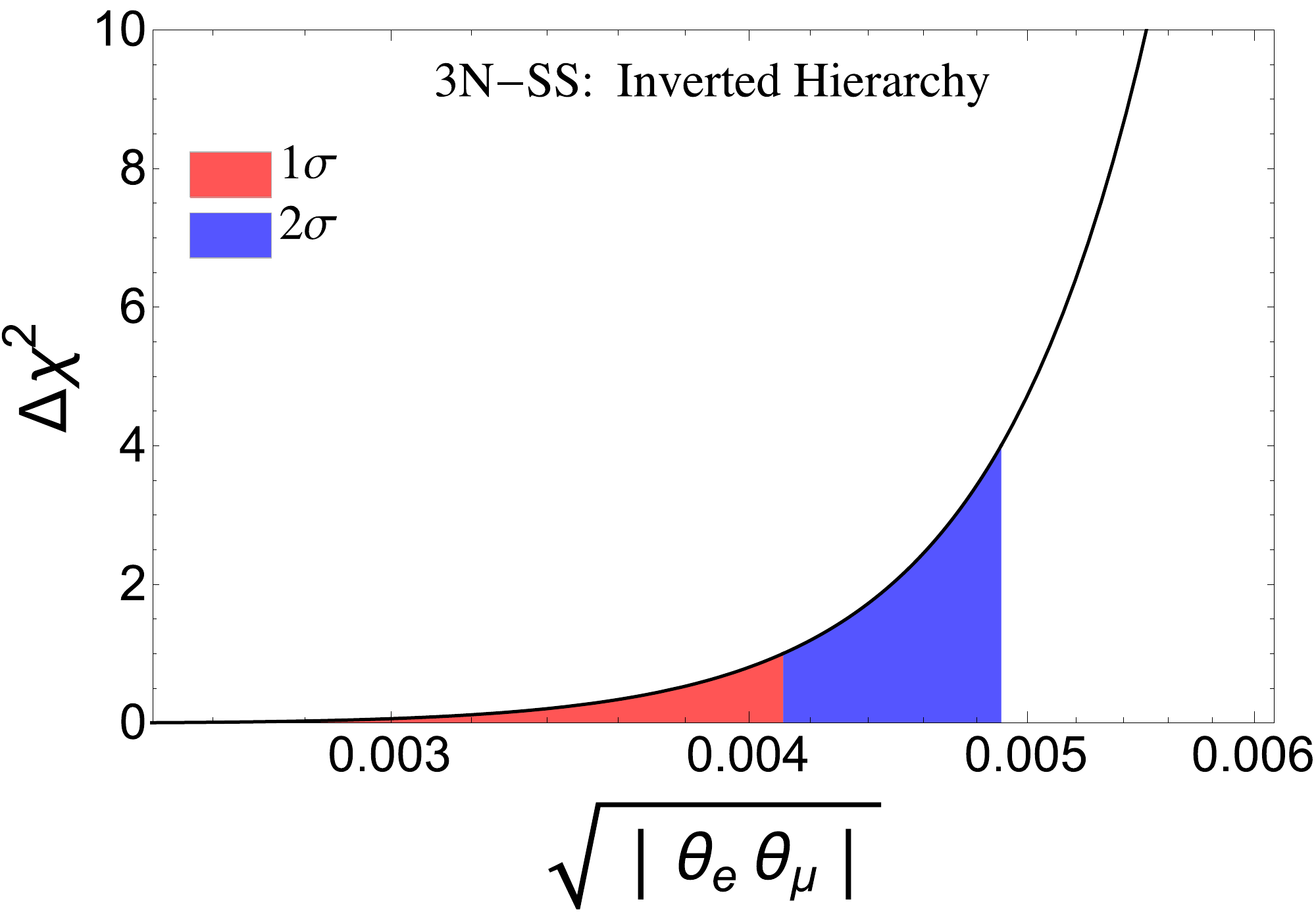} 
\includegraphics[width=0.32\textwidth]{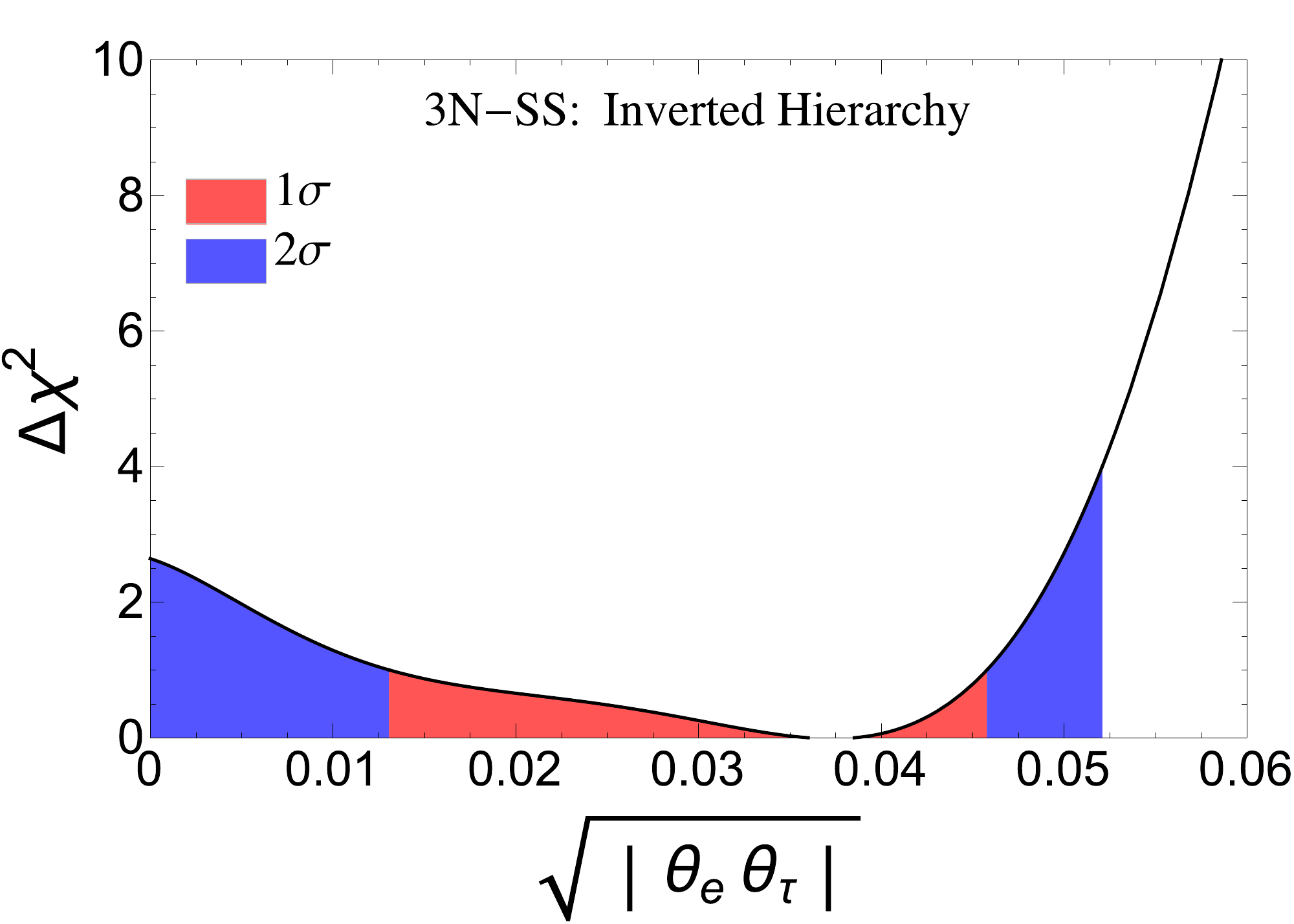} 
\includegraphics[width=0.32\textwidth]{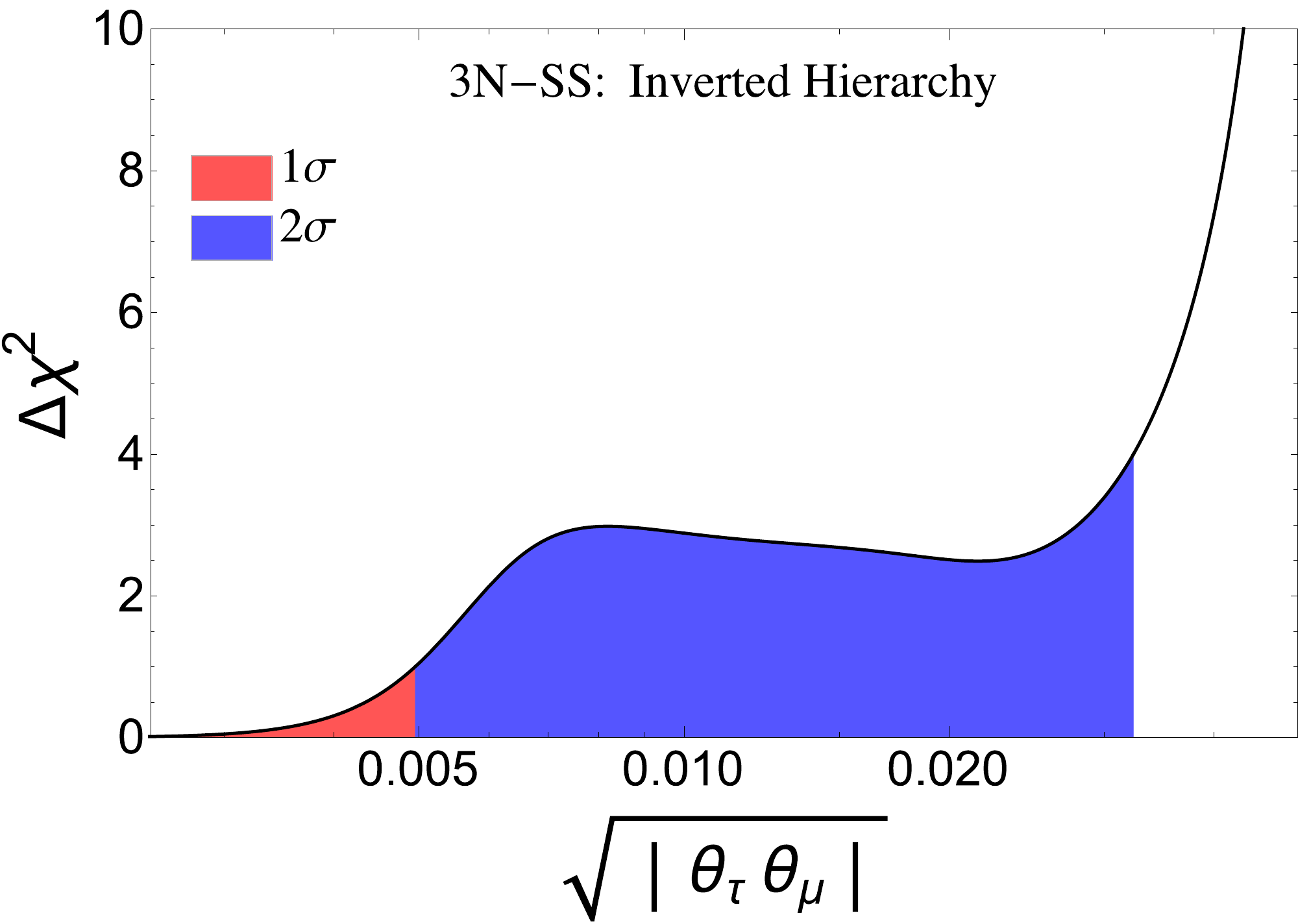}
\caption{Bounds on the off-diagonal entries of $\eta_{\alpha \beta}$ ($|\theta_\alpha \theta_\beta|$ for the 3N-SS). The upper panels are for the G-SS, 
and the middle and lower panels for the 3N-SS for a normal and inverted hierarchy respectively. For the G-SS the strongest limit between the direct bound from radiative LFV decays and the indirect limit from the diagonal entries through the Schwarz inequality is shown for each element.}
\label{fig:chi2off}
\end{figure}

Regarding the off-diagonal elements $|\eta_{\alpha \beta}|$, 
we present in Fig.~\ref{fig:chi2off} the limits obtained from the combination of all observables as
a function of $\sqrt{2|\eta_{\alpha \beta}|}$ and marginalized over all the other parameters for the G-SS 
(upper panels) and the 3N-SS for NH (middle panels) and IH (lower panels). As in Fig.~\ref{fig:chi2}, the 
1 and $2 \sigma$ regions are colored in red and blue respectively. For the G-SS the strongest limit between the direct bound from radiative LFV decays and the indirect limit from the diagonal entries through the Schwarz inequality is shown. For $|\eta_{e \mu}|$ the constraint from $\mu \to e \gamma$ gives the most stringent bound while for $|\eta_{e \tau}|$ and $|\eta_{\mu \tau}|$ the indirect constraints from the lepton flavour conserving (LFC) processes included in the global fit together with the Schwarz inequality Eq.~(\ref{eq:schwarz}) rather dominate. Moreover, the bound on the product $|\theta_e \theta_\tau|$ for the 3N-SS shows a $1\sigma$ preference for a non-zero value. This mild hint can be translated into a prediction for LFV $\tau-e$ transitions, in particular, to a branching ratio of $\tau \to e \gamma$ of $\sim 2.5 \cdot 10^{-10}$ for $|\eta_{e \tau}| \sim 6 \cdot 10^{-4}$. This is rather challenging to probe but not very far from the future sensitivities expected at Super-B factories. 

\section{Discussion and conclusions}
\label{sec:sum}

A global fit to lepton flavour and electroweak precision data has been performed to constrain the size presently 
allowed for the mixing of the extra heavy Seesaw neutrinos with the SM leptons. The analysis has been performed 
both in a completely general Seesaw (G-SS) with the effects of the extra neutrinos encoded in effective operators with no assumed correlations and for the particular case where only three heavy neutrinos are considered (3N-SS). The results of the fit are summarized in Table~\ref{tab:bounds}.  

\begin{table}[htb!]
\centering
\begin{tabular}{|c|c||c|c||c|c|}
\hline
 \multicolumn{2}{|c||}{\multirow{2}{*}{}} & \multicolumn{2}{c||}{G-SS} & \multicolumn{2}{c|}{3N-SS} \\
\cline{3-6}
 \multicolumn{2}{|c||}{}  & LFC & LFV & NH & IH \\ 
\hline
\hline
 \multirow{2}{*}{$\sqrt{2\eta_{ee}}$, $|\theta_{e}|$} & $1\sigma$ & $\mathbf{0.031^{+0.010}_{-0.020}}$ & $-$ & $0.029^{+0.012}_{-0.020}$ & $\mathbf{0.031^{+0.010}_{-0.012}}$\\
\cline{2-6}
& $2\sigma$ & $\mathbf{<0.050}$ & $-$ & $<0.050$ & $\mathbf{<0.050}$\\
\hline
 \multirow{2}{*}{$\sqrt{2\eta_{\mu \mu}}$, $|\theta_{\mu}|$} & $1\sigma$ & $\mathbf{<0.011}$ & $-$ & $\mathbf{<7.6\cdot 10^{-4}}$ & $<6.9\cdot 10^{-4}$\\
\cline{2-6}
& $2\sigma$ & $\mathbf{<0.021}$ & $-$ & $<0.020$ & $\mathbf{<0.023}$\\
\hline
 \multirow{2}{*}{$\sqrt{2\eta_{\tau \tau}}$, $|\theta_{\tau}|$} & $1\sigma$ & $\mathbf{0.044^{+0.019}_{-0.027}}$ & $-$ & $\mathbf{0.043^{+0.018}_{-0.027}}$ & $0.037^{+0.021}_{-0.032}$ \\
\cline{2-6}
& $2\sigma$ & $\mathbf{<0.075}$ & $-$ & $\mathbf{<0.074}$ & $<0.066$\\
\hline
 \multirow{2}{*}{$\sqrt{2\eta_{e \mu}}$, $\sqrt{|\theta_{e} \theta_{\mu}|}$} & $1\sigma$ & $<0.018$ & {\bf $\mathbf{<4.1\cdot 10^{-3}}$} & $\mathbf{<4.1\cdot 10^{-3}}$ & $<4.1\cdot 10^{-3}$\\
\cline{2-6}
& $2\sigma$ & $<0.026$ & {\bf $\mathbf{<4.9\cdot 10^{-3}}$} & $\mathbf{<4.9\cdot 10^{-3}}$ & $<4.9\cdot 10^{-3}$\\
\hline
 \multirow{2}{*}{$\sqrt{2\eta_{e \tau}}$, $\sqrt{|\theta_{e} \theta_{\tau}|}$} & $1\sigma$ & {\bf$\mathbf{<0.045}$} & $<0.107$ & $\mathbf{0.036^{+0.010}_{-0.016}}$ & $0.036^{+0.010}_{-0.023}$\\
\cline{2-6}
& $2\sigma$ & {\bf $\mathbf{<0.052}$} & $<0.127$ & $\mathbf{<0.054}$ & $0.052$\\
\hline
 \multirow{2}{*}{$\sqrt{2\eta_{\mu \tau}}$, $\sqrt{|\theta_{\mu} \theta_{\tau}|}$} & $1\sigma$ & {\bf $\mathbf{<0.024}$} & $<0.115$ & $\mathbf{<0.007}$ & $0.005$\\
\cline{2-6}
& $2\sigma$ & {\bf $\mathbf{<0.035}$} & $<0.137$ & $\mathbf{<0.033}$ & $0.032$\\
\hline
\end{tabular}
\caption{Comparison of all 1 and $2\sigma$ constraints on the heavy-active neutrino mixing. For the 
G-SS the bounds are expressed for $\sqrt{2 \eta_{\alpha \beta}}$ (see Eq.~(\ref{eq:sqrteta})). For the off-diagonal 
entries the indirect bounds from the LFC observables via the Schwarz inequality Eq.~(\ref{eq:schwarz}) are compared 
with the direct LFV bounds and the dominant bound is highlighted in bold face. For the 3N-SS the bounds are shown 
for $\theta_{\alpha}$ for assumptions of a normal (NH) and inverted hierarchy (IH), the less stringent bound is highlighted 
in bold face as an overall bound on the 3N-SS case.}\label{tab:bounds}
\end{table}

For the G-SS with an arbitrary number of extra heavy neutrinos the bounds are expressed in the quantity 
$\sqrt{2 |\eta_{\alpha \beta}|} = \sum_i \sqrt{ \Theta_{\alpha i} \Theta^*_{\beta i}}$ (see Eq.~(\ref{eq:sqrteta})). 
Thus, the diagonal elements $\sqrt{2 \eta_{\alpha \alpha}}$ correspond to the sum (in quadrature) of all mixings $\Theta_{\alpha i}$ of the individual extra heavy neutrinos $N_i$ to a given SM flavour $\alpha$ and represent an upper bound on each individual mixing. The off-diagonal entries, on the other hand, are the combinations that can mediate LFV transitions and even provide extra sources of CP-violation. Notice that, from this definition, $\eta$ is a positive definite matrix and its off-diagonal elements subject to the Schwarz inequality $|\eta_{\alpha \beta}| \leq \sqrt{\eta_{\alpha \alpha} \eta_{\beta \beta}}$.

In the case of the 3N-SS, only one mixing parameter $\theta_\alpha$ per SM flavour $\alpha$ can be large enough to 
saturate the bounds derived here, so as to comply with our present constraints on light neutrino masses and mixings 
from neutrino oscillation data (see discussion in Section \ref{sec:param}). Thus, the Schwarz inequality is saturated to an equality for the 3N-SS.
Furthermore, some non-trivial correlations between the parameters $\theta_\alpha$ are also present (see Eq.~(\ref{eq:Yt})).

As shown in Table~\ref{tab:bounds} the data show a mild, between $1$ and $2\sigma$ preference, for non-zero heavy-active 
mixing of order $\sim 0.03-0.04$ in the $e$ and $\tau$ sectors. At the $2 \sigma$ level, upper bounds in all mixing 
parameters are found. The most stringent one $\sim 0.02$ is found for the mixing with muons, followed by $\sim 0.05$ for 
electrons and $\sim 0.07$ for taus. Regarding the off diagonal entries, for the G-SS the indirect bounds from LFC processes 
can be compared with the direct constraints from LFV observables. Interestingly, the constraint from $\mu \to e \gamma$ strongly 
dominates over all others leading to a bound one order of magnitude better $\sim 0.005$ in the $e-\mu$ entry, while the $e-\tau$ and $\mu-\tau$ values are rather dominated by the indirect constraints from the Schwarz inequality (comparison between the LFC and LFV columns). Regarding the 3N-SS, even though the necessity of correctly reproducing the observed neutrino mass and mixing pattern introduces non-trivial correlations among the $\theta_\alpha$ and the neutrino masses and mixings (dependence on normal or inverted hierarchy assumptions shown in the comparison of the third and fourth columns), there is still enough freedom to obtain very similar bounds to those found for the G-SS. This however implies some non-trivial correlations preferred at $1\sigma$ notably among the PMNS matrix Majorana phases as well as among the phases of $\theta_e$ and $\theta_\tau$ as shown in Fig.~\ref{fig:phases}.   

The bounds derived here represent the most updated set of constraints and compare well with previous studies. Notably, it is interesting to compare with another recent global fit presented in Ref.~\cite{Antusch:2014woa} were bounds to the G-SS were also studied. We find that the agreement between the two sets of constraints is generally good. The same preference for non-zero mixing in the electron and tau sectors was found but in their case the preferred value is slightly ($\sim 20-30\%$) larger. Similarly the upper bound on muon mixing is weaker in Ref.~\cite{Antusch:2014woa}. Conversely the limits on the off-diagonal elements are slightly ($\sim 20-40\%$) stronger in Ref.~\cite{Antusch:2014woa} for the $e-\tau$ and $\mu-\tau$ sectors. The only very noticeable difference is in the $e-\mu$ sector where the limit from $\mu \to e \gamma$ is almost a factor 3 stronger than the one presented here (despite not being yet updated to the final MEG result). This difference can be attributed to not considering the propagation of the heavy neutrinos in the loop for the process which tends to restore the GIM cancellation (given the Unitarity of the full mixing matrix) and to therefore slightly weaken by the corresponding factor the bound stemming from the process. This extra contribution was not taken into account in Ref.~\cite{Antusch:2014woa} since a more agnostic source of the non-unitarity of the PMNS matrix was adopted while here we concentrate in constraining heavy neutrino mixings. The rest of the discrepancies can stem from small differences in our analyses. For example our observables for weak lepton universality and CKM unitarity are more updated and our bounds correspond to frequentist confidence regions while Ref.~\cite{Antusch:2014woa} rather presented Bayesian credible intervals. Regarding the 3N-SS, the closest study of a similar setup in the literature is that of Ref.~\cite{Antusch:2015mia}. This work is rather complementary to our results focusing instead in the region between 10 to 250~GeV, where more stringent constraints are derived since the extra neutrinos would be kinematically accessible.

It is also interesting to translate the bounds derived here to other common parametrizations, useful in particular for the analysis of neutrino non-standard interactions (see e.g. Ref.~\cite{Ohlsson:2012kf}). Indeed, the non-unitary PMNS matrix induced by the mixing with the extra heavy neutrinos modifies the neutrino production and detection processes, which can be encoded in production/detection NSI~\cite{FernandezMartinez:2007ms,Antusch:2008tz}. In particular:
\begin{equation}
|\varepsilon^{p,d}_{\alpha \beta}| = |\eta_{\alpha \beta}| \leq \left(
\begin{array}{ccc}
1.3 \cdot 10^{-3}  & 1.2 \cdot 10^{-5} & 1.4 \cdot 10^{-3} \\ 
1.2 \cdot 10^{-5}  & 2.2 \cdot 10^{-4} & 6.0 \cdot 10^{-4} \\ 
1.4 \cdot 10^{-3}  & 6.0 \cdot 10^{-4} & 2.8 \cdot 10^{-3} 
\end{array}
\right) .
\label{eq:NSI}
\end{equation} 
Furthermore, neutrino interactions with matter are also affected and these effects can also be described by matter NSI~\cite{Antusch:2008tz}:
\begin{eqnarray}
\varepsilon^m_{\alpha\beta} = 2 \eta_{\alpha e} \delta_{\beta e} + 2 \eta_{e \beta}  \delta_{e \alpha} - \frac{n_n}{n_e} \, 2 \eta_{\alpha \beta} ,
\end{eqnarray}
where $n_e$ and $n_n$ are the electron and neutron densities of the matter traversed by the neutrinos.

Finally, an alternative parametrization of the non-unitarity of the PMNS matrix of the form $N = TU$ with $T$ a lower triangular matrix~\cite{Xing:2007zj,Xing:2011ur,Escrihuela:2015wra}: 
\begin{equation}
T = \left(
\begin{array}{ccc}
\alpha_{ee}  & 0 & 0 \\ 
\alpha_{\mu e}  & \alpha_{\mu \mu} & 0\\ 
\alpha_{\tau e}  & \alpha_{\tau \mu} & \alpha_{\tau \tau} 
\end{array}
\right) 
\label{eq:triang}
\end{equation}
is also considered appropriate to study the effects of non-unitary PMNS mixing in neutrino oscillation 
searches~\cite{FernandezMartinez:2007ms,Antusch:2009pm,Parke:2015goa,Miranda:2016wdr,Ge:2016xya}. Comparing Eqs.~(\ref{eq:Neq}) and~(\ref{eq:triang}) it is easy to see 
that $\alpha_{\beta \beta} \approx 1 - \eta_{\beta \beta}$, while $|\alpha_{\beta \gamma}| \approx 2|\eta_{\beta\gamma}|=|\epsilon_{\beta\gamma}|$ so 
that the bounds derived here can be trivially translated to this parametrization too. All in all this level of 
non-unitarity (or equivalently NSI as in Eq.~(\ref{eq:NSI})) is extremely tough to probe at present or near-future 
neutrino oscillations facilities and its effects would be negligible. However, prospective very precise neutrino 
oscillation facilities such as the Neutrino factory~\cite{Geer:1997iz,DeRujula:1998umv} could probe 
beyond this very stringent present limits for some elements~\cite{FernandezMartinez:2007ms,Antusch:2009pm}.

Notice that the bounds derived here apply for any heavy neutrino mass above the electroweak scale. For 
lighter heavy neutrino masses, the LFV radiative decays start to be suppressed by 
the restoration of the GIM mechanism (see Eq.~(\ref{eq:muegamma})) and therefore the constraints shown in the LFV column of Table~\ref{tab:bounds} are not valid. 
The rest of the bounds summarized in the LFC column of Table~\ref{tab:bounds} do apply down to $\mathcal{O}$(500 MeV) with the only exception of the invisible width of the $Z$, since for masses below $\sim M_Z/2$ the heavy neutrinos can be kinematically produced and unitarity is restored. Therefore, in the region between the Kaon mass and the EW scale we do not expect any significant change in the G-SS bounds shown in the LFC column of Table~\ref{tab:bounds}. Nevertheless, at these lower energies were the extra neutrinos can be directly produced, more stringent constraints than the ones derived here, from direct searches~\cite{Atre:2009rg,Ruchayskiy:2011aa,Drewes:2015iva,Parke:2015goa,deGouvea:2015euy,Deppisch:2015qwa} and cosmology~\cite{Dolgov:2003sg,Cirelli:2004cz,Melchiorri:2008gq,Hannestad:2012ky,Ruchayskiy:2012si,Mirizzi:2012we,Jacques:2013xr,Saviano:2013ktj,Archidiacono:2013xxa,Mirizzi:2013gnd,Hernandez:2013lza,Vincent:2014rja,Hernandez:2014fha} apply.

In summary, we have combined present probes on weak lepton universality, searches for LFV processes and precision 
electroweak observables to derive updated and global constraints on the allowed mixing of heavy Seesaw neutrinos with 
the SM fermions. These bounds apply for any value of the Majorana scale larger than the electroweak scale and have been 
computed both for a completely general scenario as well as for the case in which only 3 extra heavy neutrinos are 
considered. At the $1\sigma$ level a mild preference for non-zero mixing in the electron and tau sectors around 
$0.03-0.04$ was found, which could be probed for by improving the LFC searches that currently lead to that preference, as well as through $\tau-e$ LFV transitions. 
At the $2\sigma$ level, upper bounds between $10^{-1}$ and $10^{-2}$ for all elements were 
derived with a most stringent constraint on the mixing in the $e-\mu$ sector an order of magnitude better from the 
$\mu \to e \gamma$ process. While this is by far the present dominant bound, it will be superseded in the future 
by $\mu \to eee$ and/or $\mu-e$ conversion in nuclei searches. Apart from this and other improvements in the datasets 
considered, this level of mixing is challenging but still plausible to probe at future collider~\cite{Abada:2014cca,Antusch:2015mia,Antusch:2016brq,Antusch:2016vyf} and dedicated neutrino oscillation searches~\cite{FernandezMartinez:2007ms,Antusch:2009pm}.
 
\begin{acknowledgments}
We warmly thank Luca di Luzio, Michele Lucente, Xabier Marcano, Carlos Pena and Maria Jose Herrero for very useful discussions.
We acknowledge financial support by the European Union through the ITN ELUSIVES H2020-MSCA-ITN-2015//674896 and the RISE INVISIBLESPLUS H2020-MSCA-RISE-2015//690575. EFM and JHG also acknowledge 
support from the EU through the FP7 Marie Curie Actions CIG NeuProbes PCIG11-GA-2012-321582 and the Spanish MINECO through the 
``Ram\'on y Cajal'' programme (RYC2011-07710), the HPC-Hydra cluster at IFT, the project FPA2012-31880 and through the Centro de excelencia Severo Ochoa Program 
under grant SEV-2012-0249. EFM also acknowledges the support of the COST action EuroNuNet CA15139. JLP also acknowledges support by the Marie Curie CIG program, project number 
PCIG13-GA-2013-618439.

\end{acknowledgments}


\begin{thebibliography}{117}%
\makeatletter
\providecommand \@ifxundefined [1]{%
 \@ifx{#1\undefined}
}%
\providecommand \@ifnum [1]{%
 \ifnum #1\expandafter \@firstoftwo
 \else \expandafter \@secondoftwo
 \fi
}%
\providecommand \@ifx [1]{%
 \ifx #1\expandafter \@firstoftwo
 \else \expandafter \@secondoftwo
 \fi
}%
\providecommand \natexlab [1]{#1}%
\providecommand \enquote  [1]{``#1''}%
\providecommand \bibnamefont  [1]{#1}%
\providecommand \bibfnamefont [1]{#1}%
\providecommand \citenamefont [1]{#1}%
\providecommand \href@noop [0]{\@secondoftwo}%
\providecommand \href [0]{\begingroup \@sanitize@url \@href}%
\providecommand \@href[1]{\@@startlink{#1}\@@href}%
\providecommand \@@href[1]{\endgroup#1\@@endlink}%
\providecommand \@sanitize@url [0]{\catcode `\\12\catcode `\$12\catcode
  `\&12\catcode `\#12\catcode `\^12\catcode `\_12\catcode `\%12\relax}%
\providecommand \@@startlink[1]{}%
\providecommand \@@endlink[0]{}%
\providecommand \url  [0]{\begingroup\@sanitize@url \@url }%
\providecommand \@url [1]{\endgroup\@href {#1}{\urlprefix }}%
\providecommand \urlprefix  [0]{URL }%
\providecommand \Eprint [0]{\href }%
\providecommand \doibase [0]{http://dx.doi.org/}%
\providecommand \selectlanguage [0]{\@gobble}%
\providecommand \bibinfo  [0]{\@secondoftwo}%
\providecommand \bibfield  [0]{\@secondoftwo}%
\providecommand \translation [1]{[#1]}%
\providecommand \BibitemOpen [0]{}%
\providecommand \bibitemStop [0]{}%
\providecommand \bibitemNoStop [0]{.\EOS\space}%
\providecommand \EOS [0]{\spacefactor3000\relax}%
\providecommand \BibitemShut  [1]{\csname bibitem#1\endcsname}%
\let\auto@bib@innerbib\@empty
\bibitem [{\citenamefont {Fukugita}\ and\ \citenamefont
  {Yanagida}(1986)}]{Fukugita:1986hr}%
  \BibitemOpen
  \bibfield  {author} {\bibinfo {author} {\bibfnamefont {M.}~\bibnamefont
  {Fukugita}}\ and\ \bibinfo {author} {\bibfnamefont {T.}~\bibnamefont
  {Yanagida}},\ }\href {\doibase 10.1016/0370-2693(86)91126-3} {\bibfield
  {journal} {\bibinfo  {journal} {Phys. Lett.}\ }\textbf {\bibinfo {volume}
  {B174}},\ \bibinfo {pages} {45} (\bibinfo {year} {1986})}\BibitemShut
  {NoStop}%
\bibitem [{\citenamefont {Dodelson}\ and\ \citenamefont
  {Widrow}(1994)}]{Dodelson:1993je}%
  \BibitemOpen
  \bibfield  {author} {\bibinfo {author} {\bibfnamefont {S.}~\bibnamefont
  {Dodelson}}\ and\ \bibinfo {author} {\bibfnamefont {L.~M.}\ \bibnamefont
  {Widrow}},\ }\href {\doibase 10.1103/PhysRevLett.72.17} {\bibfield  {journal}
  {\bibinfo  {journal} {Phys. Rev. Lett.}\ }\textbf {\bibinfo {volume} {72}},\
  \bibinfo {pages} {17} (\bibinfo {year} {1994})},\ \Eprint
  {http://arxiv.org/abs/hep-ph/9303287} {arXiv:hep-ph/9303287 [hep-ph]}
  \BibitemShut {NoStop}%
\bibitem [{\citenamefont {Shi}\ and\ \citenamefont
  {Fuller}(1999)}]{Shi:1998km}%
  \BibitemOpen
  \bibfield  {author} {\bibinfo {author} {\bibfnamefont {X.-D.}\ \bibnamefont
  {Shi}}\ and\ \bibinfo {author} {\bibfnamefont {G.~M.}\ \bibnamefont
  {Fuller}},\ }\href {\doibase 10.1103/PhysRevLett.82.2832} {\bibfield
  {journal} {\bibinfo  {journal} {Phys. Rev. Lett.}\ }\textbf {\bibinfo
  {volume} {82}},\ \bibinfo {pages} {2832} (\bibinfo {year} {1999})},\ \Eprint
  {http://arxiv.org/abs/astro-ph/9810076} {arXiv:astro-ph/9810076 [astro-ph]}
  \BibitemShut {NoStop}%
\bibitem [{\citenamefont {Abazajian}\ \emph {et~al.}(2001)\citenamefont
  {Abazajian}, \citenamefont {Fuller},\ and\ \citenamefont
  {Patel}}]{Abazajian:2001nj}%
  \BibitemOpen
  \bibfield  {author} {\bibinfo {author} {\bibfnamefont {K.}~\bibnamefont
  {Abazajian}}, \bibinfo {author} {\bibfnamefont {G.~M.}\ \bibnamefont
  {Fuller}}, \ and\ \bibinfo {author} {\bibfnamefont {M.}~\bibnamefont
  {Patel}},\ }\href {\doibase 10.1103/PhysRevD.64.023501} {\bibfield  {journal}
  {\bibinfo  {journal} {Phys. Rev.}\ }\textbf {\bibinfo {volume} {D64}},\
  \bibinfo {pages} {023501} (\bibinfo {year} {2001})},\ \Eprint
  {http://arxiv.org/abs/astro-ph/0101524} {arXiv:astro-ph/0101524 [astro-ph]}
  \BibitemShut {NoStop}%
\bibitem [{\citenamefont {Asaka}\ \emph {et~al.}(2005)\citenamefont {Asaka},
  \citenamefont {Blanchet},\ and\ \citenamefont {Shaposhnikov}}]{Asaka:2005an}%
  \BibitemOpen
  \bibfield  {author} {\bibinfo {author} {\bibfnamefont {T.}~\bibnamefont
  {Asaka}}, \bibinfo {author} {\bibfnamefont {S.}~\bibnamefont {Blanchet}}, \
  and\ \bibinfo {author} {\bibfnamefont {M.}~\bibnamefont {Shaposhnikov}},\
  }\href {\doibase 10.1016/j.physletb.2005.09.070} {\bibfield  {journal}
  {\bibinfo  {journal} {Phys. Lett.}\ }\textbf {\bibinfo {volume} {B631}},\
  \bibinfo {pages} {151} (\bibinfo {year} {2005})},\ \Eprint
  {http://arxiv.org/abs/hep-ph/0503065} {arXiv:hep-ph/0503065 [hep-ph]}
  \BibitemShut {NoStop}%
\bibitem [{\citenamefont {Minkowski}(1977)}]{Minkowski:1977sc}%
  \BibitemOpen
  \bibfield  {author} {\bibinfo {author} {\bibfnamefont {P.}~\bibnamefont
  {Minkowski}},\ }\href {\doibase 10.1016/0370-2693(77)90435-X} {\bibfield
  {journal} {\bibinfo  {journal} {Phys. Lett.}\ }\textbf {\bibinfo {volume}
  {B67}},\ \bibinfo {pages} {421} (\bibinfo {year} {1977})}\BibitemShut
  {NoStop}%
\bibitem [{\citenamefont {Mohapatra}\ and\ \citenamefont
  {Senjanovic}(1980)}]{Mohapatra:1979ia}%
  \BibitemOpen
  \bibfield  {author} {\bibinfo {author} {\bibfnamefont {R.~N.}\ \bibnamefont
  {Mohapatra}}\ and\ \bibinfo {author} {\bibfnamefont {G.}~\bibnamefont
  {Senjanovic}},\ }\href {\doibase 10.1103/PhysRevLett.44.912} {\bibfield
  {journal} {\bibinfo  {journal} {Phys. Rev. Lett.}\ }\textbf {\bibinfo
  {volume} {44}},\ \bibinfo {pages} {912} (\bibinfo {year} {1980})}\BibitemShut
  {NoStop}%
\bibitem [{\citenamefont {Yanagida}(1979)}]{Yanagida:1979as}%
  \BibitemOpen
  \bibfield  {author} {\bibinfo {author} {\bibfnamefont {T.}~\bibnamefont
  {Yanagida}},\ }\href@noop {} {\  (\bibinfo {year} {1979})},\ \bibinfo {note}
  {in Proceedings of the Workshop on the Baryon Number of the Universe and
  Unified Theories, Tsukuba, Japan}\BibitemShut {NoStop}%
\bibitem [{\citenamefont {Gell-Mann}\ \emph {et~al.}()\citenamefont
  {Gell-Mann}, \citenamefont {Ramond},\ and\ \citenamefont
  {Slansky}}]{GellMann:1980vs}%
  \BibitemOpen
  \bibfield  {author} {\bibinfo {author} {\bibfnamefont {M.}~\bibnamefont
  {Gell-Mann}}, \bibinfo {author} {\bibfnamefont {P.}~\bibnamefont {Ramond}}, \
  and\ \bibinfo {author} {\bibfnamefont {R.}~\bibnamefont {Slansky}},\
  }\href@noop {} {\enquote {\bibinfo {title} {{Complex Spinors and Unified
  Theories}},}\ }\bibinfo {note} {{Print-80-0576 (CERN)}}\BibitemShut {NoStop}%
\bibitem [{\citenamefont {Weinberg}(1979)}]{Weinberg:1979sa}%
  \BibitemOpen
  \bibfield  {author} {\bibinfo {author} {\bibfnamefont {S.}~\bibnamefont
  {Weinberg}},\ }\href {\doibase 10.1103/PhysRevLett.43.1566} {\bibfield
  {journal} {\bibinfo  {journal} {Phys.Rev.Lett.}\ }\textbf {\bibinfo {volume}
  {43}},\ \bibinfo {pages} {1566} (\bibinfo {year} {1979})}\BibitemShut
  {NoStop}%
\bibitem [{\citenamefont {Mohapatra}\ and\ \citenamefont
  {Valle}(1986)}]{Mohapatra:1986bd}%
  \BibitemOpen
  \bibfield  {author} {\bibinfo {author} {\bibfnamefont {R.}~\bibnamefont
  {Mohapatra}}\ and\ \bibinfo {author} {\bibfnamefont {J.}~\bibnamefont
  {Valle}},\ }\href {\doibase 10.1103/PhysRevD.34.1642} {\bibfield  {journal}
  {\bibinfo  {journal} {Phys.Rev.}\ }\textbf {\bibinfo {volume} {D34}},\
  \bibinfo {pages} {1642} (\bibinfo {year} {1986})}\BibitemShut {NoStop}%
\bibitem [{\citenamefont {Bernabeu}\ \emph {et~al.}(1987)\citenamefont
  {Bernabeu}, \citenamefont {Santamaria}, \citenamefont {Vidal}, \citenamefont
  {Mendez},\ and\ \citenamefont {Valle}}]{Bernabeu:1987gr}%
  \BibitemOpen
  \bibfield  {author} {\bibinfo {author} {\bibfnamefont {J.}~\bibnamefont
  {Bernabeu}}, \bibinfo {author} {\bibfnamefont {A.}~\bibnamefont
  {Santamaria}}, \bibinfo {author} {\bibfnamefont {J.}~\bibnamefont {Vidal}},
  \bibinfo {author} {\bibfnamefont {A.}~\bibnamefont {Mendez}}, \ and\ \bibinfo
  {author} {\bibfnamefont {J.}~\bibnamefont {Valle}},\ }\href {\doibase
  10.1016/0370-2693(87)91100-2} {\bibfield  {journal} {\bibinfo  {journal}
  {Phys.Lett.}\ }\textbf {\bibinfo {volume} {B187}},\ \bibinfo {pages} {303}
  (\bibinfo {year} {1987})}\BibitemShut {NoStop}%
\bibitem [{\citenamefont {Branco}\ \emph {et~al.}(1989)\citenamefont {Branco},
  \citenamefont {Grimus},\ and\ \citenamefont {Lavoura}}]{Branco:1988ex}%
  \BibitemOpen
  \bibfield  {author} {\bibinfo {author} {\bibfnamefont {G.~C.}\ \bibnamefont
  {Branco}}, \bibinfo {author} {\bibfnamefont {W.}~\bibnamefont {Grimus}}, \
  and\ \bibinfo {author} {\bibfnamefont {L.}~\bibnamefont {Lavoura}},\ }\href
  {\doibase 10.1016/0550-3213(89)90304-0} {\bibfield  {journal} {\bibinfo
  {journal} {Nucl. Phys.}\ }\textbf {\bibinfo {volume} {B312}},\ \bibinfo
  {pages} {492} (\bibinfo {year} {1989})}\BibitemShut {NoStop}%
\bibitem [{\citenamefont {Buchmuller}\ and\ \citenamefont
  {Wyler}(1990)}]{Buchmuller:1990du}%
  \BibitemOpen
  \bibfield  {author} {\bibinfo {author} {\bibfnamefont {W.}~\bibnamefont
  {Buchmuller}}\ and\ \bibinfo {author} {\bibfnamefont {D.}~\bibnamefont
  {Wyler}},\ }\href {\doibase 10.1016/0370-2693(90)91016-5} {\bibfield
  {journal} {\bibinfo  {journal} {Phys.Lett.}\ }\textbf {\bibinfo {volume}
  {B249}},\ \bibinfo {pages} {458} (\bibinfo {year} {1990})}\BibitemShut
  {NoStop}%
\bibitem [{\citenamefont {Pilaftsis}(1992{\natexlab{a}})}]{Pilaftsis:1991ug}%
  \BibitemOpen
  \bibfield  {author} {\bibinfo {author} {\bibfnamefont {A.}~\bibnamefont
  {Pilaftsis}},\ }\href {\doibase 10.1007/BF01482590} {\bibfield  {journal}
  {\bibinfo  {journal} {Z. Phys.}\ }\textbf {\bibinfo {volume} {C55}},\
  \bibinfo {pages} {275} (\bibinfo {year} {1992}{\natexlab{a}})},\ \Eprint
  {http://arxiv.org/abs/hep-ph/9901206} {arXiv:hep-ph/9901206 [hep-ph]}
  \BibitemShut {NoStop}%
\bibitem [{\citenamefont {Dev}\ and\ \citenamefont
  {Pilaftsis}(2012)}]{Dev:2012sg}%
  \BibitemOpen
  \bibfield  {author} {\bibinfo {author} {\bibfnamefont {P.~S.~B.}\
  \bibnamefont {Dev}}\ and\ \bibinfo {author} {\bibfnamefont {A.}~\bibnamefont
  {Pilaftsis}},\ }\href {\doibase 10.1103/PhysRevD.86.113001} {\bibfield
  {journal} {\bibinfo  {journal} {Phys. Rev.}\ }\textbf {\bibinfo {volume}
  {D86}},\ \bibinfo {pages} {113001} (\bibinfo {year} {2012})},\ \Eprint
  {http://arxiv.org/abs/1209.4051} {arXiv:1209.4051 [hep-ph]} \BibitemShut
  {NoStop}%
\bibitem [{\citenamefont {Malinsky}\ \emph {et~al.}(2005)\citenamefont
  {Malinsky}, \citenamefont {Romao},\ and\ \citenamefont
  {Valle}}]{Malinsky:2005bi}%
  \BibitemOpen
  \bibfield  {author} {\bibinfo {author} {\bibfnamefont {M.}~\bibnamefont
  {Malinsky}}, \bibinfo {author} {\bibfnamefont {J.}~\bibnamefont {Romao}}, \
  and\ \bibinfo {author} {\bibfnamefont {J.}~\bibnamefont {Valle}},\ }\href
  {\doibase 10.1103/PhysRevLett.95.161801} {\bibfield  {journal} {\bibinfo
  {journal} {Phys.Rev.Lett.}\ }\textbf {\bibinfo {volume} {95}},\ \bibinfo
  {pages} {161801} (\bibinfo {year} {2005})},\ \Eprint
  {http://arxiv.org/abs/hep-ph/0506296} {arXiv:hep-ph/0506296 [hep-ph]}
  \BibitemShut {NoStop}%
\bibitem [{\citenamefont {Lee}\ and\ \citenamefont
  {Shrock}(1977)}]{Lee:1977tib}%
  \BibitemOpen
  \bibfield  {author} {\bibinfo {author} {\bibfnamefont {B.~W.}\ \bibnamefont
  {Lee}}\ and\ \bibinfo {author} {\bibfnamefont {R.~E.}\ \bibnamefont
  {Shrock}},\ }\href {\doibase 10.1103/PhysRevD.16.1444} {\bibfield  {journal}
  {\bibinfo  {journal} {Phys. Rev.}\ }\textbf {\bibinfo {volume} {D16}},\
  \bibinfo {pages} {1444} (\bibinfo {year} {1977})}\BibitemShut {NoStop}%
\bibitem [{\citenamefont {Shrock}(1980)}]{Shrock:1980vy}%
  \BibitemOpen
  \bibfield  {author} {\bibinfo {author} {\bibfnamefont {R.~E.}\ \bibnamefont
  {Shrock}},\ }\href {\doibase 10.1016/0370-2693(80)90235-X} {\bibfield
  {journal} {\bibinfo  {journal} {Phys. Lett.}\ }\textbf {\bibinfo {volume}
  {B96}},\ \bibinfo {pages} {159} (\bibinfo {year} {1980})}\BibitemShut
  {NoStop}%
\bibitem [{\citenamefont {Schechter}\ and\ \citenamefont
  {Valle}(1980)}]{Schechter:1980gr}%
  \BibitemOpen
  \bibfield  {author} {\bibinfo {author} {\bibfnamefont {J.}~\bibnamefont
  {Schechter}}\ and\ \bibinfo {author} {\bibfnamefont {J.~W.~F.}\ \bibnamefont
  {Valle}},\ }\href {\doibase 10.1103/PhysRevD.22.2227} {\bibfield  {journal}
  {\bibinfo  {journal} {Phys. Rev.}\ }\textbf {\bibinfo {volume} {D22}},\
  \bibinfo {pages} {2227} (\bibinfo {year} {1980})}\BibitemShut {NoStop}%
\bibitem [{\citenamefont {Shrock}(1981{\natexlab{a}})}]{Shrock:1980ct}%
  \BibitemOpen
  \bibfield  {author} {\bibinfo {author} {\bibfnamefont {R.~E.}\ \bibnamefont
  {Shrock}},\ }\href {\doibase 10.1103/PhysRevD.24.1232} {\bibfield  {journal}
  {\bibinfo  {journal} {Phys. Rev.}\ }\textbf {\bibinfo {volume} {D24}},\
  \bibinfo {pages} {1232} (\bibinfo {year} {1981}{\natexlab{a}})}\BibitemShut
  {NoStop}%
\bibitem [{\citenamefont {Shrock}(1981{\natexlab{b}})}]{Shrock:1981wq}%
  \BibitemOpen
  \bibfield  {author} {\bibinfo {author} {\bibfnamefont {R.~E.}\ \bibnamefont
  {Shrock}},\ }\href {\doibase 10.1103/PhysRevD.24.1275} {\bibfield  {journal}
  {\bibinfo  {journal} {Phys. Rev.}\ }\textbf {\bibinfo {volume} {D24}},\
  \bibinfo {pages} {1275} (\bibinfo {year} {1981}{\natexlab{b}})}\BibitemShut
  {NoStop}%
\bibitem [{\citenamefont {Langacker}\ and\ \citenamefont
  {London}(1988)}]{Langacker:1988ur}%
  \BibitemOpen
  \bibfield  {author} {\bibinfo {author} {\bibfnamefont {P.}~\bibnamefont
  {Langacker}}\ and\ \bibinfo {author} {\bibfnamefont {D.}~\bibnamefont
  {London}},\ }\href {\doibase 10.1103/PhysRevD.38.886} {\bibfield  {journal}
  {\bibinfo  {journal} {Phys.Rev.}\ }\textbf {\bibinfo {volume} {D38}},\
  \bibinfo {pages} {886} (\bibinfo {year} {1988})}\BibitemShut {NoStop}%
\bibitem [{\citenamefont {Bilenky}\ and\ \citenamefont
  {Giunti}(1993)}]{Bilenky:1992wv}%
  \BibitemOpen
  \bibfield  {author} {\bibinfo {author} {\bibfnamefont {S.~M.}\ \bibnamefont
  {Bilenky}}\ and\ \bibinfo {author} {\bibfnamefont {C.}~\bibnamefont
  {Giunti}},\ }\href {\doibase 10.1016/0370-2693(93)90760-F} {\bibfield
  {journal} {\bibinfo  {journal} {Phys.Lett.}\ }\textbf {\bibinfo {volume}
  {B300}},\ \bibinfo {pages} {137} (\bibinfo {year} {1993})},\ \Eprint
  {http://arxiv.org/abs/hep-ph/9211269} {arXiv:hep-ph/9211269 [hep-ph]}
  \BibitemShut {NoStop}%
\bibitem [{\citenamefont {Nardi}\ \emph {et~al.}(1994)\citenamefont {Nardi},
  \citenamefont {Roulet},\ and\ \citenamefont {Tommasini}}]{Nardi:1994iv}%
  \BibitemOpen
  \bibfield  {author} {\bibinfo {author} {\bibfnamefont {E.}~\bibnamefont
  {Nardi}}, \bibinfo {author} {\bibfnamefont {E.}~\bibnamefont {Roulet}}, \
  and\ \bibinfo {author} {\bibfnamefont {D.}~\bibnamefont {Tommasini}},\ }\href
  {\doibase 10.1016/0370-2693(94)90736-6} {\bibfield  {journal} {\bibinfo
  {journal} {Phys.Lett.}\ }\textbf {\bibinfo {volume} {B327}},\ \bibinfo
  {pages} {319} (\bibinfo {year} {1994})},\ \Eprint
  {http://arxiv.org/abs/hep-ph/9402224} {arXiv:hep-ph/9402224 [hep-ph]}
  \BibitemShut {NoStop}%
\bibitem [{\citenamefont {Tommasini}\ \emph {et~al.}(1995)\citenamefont
  {Tommasini}, \citenamefont {Barenboim}, \citenamefont {Bernabeu},\ and\
  \citenamefont {Jarlskog}}]{Tommasini:1995ii}%
  \BibitemOpen
  \bibfield  {author} {\bibinfo {author} {\bibfnamefont {D.}~\bibnamefont
  {Tommasini}}, \bibinfo {author} {\bibfnamefont {G.}~\bibnamefont
  {Barenboim}}, \bibinfo {author} {\bibfnamefont {J.}~\bibnamefont {Bernabeu}},
  \ and\ \bibinfo {author} {\bibfnamefont {C.}~\bibnamefont {Jarlskog}},\
  }\href {\doibase 10.1016/0550-3213(95)00201-3} {\bibfield  {journal}
  {\bibinfo  {journal} {Nucl.Phys.}\ }\textbf {\bibinfo {volume} {B444}},\
  \bibinfo {pages} {451} (\bibinfo {year} {1995})},\ \Eprint
  {http://arxiv.org/abs/hep-ph/9503228} {arXiv:hep-ph/9503228 [hep-ph]}
  \BibitemShut {NoStop}%
\bibitem [{\citenamefont {Bergmann}\ and\ \citenamefont
  {Kagan}(1999)}]{Bergmann:1998rg}%
  \BibitemOpen
  \bibfield  {author} {\bibinfo {author} {\bibfnamefont {S.}~\bibnamefont
  {Bergmann}}\ and\ \bibinfo {author} {\bibfnamefont {A.}~\bibnamefont
  {Kagan}},\ }\href {\doibase 10.1016/S0550-3213(98)00686-5} {\bibfield
  {journal} {\bibinfo  {journal} {Nucl.Phys.}\ }\textbf {\bibinfo {volume}
  {B538}},\ \bibinfo {pages} {368} (\bibinfo {year} {1999})},\ \Eprint
  {http://arxiv.org/abs/hep-ph/9803305} {arXiv:hep-ph/9803305 [hep-ph]}
  \BibitemShut {NoStop}%
\bibitem [{\citenamefont {Loinaz}\ \emph
  {et~al.}(2003{\natexlab{a}})\citenamefont {Loinaz}, \citenamefont {Okamura},
  \citenamefont {Takeuchi},\ and\ \citenamefont
  {Wijewardhana}}]{Loinaz:2002ep}%
  \BibitemOpen
  \bibfield  {author} {\bibinfo {author} {\bibfnamefont {W.}~\bibnamefont
  {Loinaz}}, \bibinfo {author} {\bibfnamefont {N.}~\bibnamefont {Okamura}},
  \bibinfo {author} {\bibfnamefont {T.}~\bibnamefont {Takeuchi}}, \ and\
  \bibinfo {author} {\bibfnamefont {L.}~\bibnamefont {Wijewardhana}},\ }\href
  {\doibase 10.1103/PhysRevD.67.073012} {\bibfield  {journal} {\bibinfo
  {journal} {Phys.Rev.}\ }\textbf {\bibinfo {volume} {D67}},\ \bibinfo {pages}
  {073012} (\bibinfo {year} {2003}{\natexlab{a}})},\ \Eprint
  {http://arxiv.org/abs/hep-ph/0210193} {arXiv:hep-ph/0210193 [hep-ph]}
  \BibitemShut {NoStop}%
\bibitem [{\citenamefont {Loinaz}\ \emph
  {et~al.}(2003{\natexlab{b}})\citenamefont {Loinaz}, \citenamefont {Okamura},
  \citenamefont {Rayyan}, \citenamefont {Takeuchi},\ and\ \citenamefont
  {Wijewardhana}}]{Loinaz:2003gc}%
  \BibitemOpen
  \bibfield  {author} {\bibinfo {author} {\bibfnamefont {W.}~\bibnamefont
  {Loinaz}}, \bibinfo {author} {\bibfnamefont {N.}~\bibnamefont {Okamura}},
  \bibinfo {author} {\bibfnamefont {S.}~\bibnamefont {Rayyan}}, \bibinfo
  {author} {\bibfnamefont {T.}~\bibnamefont {Takeuchi}}, \ and\ \bibinfo
  {author} {\bibfnamefont {L.}~\bibnamefont {Wijewardhana}},\ }\href {\doibase
  10.1103/PhysRevD.68.073001} {\bibfield  {journal} {\bibinfo  {journal}
  {Phys.Rev.}\ }\textbf {\bibinfo {volume} {D68}},\ \bibinfo {pages} {073001}
  (\bibinfo {year} {2003}{\natexlab{b}})},\ \Eprint
  {http://arxiv.org/abs/hep-ph/0304004} {arXiv:hep-ph/0304004 [hep-ph]}
  \BibitemShut {NoStop}%
\bibitem [{\citenamefont {Loinaz}\ \emph {et~al.}(2004)\citenamefont {Loinaz},
  \citenamefont {Okamura}, \citenamefont {Rayyan}, \citenamefont {Takeuchi},\
  and\ \citenamefont {Wijewardhana}}]{Loinaz:2004qc}%
  \BibitemOpen
  \bibfield  {author} {\bibinfo {author} {\bibfnamefont {W.}~\bibnamefont
  {Loinaz}}, \bibinfo {author} {\bibfnamefont {N.}~\bibnamefont {Okamura}},
  \bibinfo {author} {\bibfnamefont {S.}~\bibnamefont {Rayyan}}, \bibinfo
  {author} {\bibfnamefont {T.}~\bibnamefont {Takeuchi}}, \ and\ \bibinfo
  {author} {\bibfnamefont {L.}~\bibnamefont {Wijewardhana}},\ }\href {\doibase
  10.1103/PhysRevD.70.113004} {\bibfield  {journal} {\bibinfo  {journal}
  {Phys.Rev.}\ }\textbf {\bibinfo {volume} {D70}},\ \bibinfo {pages} {113004}
  (\bibinfo {year} {2004})},\ \Eprint {http://arxiv.org/abs/hep-ph/0403306}
  {arXiv:hep-ph/0403306 [hep-ph]} \BibitemShut {NoStop}%
\bibitem [{\citenamefont {Antusch}\ \emph {et~al.}(2006)\citenamefont
  {Antusch}, \citenamefont {Biggio}, \citenamefont {Fernandez-Martinez},
  \citenamefont {Gavela},\ and\ \citenamefont {Lopez-Pavon}}]{Antusch:2006vwa}%
  \BibitemOpen
  \bibfield  {author} {\bibinfo {author} {\bibfnamefont {S.}~\bibnamefont
  {Antusch}}, \bibinfo {author} {\bibfnamefont {C.}~\bibnamefont {Biggio}},
  \bibinfo {author} {\bibfnamefont {E.}~\bibnamefont {Fernandez-Martinez}},
  \bibinfo {author} {\bibfnamefont {M.}~\bibnamefont {Gavela}}, \ and\ \bibinfo
  {author} {\bibfnamefont {J.}~\bibnamefont {Lopez-Pavon}},\ }\href {\doibase
  10.1088/1126-6708/2006/10/084} {\bibfield  {journal} {\bibinfo  {journal}
  {JHEP}\ }\textbf {\bibinfo {volume} {0610}},\ \bibinfo {pages} {084}
  (\bibinfo {year} {2006})},\ \Eprint {http://arxiv.org/abs/hep-ph/0607020}
  {arXiv:hep-ph/0607020 [hep-ph]} \BibitemShut {NoStop}%
\bibitem [{\citenamefont {Antusch}\ \emph
  {et~al.}(2009{\natexlab{a}})\citenamefont {Antusch}, \citenamefont
  {Baumann},\ and\ \citenamefont {Fernandez-Martinez}}]{Antusch:2008tz}%
  \BibitemOpen
  \bibfield  {author} {\bibinfo {author} {\bibfnamefont {S.}~\bibnamefont
  {Antusch}}, \bibinfo {author} {\bibfnamefont {J.~P.}\ \bibnamefont
  {Baumann}}, \ and\ \bibinfo {author} {\bibfnamefont {E.}~\bibnamefont
  {Fernandez-Martinez}},\ }\href {\doibase 10.1016/j.nuclphysb.2008.11.018}
  {\bibfield  {journal} {\bibinfo  {journal} {Nucl.Phys.}\ }\textbf {\bibinfo
  {volume} {B810}},\ \bibinfo {pages} {369} (\bibinfo {year}
  {2009}{\natexlab{a}})},\ \Eprint {http://arxiv.org/abs/0807.1003}
  {arXiv:0807.1003 [hep-ph]} \BibitemShut {NoStop}%
\bibitem [{\citenamefont {Biggio}(2008)}]{Biggio:2008in}%
  \BibitemOpen
  \bibfield  {author} {\bibinfo {author} {\bibfnamefont {C.}~\bibnamefont
  {Biggio}},\ }\href {\doibase 10.1016/j.physletb.2008.09.004} {\bibfield
  {journal} {\bibinfo  {journal} {Phys. Lett.}\ }\textbf {\bibinfo {volume}
  {B668}},\ \bibinfo {pages} {378} (\bibinfo {year} {2008})},\ \Eprint
  {http://arxiv.org/abs/0806.2558} {arXiv:0806.2558 [hep-ph]} \BibitemShut
  {NoStop}%
\bibitem [{\citenamefont {Alonso}\ \emph {et~al.}(2013)\citenamefont {Alonso},
  \citenamefont {Dhen}, \citenamefont {Gavela},\ and\ \citenamefont
  {Hambye}}]{Alonso:2012ji}%
  \BibitemOpen
  \bibfield  {author} {\bibinfo {author} {\bibfnamefont {R.}~\bibnamefont
  {Alonso}}, \bibinfo {author} {\bibfnamefont {M.}~\bibnamefont {Dhen}},
  \bibinfo {author} {\bibfnamefont {M.}~\bibnamefont {Gavela}}, \ and\ \bibinfo
  {author} {\bibfnamefont {T.}~\bibnamefont {Hambye}},\ }\href {\doibase
  10.1007/JHEP01(2013)118} {\bibfield  {journal} {\bibinfo  {journal} {JHEP}\
  }\textbf {\bibinfo {volume} {1301}},\ \bibinfo {pages} {118} (\bibinfo {year}
  {2013})},\ \Eprint {http://arxiv.org/abs/1209.2679} {arXiv:1209.2679
  [hep-ph]} \BibitemShut {NoStop}%
\bibitem [{\citenamefont {Abada}\ \emph {et~al.}(2013)\citenamefont {Abada},
  \citenamefont {Das}, \citenamefont {Teixeira}, \citenamefont {Vicente},\ and\
  \citenamefont {Weiland}}]{Abada:2012mc}%
  \BibitemOpen
  \bibfield  {author} {\bibinfo {author} {\bibfnamefont {A.}~\bibnamefont
  {Abada}}, \bibinfo {author} {\bibfnamefont {D.}~\bibnamefont {Das}}, \bibinfo
  {author} {\bibfnamefont {A.}~\bibnamefont {Teixeira}}, \bibinfo {author}
  {\bibfnamefont {A.}~\bibnamefont {Vicente}}, \ and\ \bibinfo {author}
  {\bibfnamefont {C.}~\bibnamefont {Weiland}},\ }\href {\doibase
  10.1007/JHEP02(2013)048} {\bibfield  {journal} {\bibinfo  {journal} {JHEP}\
  }\textbf {\bibinfo {volume} {1302}},\ \bibinfo {pages} {048} (\bibinfo {year}
  {2013})},\ \Eprint {http://arxiv.org/abs/1211.3052} {arXiv:1211.3052
  [hep-ph]} \BibitemShut {NoStop}%
\bibitem [{\citenamefont {Akhmedov}\ \emph {et~al.}(2013)\citenamefont
  {Akhmedov}, \citenamefont {Kartavtsev}, \citenamefont {Lindner},
  \citenamefont {Michaels},\ and\ \citenamefont {Smirnov}}]{Akhmedov:2013hec}%
  \BibitemOpen
  \bibfield  {author} {\bibinfo {author} {\bibfnamefont {E.}~\bibnamefont
  {Akhmedov}}, \bibinfo {author} {\bibfnamefont {A.}~\bibnamefont
  {Kartavtsev}}, \bibinfo {author} {\bibfnamefont {M.}~\bibnamefont {Lindner}},
  \bibinfo {author} {\bibfnamefont {L.}~\bibnamefont {Michaels}}, \ and\
  \bibinfo {author} {\bibfnamefont {J.}~\bibnamefont {Smirnov}},\ }\href
  {\doibase 10.1007/JHEP05(2013)081} {\bibfield  {journal} {\bibinfo  {journal}
  {JHEP}\ }\textbf {\bibinfo {volume} {1305}},\ \bibinfo {pages} {081}
  (\bibinfo {year} {2013})},\ \Eprint {http://arxiv.org/abs/1302.1872}
  {arXiv:1302.1872 [hep-ph]} \BibitemShut {NoStop}%
\bibitem [{\citenamefont {Basso}\ \emph {et~al.}(2014)\citenamefont {Basso},
  \citenamefont {Fischer},\ and\ \citenamefont {van~der Bij}}]{Basso:2013jka}%
  \BibitemOpen
  \bibfield  {author} {\bibinfo {author} {\bibfnamefont {L.}~\bibnamefont
  {Basso}}, \bibinfo {author} {\bibfnamefont {O.}~\bibnamefont {Fischer}}, \
  and\ \bibinfo {author} {\bibfnamefont {J.~J.}\ \bibnamefont {van~der Bij}},\
  }\href {\doibase 10.1209/0295-5075/105/11001} {\bibfield  {journal} {\bibinfo
   {journal} {Europhys.Lett.}\ }\textbf {\bibinfo {volume} {105}},\ \bibinfo
  {pages} {11001} (\bibinfo {year} {2014})},\ \Eprint
  {http://arxiv.org/abs/1310.2057} {arXiv:1310.2057 [hep-ph]} \BibitemShut
  {NoStop}%
\bibitem [{\citenamefont {Abada}\ \emph {et~al.}(2014)\citenamefont {Abada},
  \citenamefont {Teixeira}, \citenamefont {Vicente},\ and\ \citenamefont
  {Weiland}}]{Abada:2013aba}%
  \BibitemOpen
  \bibfield  {author} {\bibinfo {author} {\bibfnamefont {A.}~\bibnamefont
  {Abada}}, \bibinfo {author} {\bibfnamefont {A.}~\bibnamefont {Teixeira}},
  \bibinfo {author} {\bibfnamefont {A.}~\bibnamefont {Vicente}}, \ and\
  \bibinfo {author} {\bibfnamefont {C.}~\bibnamefont {Weiland}},\ }\href
  {\doibase 10.1007/JHEP02(2014)091} {\bibfield  {journal} {\bibinfo  {journal}
  {JHEP}\ }\textbf {\bibinfo {volume} {1402}},\ \bibinfo {pages} {091}
  (\bibinfo {year} {2014})},\ \Eprint {http://arxiv.org/abs/1311.2830}
  {arXiv:1311.2830 [hep-ph]} \BibitemShut {NoStop}%
\bibitem [{\citenamefont {Antusch}\ and\ \citenamefont
  {Fischer}(2014)}]{Antusch:2014woa}%
  \BibitemOpen
  \bibfield  {author} {\bibinfo {author} {\bibfnamefont {S.}~\bibnamefont
  {Antusch}}\ and\ \bibinfo {author} {\bibfnamefont {O.}~\bibnamefont
  {Fischer}},\ }\href {\doibase 10.1007/JHEP10(2014)094} {\bibfield  {journal}
  {\bibinfo  {journal} {JHEP}\ }\textbf {\bibinfo {volume} {1410}},\ \bibinfo
  {pages} {94} (\bibinfo {year} {2014})},\ \Eprint
  {http://arxiv.org/abs/1407.6607} {arXiv:1407.6607 [hep-ph]} \BibitemShut
  {NoStop}%
\bibitem [{\citenamefont {Antusch}\ and\ \citenamefont
  {Fischer}(2015)}]{Antusch:2015mia}%
  \BibitemOpen
  \bibfield  {author} {\bibinfo {author} {\bibfnamefont {S.}~\bibnamefont
  {Antusch}}\ and\ \bibinfo {author} {\bibfnamefont {O.}~\bibnamefont
  {Fischer}},\ }\href {\doibase 10.1007/JHEP05(2015)053} {\bibfield  {journal}
  {\bibinfo  {journal} {JHEP}\ }\textbf {\bibinfo {volume} {05}},\ \bibinfo
  {pages} {053} (\bibinfo {year} {2015})},\ \Eprint
  {http://arxiv.org/abs/1502.05915} {arXiv:1502.05915 [hep-ph]} \BibitemShut
  {NoStop}%
\bibitem [{\citenamefont {Abada}\ \emph {et~al.}(2016)\citenamefont {Abada},
  \citenamefont {De~Romeri},\ and\ \citenamefont {Teixeira}}]{Abada:2015oba}%
  \BibitemOpen
  \bibfield  {author} {\bibinfo {author} {\bibfnamefont {A.}~\bibnamefont
  {Abada}}, \bibinfo {author} {\bibfnamefont {V.}~\bibnamefont {De~Romeri}}, \
  and\ \bibinfo {author} {\bibfnamefont {A.~M.}\ \bibnamefont {Teixeira}},\
  }\href {\doibase 10.1007/JHEP02(2016)083} {\bibfield  {journal} {\bibinfo
  {journal} {JHEP}\ }\textbf {\bibinfo {volume} {02}},\ \bibinfo {pages} {083}
  (\bibinfo {year} {2016})},\ \Eprint {http://arxiv.org/abs/1510.06657}
  {arXiv:1510.06657 [hep-ph]} \BibitemShut {NoStop}%
\bibitem [{\citenamefont {Abada}\ and\ \citenamefont
  {Toma}(2016{\natexlab{a}})}]{Abada:2015trh}%
  \BibitemOpen
  \bibfield  {author} {\bibinfo {author} {\bibfnamefont {A.}~\bibnamefont
  {Abada}}\ and\ \bibinfo {author} {\bibfnamefont {T.}~\bibnamefont {Toma}},\
  }\href {\doibase 10.1007/JHEP02(2016)174} {\bibfield  {journal} {\bibinfo
  {journal} {JHEP}\ }\textbf {\bibinfo {volume} {02}},\ \bibinfo {pages} {174}
  (\bibinfo {year} {2016}{\natexlab{a}})},\ \Eprint
  {http://arxiv.org/abs/1511.03265} {arXiv:1511.03265 [hep-ph]} \BibitemShut
  {NoStop}%
\bibitem [{\citenamefont {Abada}\ and\ \citenamefont
  {Toma}(2016{\natexlab{b}})}]{Abada:2016awd}%
  \BibitemOpen
  \bibfield  {author} {\bibinfo {author} {\bibfnamefont {A.}~\bibnamefont
  {Abada}}\ and\ \bibinfo {author} {\bibfnamefont {T.}~\bibnamefont {Toma}},\
  }\href@noop {} {\  (\bibinfo {year} {2016}{\natexlab{b}})},\ \Eprint
  {http://arxiv.org/abs/1605.07643} {arXiv:1605.07643 [hep-ph]} \BibitemShut
  {NoStop}%
\bibitem [{\citenamefont {Broncano}\ \emph
  {et~al.}(2003{\natexlab{a}})\citenamefont {Broncano}, \citenamefont
  {Gavela},\ and\ \citenamefont {Jenkins}}]{Broncano:2002rw}%
  \BibitemOpen
  \bibfield  {author} {\bibinfo {author} {\bibfnamefont {A.}~\bibnamefont
  {Broncano}}, \bibinfo {author} {\bibfnamefont {M.~B.}\ \bibnamefont
  {Gavela}}, \ and\ \bibinfo {author} {\bibfnamefont {E.~E.}\ \bibnamefont
  {Jenkins}},\ }\href {\doibase 10.1016/j.physletb.2006.04.003} {\bibfield
  {journal} {\bibinfo  {journal} {Phys. Lett.}\ }\textbf {\bibinfo {volume}
  {B552}},\ \bibinfo {pages} {177} (\bibinfo {year} {2003}{\natexlab{a}})},\
  \Eprint {http://arxiv.org/abs/hep-ph/0210271} {arXiv:hep-ph/0210271}
  \BibitemShut {NoStop}%
\bibitem [{\citenamefont {Fernandez-Martinez}\ \emph
  {et~al.}(2007)\citenamefont {Fernandez-Martinez}, \citenamefont {Gavela},
  \citenamefont {Lopez-Pavon},\ and\ \citenamefont
  {Yasuda}}]{FernandezMartinez:2007ms}%
  \BibitemOpen
  \bibfield  {author} {\bibinfo {author} {\bibfnamefont {E.}~\bibnamefont
  {Fernandez-Martinez}}, \bibinfo {author} {\bibfnamefont {M.~B.}\ \bibnamefont
  {Gavela}}, \bibinfo {author} {\bibfnamefont {J.}~\bibnamefont {Lopez-Pavon}},
  \ and\ \bibinfo {author} {\bibfnamefont {O.}~\bibnamefont {Yasuda}},\ }\href
  {\doibase 10.1016/j.physletb.2007.03.069} {\bibfield  {journal} {\bibinfo
  {journal} {Phys. Lett.}\ }\textbf {\bibinfo {volume} {B649}},\ \bibinfo
  {pages} {427} (\bibinfo {year} {2007})},\ \Eprint
  {http://arxiv.org/abs/hep-ph/0703098} {arXiv:hep-ph/0703098} \BibitemShut
  {NoStop}%
\bibitem [{\citenamefont {Blennow}\ and\ \citenamefont
  {Fernandez-Martinez}(2011)}]{Blennow:2011vn}%
  \BibitemOpen
  \bibfield  {author} {\bibinfo {author} {\bibfnamefont {M.}~\bibnamefont
  {Blennow}}\ and\ \bibinfo {author} {\bibfnamefont {E.}~\bibnamefont
  {Fernandez-Martinez}},\ }\href {\doibase 10.1016/j.physletb.2011.09.028}
  {\bibfield  {journal} {\bibinfo  {journal} {Phys.Lett.}\ }\textbf {\bibinfo
  {volume} {B704}},\ \bibinfo {pages} {223} (\bibinfo {year} {2011})},\ \Eprint
  {http://arxiv.org/abs/1107.3992} {arXiv:1107.3992 [hep-ph]} \BibitemShut
  {NoStop}%
\bibitem [{\citenamefont {Broncano}\ \emph
  {et~al.}(2003{\natexlab{b}})\citenamefont {Broncano}, \citenamefont
  {Gavela},\ and\ \citenamefont {Jenkins}}]{Broncano:2003fq}%
  \BibitemOpen
  \bibfield  {author} {\bibinfo {author} {\bibfnamefont {A.}~\bibnamefont
  {Broncano}}, \bibinfo {author} {\bibfnamefont {M.~B.}\ \bibnamefont
  {Gavela}}, \ and\ \bibinfo {author} {\bibfnamefont {E.~E.}\ \bibnamefont
  {Jenkins}},\ }\href {\doibase 10.1016/j.nuclphysb.2003.09.011} {\bibfield
  {journal} {\bibinfo  {journal} {Nucl. Phys.}\ }\textbf {\bibinfo {volume}
  {B672}},\ \bibinfo {pages} {163} (\bibinfo {year} {2003}{\natexlab{b}})},\
  \Eprint {http://arxiv.org/abs/hep-ph/0307058} {arXiv:hep-ph/0307058}
  \BibitemShut {NoStop}%
\bibitem [{\citenamefont {Antusch}\ \emph {et~al.}(2010)\citenamefont
  {Antusch}, \citenamefont {Blanchet}, \citenamefont {Blennow},\ and\
  \citenamefont {Fernandez-Martinez}}]{Antusch:2009gn}%
  \BibitemOpen
  \bibfield  {author} {\bibinfo {author} {\bibfnamefont {S.}~\bibnamefont
  {Antusch}}, \bibinfo {author} {\bibfnamefont {S.}~\bibnamefont {Blanchet}},
  \bibinfo {author} {\bibfnamefont {M.}~\bibnamefont {Blennow}}, \ and\
  \bibinfo {author} {\bibfnamefont {E.}~\bibnamefont {Fernandez-Martinez}},\
  }\href {\doibase 10.1007/JHEP01(2010)017} {\bibfield  {journal} {\bibinfo
  {journal} {JHEP}\ }\textbf {\bibinfo {volume} {01}},\ \bibinfo {pages} {017}
  (\bibinfo {year} {2010})},\ \Eprint {http://arxiv.org/abs/0910.5957}
  {arXiv:0910.5957 [hep-ph]} \BibitemShut {NoStop}%
\bibitem [{\citenamefont {Kersten}\ and\ \citenamefont
  {Smirnov}(2007)}]{Kersten:2007vk}%
  \BibitemOpen
  \bibfield  {author} {\bibinfo {author} {\bibfnamefont {J.}~\bibnamefont
  {Kersten}}\ and\ \bibinfo {author} {\bibfnamefont {A.~Y.}\ \bibnamefont
  {Smirnov}},\ }\href {\doibase 10.1103/PhysRevD.76.073005} {\bibfield
  {journal} {\bibinfo  {journal} {Phys.Rev.}\ }\textbf {\bibinfo {volume}
  {D76}},\ \bibinfo {pages} {073005} (\bibinfo {year} {2007})},\ \Eprint
  {http://arxiv.org/abs/0705.3221} {arXiv:0705.3221 [hep-ph]} \BibitemShut
  {NoStop}%
\bibitem [{\citenamefont {Abada}\ \emph {et~al.}(2007)\citenamefont {Abada},
  \citenamefont {Biggio}, \citenamefont {Bonnet}, \citenamefont {Gavela},\ and\
  \citenamefont {Hambye}}]{Abada:2007ux}%
  \BibitemOpen
  \bibfield  {author} {\bibinfo {author} {\bibfnamefont {A.}~\bibnamefont
  {Abada}}, \bibinfo {author} {\bibfnamefont {C.}~\bibnamefont {Biggio}},
  \bibinfo {author} {\bibfnamefont {F.}~\bibnamefont {Bonnet}}, \bibinfo
  {author} {\bibfnamefont {M.~B.}\ \bibnamefont {Gavela}}, \ and\ \bibinfo
  {author} {\bibfnamefont {T.}~\bibnamefont {Hambye}},\ }\href {\doibase
  10.1088/1126-6708/2007/12/061} {\bibfield  {journal} {\bibinfo  {journal}
  {JHEP}\ }\textbf {\bibinfo {volume} {12}},\ \bibinfo {pages} {061} (\bibinfo
  {year} {2007})},\ \Eprint {http://arxiv.org/abs/0707.4058} {arXiv:0707.4058
  [hep-ph]} \BibitemShut {NoStop}%
\bibitem [{\citenamefont {Adhikari}\ and\ \citenamefont
  {Raychaudhuri}(2011)}]{Adhikari:2010yt}%
  \BibitemOpen
  \bibfield  {author} {\bibinfo {author} {\bibfnamefont {R.}~\bibnamefont
  {Adhikari}}\ and\ \bibinfo {author} {\bibfnamefont {A.}~\bibnamefont
  {Raychaudhuri}},\ }\href {\doibase 10.1103/PhysRevD.84.033002} {\bibfield
  {journal} {\bibinfo  {journal} {Phys. Rev.}\ }\textbf {\bibinfo {volume}
  {D84}},\ \bibinfo {pages} {033002} (\bibinfo {year} {2011})},\ \Eprint
  {http://arxiv.org/abs/1004.5111} {arXiv:1004.5111 [hep-ph]} \BibitemShut
  {NoStop}%
\bibitem [{\citenamefont {Lee}\ \emph {et~al.}(2013)\citenamefont {Lee},
  \citenamefont {Bhupal~Dev},\ and\ \citenamefont {Mohapatra}}]{Dev:2013oxa}%
  \BibitemOpen
  \bibfield  {author} {\bibinfo {author} {\bibfnamefont {C.-H.}\ \bibnamefont
  {Lee}}, \bibinfo {author} {\bibfnamefont {P.~S.}\ \bibnamefont {Bhupal~Dev}},
  \ and\ \bibinfo {author} {\bibfnamefont {R.~N.}\ \bibnamefont {Mohapatra}},\
  }\href {\doibase 10.1103/PhysRevD.88.093010} {\bibfield  {journal} {\bibinfo
  {journal} {Phys. Rev.}\ }\textbf {\bibinfo {volume} {D88}},\ \bibinfo {pages}
  {093010} (\bibinfo {year} {2013})},\ \Eprint {http://arxiv.org/abs/1309.0774}
  {arXiv:1309.0774 [hep-ph]} \BibitemShut {NoStop}%
\bibitem [{\citenamefont {Fernandez-Martinez}\ \emph
  {et~al.}(2015)\citenamefont {Fernandez-Martinez}, \citenamefont
  {Hernandez-Garcia}, \citenamefont {Lopez-Pavon},\ and\ \citenamefont
  {Lucente}}]{Fernandez-Martinez:2015hxa}%
  \BibitemOpen
  \bibfield  {author} {\bibinfo {author} {\bibfnamefont {E.}~\bibnamefont
  {Fernandez-Martinez}}, \bibinfo {author} {\bibfnamefont {J.}~\bibnamefont
  {Hernandez-Garcia}}, \bibinfo {author} {\bibfnamefont {J.}~\bibnamefont
  {Lopez-Pavon}}, \ and\ \bibinfo {author} {\bibfnamefont {M.}~\bibnamefont
  {Lucente}},\ }\href {\doibase 10.1007/JHEP10(2015)130} {\bibfield  {journal}
  {\bibinfo  {journal} {JHEP}\ }\textbf {\bibinfo {volume} {10}},\ \bibinfo
  {pages} {130} (\bibinfo {year} {2015})},\ \Eprint
  {http://arxiv.org/abs/1508.03051} {arXiv:1508.03051 [hep-ph]} \BibitemShut
  {NoStop}%
\bibitem [{\citenamefont {Ade}\ \emph {et~al.}(2015)\citenamefont {Ade} \emph
  {et~al.}}]{Ade:2015xua}%
  \BibitemOpen
  \bibfield  {author} {\bibinfo {author} {\bibfnamefont {P.~A.~R.}\
  \bibnamefont {Ade}} \emph {et~al.} (\bibinfo {collaboration} {Planck}),\
  }\href@noop {} {\  (\bibinfo {year} {2015})},\ \Eprint
  {http://arxiv.org/abs/1502.01589} {arXiv:1502.01589 [astro-ph.CO]}
  \BibitemShut {NoStop}%
\bibitem [{\citenamefont {Gonzalez-Garcia}\ \emph {et~al.}(2014)\citenamefont
  {Gonzalez-Garcia}, \citenamefont {Maltoni},\ and\ \citenamefont
  {Schwetz}}]{Gonzalez-Garcia:2014bfa}%
  \BibitemOpen
  \bibfield  {author} {\bibinfo {author} {\bibfnamefont {M.}~\bibnamefont
  {Gonzalez-Garcia}}, \bibinfo {author} {\bibfnamefont {M.}~\bibnamefont
  {Maltoni}}, \ and\ \bibinfo {author} {\bibfnamefont {T.}~\bibnamefont
  {Schwetz}},\ }\href {\doibase 10.1007/JHEP11(2014)052} {\bibfield  {journal}
  {\bibinfo  {journal} {JHEP}\ }\textbf {\bibinfo {volume} {1411}},\ \bibinfo
  {pages} {052} (\bibinfo {year} {2014})},\ \Eprint
  {http://arxiv.org/abs/1409.5439} {arXiv:1409.5439 [hep-ph]} \BibitemShut
  {NoStop}%
\bibitem [{\citenamefont {Olive}\ \emph {et~al.}(2014)\citenamefont {Olive}
  \emph {et~al.}}]{Agashe:2014kda}%
  \BibitemOpen
  \bibfield  {author} {\bibinfo {author} {\bibfnamefont {K.~A.}\ \bibnamefont
  {Olive}} \emph {et~al.} (\bibinfo {collaboration} {Particle Data Group}),\
  }\href {\doibase 10.1088/1674-1137/38/9/090001} {\bibfield  {journal}
  {\bibinfo  {journal} {Chin. Phys.}\ }\textbf {\bibinfo {volume} {C38}},\
  \bibinfo {pages} {090001} (\bibinfo {year} {2014})}\BibitemShut {NoStop}%
\bibitem [{\citenamefont {Schael}\ \emph {et~al.}(2006)\citenamefont {Schael}
  \emph {et~al.}}]{ALEPH:2005ab}%
  \BibitemOpen
  \bibfield  {author} {\bibinfo {author} {\bibfnamefont {S.}~\bibnamefont
  {Schael}} \emph {et~al.} (\bibinfo {collaboration} {SLD Electroweak Group,
  DELPHI, ALEPH, SLD, SLD Heavy Flavour Group, OPAL, LEP Electroweak Working
  Group, L3}),\ }\href {\doibase 10.1016/j.physrep.2005.12.006} {\bibfield
  {journal} {\bibinfo  {journal} {Phys. Rept.}\ }\textbf {\bibinfo {volume}
  {427}},\ \bibinfo {pages} {257} (\bibinfo {year} {2006})},\ \Eprint
  {http://arxiv.org/abs/hep-ex/0509008} {arXiv:hep-ex/0509008 [hep-ex]}
  \BibitemShut {NoStop}%
\bibitem [{\citenamefont {Pich}(2014)}]{Pich:2013lsa}%
  \BibitemOpen
  \bibfield  {author} {\bibinfo {author} {\bibfnamefont {A.}~\bibnamefont
  {Pich}},\ }\href {\doibase 10.1016/j.ppnp.2013.11.002} {\bibfield  {journal}
  {\bibinfo  {journal} {Prog. Part. Nucl. Phys.}\ }\textbf {\bibinfo {volume}
  {75}},\ \bibinfo {pages} {41} (\bibinfo {year} {2014})},\ \Eprint
  {http://arxiv.org/abs/1310.7922} {arXiv:1310.7922 [hep-ph]} \BibitemShut
  {NoStop}%
\bibitem [{\citenamefont {Hardy}\ and\ \citenamefont
  {Towner}(2015)}]{Hardy:2014qxa}%
  \BibitemOpen
  \bibfield  {author} {\bibinfo {author} {\bibfnamefont {J.~C.}\ \bibnamefont
  {Hardy}}\ and\ \bibinfo {author} {\bibfnamefont {I.~S.}\ \bibnamefont
  {Towner}},\ }\href {\doibase 10.1103/PhysRevC.91.025501} {\bibfield
  {journal} {\bibinfo  {journal} {Phys. Rev.}\ }\textbf {\bibinfo {volume}
  {C91}},\ \bibinfo {pages} {025501} (\bibinfo {year} {2015})},\ \Eprint
  {http://arxiv.org/abs/1411.5987} {arXiv:1411.5987 [nucl-ex]} \BibitemShut
  {NoStop}%
\bibitem [{\citenamefont {Aoki}\ \emph {et~al.}(2014)\citenamefont {Aoki} \emph
  {et~al.}}]{Aoki:2013ldr}%
  \BibitemOpen
  \bibfield  {author} {\bibinfo {author} {\bibfnamefont {S.}~\bibnamefont
  {Aoki}} \emph {et~al.},\ }\href {\doibase 10.1140/epjc/s10052-014-2890-7}
  {\bibfield  {journal} {\bibinfo  {journal} {Eur. Phys. J.}\ }\textbf
  {\bibinfo {volume} {C74}},\ \bibinfo {pages} {2890} (\bibinfo {year}
  {2014})},\ \Eprint {http://arxiv.org/abs/1310.8555} {arXiv:1310.8555
  [hep-lat]} \BibitemShut {NoStop}%
\bibitem [{\citenamefont {Antonelli}\ \emph {et~al.}(2010)\citenamefont
  {Antonelli} \emph {et~al.}}]{Antonelli:2010yf}%
  \BibitemOpen
  \bibfield  {author} {\bibinfo {author} {\bibfnamefont {M.}~\bibnamefont
  {Antonelli}} \emph {et~al.} (\bibinfo {collaboration} {FlaviaNet Working
  Group on Kaon Decays}),\ }\href {\doibase 10.1140/epjc/s10052-010-1406-3}
  {\bibfield  {journal} {\bibinfo  {journal} {Eur. Phys. J.}\ }\textbf
  {\bibinfo {volume} {C69}},\ \bibinfo {pages} {399} (\bibinfo {year}
  {2010})},\ \Eprint {http://arxiv.org/abs/1005.2323} {arXiv:1005.2323
  [hep-ph]} \BibitemShut {NoStop}%
\bibitem [{\citenamefont {Moulson}(2014)}]{Moulson:2014cra}%
  \BibitemOpen
  \bibfield  {author} {\bibinfo {author} {\bibfnamefont {M.}~\bibnamefont
  {Moulson}},\ }in\ \href
  {http://inspirehep.net/record/1328784/files/arXiv:1411.5252.pdf} {\emph
  {\bibinfo {booktitle} {{8th International Workshop on the CKM Unitarity
  Triangle (CKM2014) Vienna, Austria, September 8-12, 2014}}}}\ (\bibinfo
  {year} {2014})\ \Eprint {http://arxiv.org/abs/1411.5252} {arXiv:1411.5252
  [hep-ex]} \BibitemShut {NoStop}%
\bibitem [{\citenamefont {Amhis}\ \emph {et~al.}(2014)\citenamefont {Amhis}
  \emph {et~al.}}]{Amhis:2014hma}%
  \BibitemOpen
  \bibfield  {author} {\bibinfo {author} {\bibfnamefont {Y.}~\bibnamefont
  {Amhis}} \emph {et~al.} (\bibinfo {collaboration} {Heavy Flavor Averaging
  Group (HFAG)}),\ }\href@noop {} {\  (\bibinfo {year} {2014})},\ \Eprint
  {http://arxiv.org/abs/1412.7515} {arXiv:1412.7515 [hep-ex]} \BibitemShut
  {NoStop}%
\bibitem [{\citenamefont {Glashow}\ \emph {et~al.}(1970)\citenamefont
  {Glashow}, \citenamefont {Iliopoulos},\ and\ \citenamefont
  {Maiani}}]{Glashow:1970gm}%
  \BibitemOpen
  \bibfield  {author} {\bibinfo {author} {\bibfnamefont {S.~L.}\ \bibnamefont
  {Glashow}}, \bibinfo {author} {\bibfnamefont {J.}~\bibnamefont {Iliopoulos}},
  \ and\ \bibinfo {author} {\bibfnamefont {L.}~\bibnamefont {Maiani}},\ }\href
  {\doibase 10.1103/PhysRevD.2.1285} {\bibfield  {journal} {\bibinfo  {journal}
  {Phys. Rev.}\ }\textbf {\bibinfo {volume} {D2}},\ \bibinfo {pages} {1285}
  (\bibinfo {year} {1970})}\BibitemShut {NoStop}%
\bibitem [{\citenamefont {Khachatryan}\ \emph {et~al.}(2015)\citenamefont
  {Khachatryan} \emph {et~al.}}]{Khachatryan:2015kon}%
  \BibitemOpen
  \bibfield  {author} {\bibinfo {author} {\bibfnamefont {V.}~\bibnamefont
  {Khachatryan}} \emph {et~al.} (\bibinfo {collaboration} {CMS}),\ }\href
  {\doibase 10.1016/j.physletb.2015.07.053} {\bibfield  {journal} {\bibinfo
  {journal} {Phys. Lett.}\ }\textbf {\bibinfo {volume} {B749}},\ \bibinfo
  {pages} {337} (\bibinfo {year} {2015})},\ \Eprint
  {http://arxiv.org/abs/1502.07400} {arXiv:1502.07400 [hep-ex]} \BibitemShut
  {NoStop}%
\bibitem [{\citenamefont {Aad}\ \emph {et~al.}(2015)\citenamefont {Aad} \emph
  {et~al.}}]{Aad:2015gha}%
  \BibitemOpen
  \bibfield  {author} {\bibinfo {author} {\bibfnamefont {G.}~\bibnamefont
  {Aad}} \emph {et~al.} (\bibinfo {collaboration} {ATLAS}),\ }\href {\doibase
  10.1007/JHEP11(2015)211} {\bibfield  {journal} {\bibinfo  {journal} {JHEP}\
  }\textbf {\bibinfo {volume} {11}},\ \bibinfo {pages} {211} (\bibinfo {year}
  {2015})},\ \Eprint {http://arxiv.org/abs/1508.03372} {arXiv:1508.03372
  [hep-ex]} \BibitemShut {NoStop}%
\bibitem [{\citenamefont {Baldini}\ \emph {et~al.}(2013)\citenamefont {Baldini}
  \emph {et~al.}}]{Baldini:2013ke}%
  \BibitemOpen
  \bibfield  {author} {\bibinfo {author} {\bibfnamefont {A.~M.}\ \bibnamefont
  {Baldini}} \emph {et~al.},\ }\href@noop {} {\  (\bibinfo {year} {2013})},\
  \Eprint {http://arxiv.org/abs/1301.7225} {arXiv:1301.7225 [physics.ins-det]}
  \BibitemShut {NoStop}%
\bibitem [{\citenamefont {Bona}\ \emph {et~al.}(2007)\citenamefont {Bona} \emph
  {et~al.}}]{Bona:2007qt}%
  \BibitemOpen
  \bibfield  {author} {\bibinfo {author} {\bibfnamefont {M.}~\bibnamefont
  {Bona}} \emph {et~al.} (\bibinfo {collaboration} {SuperB}),\ }\href@noop {}
  {\  (\bibinfo {year} {2007})},\ \Eprint {http://arxiv.org/abs/0709.0451}
  {arXiv:0709.0451 [hep-ex]} \BibitemShut {NoStop}%
\bibitem [{\citenamefont {Aad}\ \emph {et~al.}(2014)\citenamefont {Aad} \emph
  {et~al.}}]{Aad:2014bca}%
  \BibitemOpen
  \bibfield  {author} {\bibinfo {author} {\bibfnamefont {G.}~\bibnamefont
  {Aad}} \emph {et~al.} (\bibinfo {collaboration} {ATLAS}),\ }\href {\doibase
  10.1103/PhysRevD.90.072010} {\bibfield  {journal} {\bibinfo  {journal} {Phys.
  Rev.}\ }\textbf {\bibinfo {volume} {D90}},\ \bibinfo {pages} {072010}
  (\bibinfo {year} {2014})},\ \Eprint {http://arxiv.org/abs/1408.5774}
  {arXiv:1408.5774 [hep-ex]} \BibitemShut {NoStop}%
\bibitem [{\citenamefont {Abada}\ \emph {et~al.}(2015)\citenamefont {Abada},
  \citenamefont {De~Romeri}, \citenamefont {Monteil}, \citenamefont {Orloff},\
  and\ \citenamefont {Teixeira}}]{Abada:2014cca}%
  \BibitemOpen
  \bibfield  {author} {\bibinfo {author} {\bibfnamefont {A.}~\bibnamefont
  {Abada}}, \bibinfo {author} {\bibfnamefont {V.}~\bibnamefont {De~Romeri}},
  \bibinfo {author} {\bibfnamefont {S.}~\bibnamefont {Monteil}}, \bibinfo
  {author} {\bibfnamefont {J.}~\bibnamefont {Orloff}}, \ and\ \bibinfo {author}
  {\bibfnamefont {A.~M.}\ \bibnamefont {Teixeira}},\ }\href {\doibase
  10.1007/JHEP04(2015)051} {\bibfield  {journal} {\bibinfo  {journal} {JHEP}\
  }\textbf {\bibinfo {volume} {04}},\ \bibinfo {pages} {051} (\bibinfo {year}
  {2015})},\ \Eprint {http://arxiv.org/abs/1412.6322} {arXiv:1412.6322
  [hep-ph]} \BibitemShut {NoStop}%
\bibitem [{\citenamefont {Adriani}\ \emph {et~al.}(1993)\citenamefont {Adriani}
  \emph {et~al.}}]{Adriani:1993sy}%
  \BibitemOpen
  \bibfield  {author} {\bibinfo {author} {\bibfnamefont {O.}~\bibnamefont
  {Adriani}} \emph {et~al.} (\bibinfo {collaboration} {L3}),\ }\href {\doibase
  10.1016/0370-2693(93)90348-L} {\bibfield  {journal} {\bibinfo  {journal}
  {Phys. Lett.}\ }\textbf {\bibinfo {volume} {B316}},\ \bibinfo {pages} {427}
  (\bibinfo {year} {1993})}\BibitemShut {NoStop}%
\bibitem [{\citenamefont {Akers}\ \emph {et~al.}(1995)\citenamefont {Akers}
  \emph {et~al.}}]{Akers:1995gz}%
  \BibitemOpen
  \bibfield  {author} {\bibinfo {author} {\bibfnamefont {R.}~\bibnamefont
  {Akers}} \emph {et~al.} (\bibinfo {collaboration} {OPAL}),\ }\href {\doibase
  10.1007/BF01553981} {\bibfield  {journal} {\bibinfo  {journal} {Z. Phys.}\
  }\textbf {\bibinfo {volume} {C67}},\ \bibinfo {pages} {555} (\bibinfo {year}
  {1995})}\BibitemShut {NoStop}%
\bibitem [{\citenamefont {Abreu}\ \emph {et~al.}(1997)\citenamefont {Abreu}
  \emph {et~al.}}]{Abreu:1996mj}%
  \BibitemOpen
  \bibfield  {author} {\bibinfo {author} {\bibfnamefont {P.}~\bibnamefont
  {Abreu}} \emph {et~al.} (\bibinfo {collaboration} {DELPHI}),\ }\href
  {\doibase 10.1007/s002880050313} {\bibfield  {journal} {\bibinfo  {journal}
  {Z. Phys.}\ }\textbf {\bibinfo {volume} {C73}},\ \bibinfo {pages} {243}
  (\bibinfo {year} {1997})}\BibitemShut {NoStop}%
\bibitem [{\citenamefont {Collaboration}(2015)}]{CMS:2015udp}%
  \BibitemOpen
  \bibfield  {author} {\bibinfo {author} {\bibfnamefont {C.}~\bibnamefont
  {Collaboration}} (\bibinfo {collaboration} {CMS}),\ }\href@noop {} {\
  (\bibinfo {year} {2015})}\BibitemShut {NoStop}%
\bibitem [{\citenamefont {Bellgardt}\ \emph {et~al.}(1988)\citenamefont
  {Bellgardt} \emph {et~al.}}]{Bellgardt:1987du}%
  \BibitemOpen
  \bibfield  {author} {\bibinfo {author} {\bibfnamefont {U.}~\bibnamefont
  {Bellgardt}} \emph {et~al.} (\bibinfo {collaboration} {SINDRUM}),\ }\href
  {\doibase 10.1016/0550-3213(88)90462-2} {\bibfield  {journal} {\bibinfo
  {journal} {Nucl. Phys.}\ }\textbf {\bibinfo {volume} {B299}},\ \bibinfo
  {pages} {1} (\bibinfo {year} {1988})}\BibitemShut {NoStop}%
\bibitem [{\citenamefont {Blondel}\ \emph {et~al.}(2013)\citenamefont {Blondel}
  \emph {et~al.}}]{Blondel:2013ia}%
  \BibitemOpen
  \bibfield  {author} {\bibinfo {author} {\bibfnamefont {A.}~\bibnamefont
  {Blondel}} \emph {et~al.},\ }\href@noop {} {\  (\bibinfo {year} {2013})},\
  \Eprint {http://arxiv.org/abs/1301.6113} {arXiv:1301.6113 [physics.ins-det]}
  \BibitemShut {NoStop}%
\bibitem [{\citenamefont {Hayasaka}\ \emph {et~al.}(2010)\citenamefont
  {Hayasaka} \emph {et~al.}}]{Hayasaka:2010np}%
  \BibitemOpen
  \bibfield  {author} {\bibinfo {author} {\bibfnamefont {K.}~\bibnamefont
  {Hayasaka}} \emph {et~al.},\ }\href {\doibase 10.1016/j.physletb.2010.03.037}
  {\bibfield  {journal} {\bibinfo  {journal} {Phys. Lett.}\ }\textbf {\bibinfo
  {volume} {B687}},\ \bibinfo {pages} {139} (\bibinfo {year} {2010})},\ \Eprint
  {http://arxiv.org/abs/1001.3221} {arXiv:1001.3221 [hep-ex]} \BibitemShut
  {NoStop}%
\bibitem [{\citenamefont {Kutschke}(2011)}]{Kutschke:2011ux}%
  \BibitemOpen
  \bibfield  {author} {\bibinfo {author} {\bibfnamefont {R.~K.}\ \bibnamefont
  {Kutschke}},\ }in\ \href
  {http://inspirehep.net/record/1079590/files/arXiv:1112.0242.pdf} {\emph
  {\bibinfo {booktitle} {{Proceedings, 31st International Conference on Physics
  in collisions (PIC 2011)}}}}\ (\bibinfo {year} {2011})\ \Eprint
  {http://arxiv.org/abs/1112.0242} {arXiv:1112.0242 [hep-ex]} \BibitemShut
  {NoStop}%
\bibitem [{\citenamefont {Dohmen}\ \emph {et~al.}(1993)\citenamefont {Dohmen}
  \emph {et~al.}}]{Dohmen:1993mp}%
  \BibitemOpen
  \bibfield  {author} {\bibinfo {author} {\bibfnamefont {C.}~\bibnamefont
  {Dohmen}} \emph {et~al.} (\bibinfo {collaboration} {SINDRUM II}),\ }\href
  {\doibase 10.1016/0370-2693(93)91383-X} {\bibfield  {journal} {\bibinfo
  {journal} {Phys. Lett.}\ }\textbf {\bibinfo {volume} {B317}},\ \bibinfo
  {pages} {631} (\bibinfo {year} {1993})}\BibitemShut {NoStop}%
\bibitem [{\citenamefont {Barlow}(2011)}]{Barlow:2011zza}%
  \BibitemOpen
  \bibfield  {author} {\bibinfo {author} {\bibfnamefont {R.~J.}\ \bibnamefont
  {Barlow}},\ }\href {\doibase 10.1016/j.nuclphysbps.2011.06.009} {\bibfield
  {journal} {\bibinfo  {journal} {Nucl. Phys. Proc. Suppl.}\ }\textbf {\bibinfo
  {volume} {218}},\ \bibinfo {pages} {44} (\bibinfo {year} {2011})}\BibitemShut
  {NoStop}%
\bibitem [{\citenamefont {Illana}\ \emph {et~al.}(1999)\citenamefont {Illana},
  \citenamefont {Jack},\ and\ \citenamefont {Riemann}}]{Illana:1999ww}%
  \BibitemOpen
  \bibfield  {author} {\bibinfo {author} {\bibfnamefont {J.~I.}\ \bibnamefont
  {Illana}}, \bibinfo {author} {\bibfnamefont {M.}~\bibnamefont {Jack}}, \ and\
  \bibinfo {author} {\bibfnamefont {T.}~\bibnamefont {Riemann}},\ }in\ \href
  {http://www-library.desy.de/cgi-bin/showprep.pl?LC-TH-2000-007} {\emph
  {\bibinfo {booktitle} {{5th Workshop of the 2nd ECFA - DESY Study on Physics
  and Detectors for a Linear Electron - Positron Collider Obernai, France,
  October 16-19, 1999}}}}\ (\bibinfo {year} {1999})\ pp.\ \bibinfo {pages}
  {490--524},\ \bibinfo {note} {[,490(1999)]},\ \Eprint
  {http://arxiv.org/abs/hep-ph/0001273} {arXiv:hep-ph/0001273 [hep-ph]}
  \BibitemShut {NoStop}%
\bibitem [{\citenamefont {Pilaftsis}(1992{\natexlab{b}})}]{Pilaftsis:1992st}%
  \BibitemOpen
  \bibfield  {author} {\bibinfo {author} {\bibfnamefont {A.}~\bibnamefont
  {Pilaftsis}},\ }\href {\doibase 10.1016/0370-2693(92)91301-O} {\bibfield
  {journal} {\bibinfo  {journal} {Phys. Lett.}\ }\textbf {\bibinfo {volume}
  {B285}},\ \bibinfo {pages} {68} (\bibinfo {year}
  {1992}{\natexlab{b}})}\BibitemShut {NoStop}%
\bibitem [{\citenamefont {Arganda}\ \emph {et~al.}(2005)\citenamefont
  {Arganda}, \citenamefont {Curiel}, \citenamefont {Herrero},\ and\
  \citenamefont {Temes}}]{Arganda:2004bz}%
  \BibitemOpen
  \bibfield  {author} {\bibinfo {author} {\bibfnamefont {E.}~\bibnamefont
  {Arganda}}, \bibinfo {author} {\bibfnamefont {A.~M.}\ \bibnamefont {Curiel}},
  \bibinfo {author} {\bibfnamefont {M.~J.}\ \bibnamefont {Herrero}}, \ and\
  \bibinfo {author} {\bibfnamefont {D.}~\bibnamefont {Temes}},\ }\href
  {\doibase 10.1103/PhysRevD.71.035011} {\bibfield  {journal} {\bibinfo
  {journal} {Phys. Rev.}\ }\textbf {\bibinfo {volume} {D71}},\ \bibinfo {pages}
  {035011} (\bibinfo {year} {2005})},\ \Eprint
  {http://arxiv.org/abs/hep-ph/0407302} {arXiv:hep-ph/0407302 [hep-ph]}
  \BibitemShut {NoStop}%
\bibitem [{\citenamefont {Arganda}\ \emph {et~al.}(2015)\citenamefont
  {Arganda}, \citenamefont {Herrero}, \citenamefont {Marcano},\ and\
  \citenamefont {Weiland}}]{Arganda:2014dta}%
  \BibitemOpen
  \bibfield  {author} {\bibinfo {author} {\bibfnamefont {E.}~\bibnamefont
  {Arganda}}, \bibinfo {author} {\bibfnamefont {M.~J.}\ \bibnamefont
  {Herrero}}, \bibinfo {author} {\bibfnamefont {X.}~\bibnamefont {Marcano}}, \
  and\ \bibinfo {author} {\bibfnamefont {C.}~\bibnamefont {Weiland}},\ }\href
  {\doibase 10.1103/PhysRevD.91.015001} {\bibfield  {journal} {\bibinfo
  {journal} {Phys. Rev.}\ }\textbf {\bibinfo {volume} {D91}},\ \bibinfo {pages}
  {015001} (\bibinfo {year} {2015})},\ \Eprint {http://arxiv.org/abs/1405.4300}
  {arXiv:1405.4300 [hep-ph]} \BibitemShut {NoStop}%
\bibitem [{\citenamefont {Ilakovac}\ and\ \citenamefont
  {Pilaftsis}(1995)}]{Ilakovac:1994kj}%
  \BibitemOpen
  \bibfield  {author} {\bibinfo {author} {\bibfnamefont {A.}~\bibnamefont
  {Ilakovac}}\ and\ \bibinfo {author} {\bibfnamefont {A.}~\bibnamefont
  {Pilaftsis}},\ }\href {\doibase 10.1016/0550-3213(94)00567-X} {\bibfield
  {journal} {\bibinfo  {journal} {Nucl. Phys.}\ }\textbf {\bibinfo {volume}
  {B437}},\ \bibinfo {pages} {491} (\bibinfo {year} {1995})},\ \Eprint
  {http://arxiv.org/abs/hep-ph/9403398} {arXiv:hep-ph/9403398 [hep-ph]}
  \BibitemShut {NoStop}%
\bibitem [{\citenamefont {Kitano}\ \emph {et~al.}(2002)\citenamefont {Kitano},
  \citenamefont {Koike},\ and\ \citenamefont {Okada}}]{Kitano:2002mt}%
  \BibitemOpen
  \bibfield  {author} {\bibinfo {author} {\bibfnamefont {R.}~\bibnamefont
  {Kitano}}, \bibinfo {author} {\bibfnamefont {M.}~\bibnamefont {Koike}}, \
  and\ \bibinfo {author} {\bibfnamefont {Y.}~\bibnamefont {Okada}},\ }\href
  {\doibase 10.1103/PhysRevD.76.059902, 10.1103/PhysRevD.66.096002} {\bibfield
  {journal} {\bibinfo  {journal} {Phys. Rev.}\ }\textbf {\bibinfo {volume}
  {D66}},\ \bibinfo {pages} {096002} (\bibinfo {year} {2002})},\ \bibinfo
  {note} {[Erratum: Phys. Rev.D76,059902(2007)]},\ \Eprint
  {http://arxiv.org/abs/hep-ph/0203110} {arXiv:hep-ph/0203110 [hep-ph]}
  \BibitemShut {NoStop}%
\bibitem [{\citenamefont {Suzuki}\ \emph {et~al.}(1987)\citenamefont {Suzuki},
  \citenamefont {Measday},\ and\ \citenamefont {Roalsvig}}]{Suzuki:1987jf}%
  \BibitemOpen
  \bibfield  {author} {\bibinfo {author} {\bibfnamefont {T.}~\bibnamefont
  {Suzuki}}, \bibinfo {author} {\bibfnamefont {D.~F.}\ \bibnamefont {Measday}},
  \ and\ \bibinfo {author} {\bibfnamefont {J.~P.}\ \bibnamefont {Roalsvig}},\
  }\href {\doibase 10.1103/PhysRevC.35.2212} {\bibfield  {journal} {\bibinfo
  {journal} {Phys. Rev.}\ }\textbf {\bibinfo {volume} {C35}},\ \bibinfo {pages}
  {2212} (\bibinfo {year} {1987})}\BibitemShut {NoStop}%
\bibitem [{\citenamefont {Ohlsson}(2013)}]{Ohlsson:2012kf}%
  \BibitemOpen
  \bibfield  {author} {\bibinfo {author} {\bibfnamefont {T.}~\bibnamefont
  {Ohlsson}},\ }\href {\doibase 10.1088/0034-4885/76/4/044201} {\bibfield
  {journal} {\bibinfo  {journal} {Rept. Prog. Phys.}\ }\textbf {\bibinfo
  {volume} {76}},\ \bibinfo {pages} {044201} (\bibinfo {year} {2013})},\
  \Eprint {http://arxiv.org/abs/1209.2710} {arXiv:1209.2710 [hep-ph]}
  \BibitemShut {NoStop}%
\bibitem [{\citenamefont {Xing}(2008)}]{Xing:2007zj}%
  \BibitemOpen
  \bibfield  {author} {\bibinfo {author} {\bibfnamefont {Z.-z.}\ \bibnamefont
  {Xing}},\ }\href {\doibase 10.1016/j.physletb.2008.01.038} {\bibfield
  {journal} {\bibinfo  {journal} {Phys. Lett.}\ }\textbf {\bibinfo {volume}
  {B660}},\ \bibinfo {pages} {515} (\bibinfo {year} {2008})},\ \Eprint
  {http://arxiv.org/abs/0709.2220} {arXiv:0709.2220 [hep-ph]} \BibitemShut
  {NoStop}%
\bibitem [{\citenamefont {Xing}(2012)}]{Xing:2011ur}%
  \BibitemOpen
  \bibfield  {author} {\bibinfo {author} {\bibfnamefont {Z.-z.}\ \bibnamefont
  {Xing}},\ }\href {\doibase 10.1103/PhysRevD.85.013008} {\bibfield  {journal}
  {\bibinfo  {journal} {Phys. Rev.}\ }\textbf {\bibinfo {volume} {D85}},\
  \bibinfo {pages} {013008} (\bibinfo {year} {2012})},\ \Eprint
  {http://arxiv.org/abs/1110.0083} {arXiv:1110.0083 [hep-ph]} \BibitemShut
  {NoStop}%
\bibitem [{\citenamefont {Escrihuela}\ \emph {et~al.}(2015)\citenamefont
  {Escrihuela}, \citenamefont {Forero}, \citenamefont {Miranda}, \citenamefont
  {Tortola},\ and\ \citenamefont {Valle}}]{Escrihuela:2015wra}%
  \BibitemOpen
  \bibfield  {author} {\bibinfo {author} {\bibfnamefont {F.~J.}\ \bibnamefont
  {Escrihuela}}, \bibinfo {author} {\bibfnamefont {D.~V.}\ \bibnamefont
  {Forero}}, \bibinfo {author} {\bibfnamefont {O.~G.}\ \bibnamefont {Miranda}},
  \bibinfo {author} {\bibfnamefont {M.}~\bibnamefont {Tortola}}, \ and\
  \bibinfo {author} {\bibfnamefont {J.~W.~F.}\ \bibnamefont {Valle}},\ }\href
  {\doibase 10.1103/PhysRevD.92.053009} {\bibfield  {journal} {\bibinfo
  {journal} {Phys. Rev.}\ }\textbf {\bibinfo {volume} {D92}},\ \bibinfo {pages}
  {053009} (\bibinfo {year} {2015})},\ \Eprint
  {http://arxiv.org/abs/1503.08879} {arXiv:1503.08879 [hep-ph]} \BibitemShut
  {NoStop}%
\bibitem [{\citenamefont {Antusch}\ \emph
  {et~al.}(2009{\natexlab{b}})\citenamefont {Antusch}, \citenamefont {Blennow},
  \citenamefont {Fernandez-Martinez},\ and\ \citenamefont
  {Lopez-Pavon}}]{Antusch:2009pm}%
  \BibitemOpen
  \bibfield  {author} {\bibinfo {author} {\bibfnamefont {S.}~\bibnamefont
  {Antusch}}, \bibinfo {author} {\bibfnamefont {M.}~\bibnamefont {Blennow}},
  \bibinfo {author} {\bibfnamefont {E.}~\bibnamefont {Fernandez-Martinez}}, \
  and\ \bibinfo {author} {\bibfnamefont {J.}~\bibnamefont {Lopez-Pavon}},\
  }\href {\doibase 10.1103/PhysRevD.80.033002} {\bibfield  {journal} {\bibinfo
  {journal} {Phys. Rev.}\ }\textbf {\bibinfo {volume} {D80}},\ \bibinfo {pages}
  {033002} (\bibinfo {year} {2009}{\natexlab{b}})},\ \Eprint
  {http://arxiv.org/abs/0903.3986} {arXiv:0903.3986 [hep-ph]} \BibitemShut
  {NoStop}%
\bibitem [{\citenamefont {Parke}\ and\ \citenamefont
  {Ross-Lonergan}(2015)}]{Parke:2015goa}%
  \BibitemOpen
  \bibfield  {author} {\bibinfo {author} {\bibfnamefont {S.}~\bibnamefont
  {Parke}}\ and\ \bibinfo {author} {\bibfnamefont {M.}~\bibnamefont
  {Ross-Lonergan}},\ }\href@noop {} {\  (\bibinfo {year} {2015})},\ \Eprint
  {http://arxiv.org/abs/1508.05095} {arXiv:1508.05095 [hep-ph]} \BibitemShut
  {NoStop}%
\bibitem [{\citenamefont {Miranda}\ \emph {et~al.}(2016)\citenamefont
  {Miranda}, \citenamefont {Tortola},\ and\ \citenamefont
  {Valle}}]{Miranda:2016wdr}%
  \BibitemOpen
  \bibfield  {author} {\bibinfo {author} {\bibfnamefont {O.~G.}\ \bibnamefont
  {Miranda}}, \bibinfo {author} {\bibfnamefont {M.}~\bibnamefont {Tortola}}, \
  and\ \bibinfo {author} {\bibfnamefont {J.~W.~F.}\ \bibnamefont {Valle}},\
  }\href@noop {} {\  (\bibinfo {year} {2016})},\ \Eprint
  {http://arxiv.org/abs/1604.05690} {arXiv:1604.05690 [hep-ph]} \BibitemShut
  {NoStop}%
\bibitem [{\citenamefont {Ge}\ \emph {et~al.}(2016)\citenamefont {Ge},
  \citenamefont {Pasquini}, \citenamefont {Tortola},\ and\ \citenamefont
  {Valle}}]{Ge:2016xya}%
  \BibitemOpen
  \bibfield  {author} {\bibinfo {author} {\bibfnamefont {S.-F.}\ \bibnamefont
  {Ge}}, \bibinfo {author} {\bibfnamefont {P.}~\bibnamefont {Pasquini}},
  \bibinfo {author} {\bibfnamefont {M.}~\bibnamefont {Tortola}}, \ and\
  \bibinfo {author} {\bibfnamefont {J.~W.~F.}\ \bibnamefont {Valle}},\
  }\href@noop {} {\  (\bibinfo {year} {2016})},\ \Eprint
  {http://arxiv.org/abs/1605.01670} {arXiv:1605.01670 [hep-ph]} \BibitemShut
  {NoStop}%
\bibitem [{\citenamefont {Geer}(1998)}]{Geer:1997iz}%
  \BibitemOpen
  \bibfield  {author} {\bibinfo {author} {\bibfnamefont {S.}~\bibnamefont
  {Geer}},\ }\href {\doibase 10.1103/PhysRevD.57.6989,
  10.1103/PhysRevD.59.039903} {\bibfield  {journal} {\bibinfo  {journal} {Phys.
  Rev.}\ }\textbf {\bibinfo {volume} {D57}},\ \bibinfo {pages} {6989} (\bibinfo
  {year} {1998})},\ \bibinfo {note} {[Erratum: Phys. Rev.D59,039903(1999)]},\
  \Eprint {http://arxiv.org/abs/hep-ph/9712290} {arXiv:hep-ph/9712290 [hep-ph]}
  \BibitemShut {NoStop}%
\bibitem [{\citenamefont {De~Rujula}\ \emph {et~al.}(1999)\citenamefont
  {De~Rujula}, \citenamefont {Gavela},\ and\ \citenamefont
  {Hernandez}}]{DeRujula:1998umv}%
  \BibitemOpen
  \bibfield  {author} {\bibinfo {author} {\bibfnamefont {A.}~\bibnamefont
  {De~Rujula}}, \bibinfo {author} {\bibfnamefont {M.~B.}\ \bibnamefont
  {Gavela}}, \ and\ \bibinfo {author} {\bibfnamefont {P.}~\bibnamefont
  {Hernandez}},\ }\href {\doibase 10.1016/S0550-3213(99)00070-X} {\bibfield
  {journal} {\bibinfo  {journal} {Nucl. Phys.}\ }\textbf {\bibinfo {volume}
  {B547}},\ \bibinfo {pages} {21} (\bibinfo {year} {1999})},\ \Eprint
  {http://arxiv.org/abs/hep-ph/9811390} {arXiv:hep-ph/9811390 [hep-ph]}
  \BibitemShut {NoStop}%
\bibitem [{\citenamefont {Atre}\ \emph {et~al.}(2009)\citenamefont {Atre},
  \citenamefont {Han}, \citenamefont {Pascoli},\ and\ \citenamefont
  {Zhang}}]{Atre:2009rg}%
  \BibitemOpen
  \bibfield  {author} {\bibinfo {author} {\bibfnamefont {A.}~\bibnamefont
  {Atre}}, \bibinfo {author} {\bibfnamefont {T.}~\bibnamefont {Han}}, \bibinfo
  {author} {\bibfnamefont {S.}~\bibnamefont {Pascoli}}, \ and\ \bibinfo
  {author} {\bibfnamefont {B.}~\bibnamefont {Zhang}},\ }\href {\doibase
  10.1088/1126-6708/2009/05/030} {\bibfield  {journal} {\bibinfo  {journal}
  {JHEP}\ }\textbf {\bibinfo {volume} {05}},\ \bibinfo {pages} {030} (\bibinfo
  {year} {2009})},\ \Eprint {http://arxiv.org/abs/0901.3589} {arXiv:0901.3589
  [hep-ph]} \BibitemShut {NoStop}%
\bibitem [{\citenamefont {Ruchayskiy}\ and\ \citenamefont
  {Ivashko}(2012{\natexlab{a}})}]{Ruchayskiy:2011aa}%
  \BibitemOpen
  \bibfield  {author} {\bibinfo {author} {\bibfnamefont {O.}~\bibnamefont
  {Ruchayskiy}}\ and\ \bibinfo {author} {\bibfnamefont {A.}~\bibnamefont
  {Ivashko}},\ }\href {\doibase 10.1007/JHEP06(2012)100} {\bibfield  {journal}
  {\bibinfo  {journal} {JHEP}\ }\textbf {\bibinfo {volume} {06}},\ \bibinfo
  {pages} {100} (\bibinfo {year} {2012}{\natexlab{a}})},\ \Eprint
  {http://arxiv.org/abs/1112.3319} {arXiv:1112.3319 [hep-ph]} \BibitemShut
  {NoStop}%
\bibitem [{\citenamefont {Drewes}\ and\ \citenamefont
  {Garbrecht}(2015)}]{Drewes:2015iva}%
  \BibitemOpen
  \bibfield  {author} {\bibinfo {author} {\bibfnamefont {M.}~\bibnamefont
  {Drewes}}\ and\ \bibinfo {author} {\bibfnamefont {B.}~\bibnamefont
  {Garbrecht}},\ }\href@noop {} {\  (\bibinfo {year} {2015})},\ \Eprint
  {http://arxiv.org/abs/1502.00477} {arXiv:1502.00477 [hep-ph]} \BibitemShut
  {NoStop}%
\bibitem [{\citenamefont {de~Gouvêa}\ and\ \citenamefont
  {Kobach}(2016)}]{deGouvea:2015euy}%
  \BibitemOpen
  \bibfield  {author} {\bibinfo {author} {\bibfnamefont {A.}~\bibnamefont
  {de~Gouvêa}}\ and\ \bibinfo {author} {\bibfnamefont {A.}~\bibnamefont
  {Kobach}},\ }\href {\doibase 10.1103/PhysRevD.93.033005} {\bibfield
  {journal} {\bibinfo  {journal} {Phys. Rev.}\ }\textbf {\bibinfo {volume}
  {D93}},\ \bibinfo {pages} {033005} (\bibinfo {year} {2016})},\ \Eprint
  {http://arxiv.org/abs/1511.00683} {arXiv:1511.00683 [hep-ph]} \BibitemShut
  {NoStop}%
\bibitem [{\citenamefont {Deppisch}\ \emph {et~al.}(2015)\citenamefont
  {Deppisch}, \citenamefont {Bhupal~Dev},\ and\ \citenamefont
  {Pilaftsis}}]{Deppisch:2015qwa}%
  \BibitemOpen
  \bibfield  {author} {\bibinfo {author} {\bibfnamefont {F.~F.}\ \bibnamefont
  {Deppisch}}, \bibinfo {author} {\bibfnamefont {P.~S.}\ \bibnamefont
  {Bhupal~Dev}}, \ and\ \bibinfo {author} {\bibfnamefont {A.}~\bibnamefont
  {Pilaftsis}},\ }\href {\doibase 10.1088/1367-2630/17/7/075019} {\bibfield
  {journal} {\bibinfo  {journal} {New J. Phys.}\ }\textbf {\bibinfo {volume}
  {17}},\ \bibinfo {pages} {075019} (\bibinfo {year} {2015})},\ \Eprint
  {http://arxiv.org/abs/1502.06541} {arXiv:1502.06541 [hep-ph]} \BibitemShut
  {NoStop}%
\bibitem [{\citenamefont {Dolgov}\ and\ \citenamefont
  {Villante}(2004)}]{Dolgov:2003sg}%
  \BibitemOpen
  \bibfield  {author} {\bibinfo {author} {\bibfnamefont {A.~D.}\ \bibnamefont
  {Dolgov}}\ and\ \bibinfo {author} {\bibfnamefont {F.~L.}\ \bibnamefont
  {Villante}},\ }\href {\doibase 10.1016/j.nuclphysb.2003.11.031} {\bibfield
  {journal} {\bibinfo  {journal} {Nucl. Phys.}\ }\textbf {\bibinfo {volume}
  {B679}},\ \bibinfo {pages} {261} (\bibinfo {year} {2004})},\ \Eprint
  {http://arxiv.org/abs/hep-ph/0308083} {arXiv:hep-ph/0308083 [hep-ph]}
  \BibitemShut {NoStop}%
\bibitem [{\citenamefont {Cirelli}\ \emph {et~al.}(2005)\citenamefont
  {Cirelli}, \citenamefont {Marandella}, \citenamefont {Strumia},\ and\
  \citenamefont {Vissani}}]{Cirelli:2004cz}%
  \BibitemOpen
  \bibfield  {author} {\bibinfo {author} {\bibfnamefont {M.}~\bibnamefont
  {Cirelli}}, \bibinfo {author} {\bibfnamefont {G.}~\bibnamefont {Marandella}},
  \bibinfo {author} {\bibfnamefont {A.}~\bibnamefont {Strumia}}, \ and\
  \bibinfo {author} {\bibfnamefont {F.}~\bibnamefont {Vissani}},\ }\href
  {\doibase 10.1016/j.nuclphysb.2004.11.056} {\bibfield  {journal} {\bibinfo
  {journal} {Nucl. Phys.}\ }\textbf {\bibinfo {volume} {B708}},\ \bibinfo
  {pages} {215} (\bibinfo {year} {2005})},\ \Eprint
  {http://arxiv.org/abs/hep-ph/0403158} {arXiv:hep-ph/0403158 [hep-ph]}
  \BibitemShut {NoStop}%
\bibitem [{\citenamefont {Melchiorri}\ \emph {et~al.}(2009)\citenamefont
  {Melchiorri}, \citenamefont {Mena}, \citenamefont {Palomares-Ruiz},
  \citenamefont {Pascoli}, \citenamefont {Slosar},\ and\ \citenamefont
  {Sorel}}]{Melchiorri:2008gq}%
  \BibitemOpen
  \bibfield  {author} {\bibinfo {author} {\bibfnamefont {A.}~\bibnamefont
  {Melchiorri}}, \bibinfo {author} {\bibfnamefont {O.}~\bibnamefont {Mena}},
  \bibinfo {author} {\bibfnamefont {S.}~\bibnamefont {Palomares-Ruiz}},
  \bibinfo {author} {\bibfnamefont {S.}~\bibnamefont {Pascoli}}, \bibinfo
  {author} {\bibfnamefont {A.}~\bibnamefont {Slosar}}, \ and\ \bibinfo {author}
  {\bibfnamefont {M.}~\bibnamefont {Sorel}},\ }\href {\doibase
  10.1088/1475-7516/2009/01/036} {\bibfield  {journal} {\bibinfo  {journal}
  {JCAP}\ }\textbf {\bibinfo {volume} {0901}},\ \bibinfo {pages} {036}
  (\bibinfo {year} {2009})},\ \Eprint {http://arxiv.org/abs/0810.5133}
  {arXiv:0810.5133 [hep-ph]} \BibitemShut {NoStop}%
\bibitem [{\citenamefont {Hannestad}\ \emph {et~al.}(2012)\citenamefont
  {Hannestad}, \citenamefont {Tamborra},\ and\ \citenamefont
  {Tram}}]{Hannestad:2012ky}%
  \BibitemOpen
  \bibfield  {author} {\bibinfo {author} {\bibfnamefont {S.}~\bibnamefont
  {Hannestad}}, \bibinfo {author} {\bibfnamefont {I.}~\bibnamefont {Tamborra}},
  \ and\ \bibinfo {author} {\bibfnamefont {T.}~\bibnamefont {Tram}},\ }\href
  {\doibase 10.1088/1475-7516/2012/07/025} {\bibfield  {journal} {\bibinfo
  {journal} {JCAP}\ }\textbf {\bibinfo {volume} {1207}},\ \bibinfo {pages}
  {025} (\bibinfo {year} {2012})},\ \Eprint {http://arxiv.org/abs/1204.5861}
  {arXiv:1204.5861 [astro-ph.CO]} \BibitemShut {NoStop}%
\bibitem [{\citenamefont {Ruchayskiy}\ and\ \citenamefont
  {Ivashko}(2012{\natexlab{b}})}]{Ruchayskiy:2012si}%
  \BibitemOpen
  \bibfield  {author} {\bibinfo {author} {\bibfnamefont {O.}~\bibnamefont
  {Ruchayskiy}}\ and\ \bibinfo {author} {\bibfnamefont {A.}~\bibnamefont
  {Ivashko}},\ }\href {\doibase 10.1088/1475-7516/2012/10/014} {\bibfield
  {journal} {\bibinfo  {journal} {JCAP}\ }\textbf {\bibinfo {volume} {1210}},\
  \bibinfo {pages} {014} (\bibinfo {year} {2012}{\natexlab{b}})},\ \Eprint
  {http://arxiv.org/abs/1202.2841} {arXiv:1202.2841 [hep-ph]} \BibitemShut
  {NoStop}%
\bibitem [{\citenamefont {Mirizzi}\ \emph {et~al.}(2012)\citenamefont
  {Mirizzi}, \citenamefont {Saviano}, \citenamefont {Miele},\ and\
  \citenamefont {Serpico}}]{Mirizzi:2012we}%
  \BibitemOpen
  \bibfield  {author} {\bibinfo {author} {\bibfnamefont {A.}~\bibnamefont
  {Mirizzi}}, \bibinfo {author} {\bibfnamefont {N.}~\bibnamefont {Saviano}},
  \bibinfo {author} {\bibfnamefont {G.}~\bibnamefont {Miele}}, \ and\ \bibinfo
  {author} {\bibfnamefont {P.~D.}\ \bibnamefont {Serpico}},\ }\href {\doibase
  10.1103/PhysRevD.86.053009} {\bibfield  {journal} {\bibinfo  {journal} {Phys.
  Rev.}\ }\textbf {\bibinfo {volume} {D86}},\ \bibinfo {pages} {053009}
  (\bibinfo {year} {2012})},\ \Eprint {http://arxiv.org/abs/1206.1046}
  {arXiv:1206.1046 [hep-ph]} \BibitemShut {NoStop}%
\bibitem [{\citenamefont {Jacques}\ \emph {et~al.}(2013)\citenamefont
  {Jacques}, \citenamefont {Krauss},\ and\ \citenamefont
  {Lunardini}}]{Jacques:2013xr}%
  \BibitemOpen
  \bibfield  {author} {\bibinfo {author} {\bibfnamefont {T.~D.}\ \bibnamefont
  {Jacques}}, \bibinfo {author} {\bibfnamefont {L.~M.}\ \bibnamefont {Krauss}},
  \ and\ \bibinfo {author} {\bibfnamefont {C.}~\bibnamefont {Lunardini}},\
  }\href {\doibase 10.1103/PhysRevD.87.083515, 10.1103/PhysRevD.88.109901}
  {\bibfield  {journal} {\bibinfo  {journal} {Phys. Rev.}\ }\textbf {\bibinfo
  {volume} {D87}},\ \bibinfo {pages} {083515} (\bibinfo {year} {2013})},\
  \bibinfo {note} {[Erratum: Phys. Rev.D88,no.10,109901(2013)]},\ \Eprint
  {http://arxiv.org/abs/1301.3119} {arXiv:1301.3119 [astro-ph.CO]} \BibitemShut
  {NoStop}%
\bibitem [{\citenamefont {Saviano}\ \emph {et~al.}(2013)\citenamefont
  {Saviano}, \citenamefont {Mirizzi}, \citenamefont {Pisanti}, \citenamefont
  {Serpico}, \citenamefont {Mangano},\ and\ \citenamefont
  {Miele}}]{Saviano:2013ktj}%
  \BibitemOpen
  \bibfield  {author} {\bibinfo {author} {\bibfnamefont {N.}~\bibnamefont
  {Saviano}}, \bibinfo {author} {\bibfnamefont {A.}~\bibnamefont {Mirizzi}},
  \bibinfo {author} {\bibfnamefont {O.}~\bibnamefont {Pisanti}}, \bibinfo
  {author} {\bibfnamefont {P.~D.}\ \bibnamefont {Serpico}}, \bibinfo {author}
  {\bibfnamefont {G.}~\bibnamefont {Mangano}}, \ and\ \bibinfo {author}
  {\bibfnamefont {G.}~\bibnamefont {Miele}},\ }\href {\doibase
  10.1103/PhysRevD.87.073006} {\bibfield  {journal} {\bibinfo  {journal} {Phys.
  Rev.}\ }\textbf {\bibinfo {volume} {D87}},\ \bibinfo {pages} {073006}
  (\bibinfo {year} {2013})},\ \Eprint {http://arxiv.org/abs/1302.1200}
  {arXiv:1302.1200 [astro-ph.CO]} \BibitemShut {NoStop}%
\bibitem [{\citenamefont {Archidiacono}\ \emph {et~al.}(2013)\citenamefont
  {Archidiacono}, \citenamefont {Fornengo}, \citenamefont {Giunti},
  \citenamefont {Hannestad},\ and\ \citenamefont
  {Melchiorri}}]{Archidiacono:2013xxa}%
  \BibitemOpen
  \bibfield  {author} {\bibinfo {author} {\bibfnamefont {M.}~\bibnamefont
  {Archidiacono}}, \bibinfo {author} {\bibfnamefont {N.}~\bibnamefont
  {Fornengo}}, \bibinfo {author} {\bibfnamefont {C.}~\bibnamefont {Giunti}},
  \bibinfo {author} {\bibfnamefont {S.}~\bibnamefont {Hannestad}}, \ and\
  \bibinfo {author} {\bibfnamefont {A.}~\bibnamefont {Melchiorri}},\ }\href
  {\doibase 10.1103/PhysRevD.87.125034} {\bibfield  {journal} {\bibinfo
  {journal} {Phys. Rev.}\ }\textbf {\bibinfo {volume} {D87}},\ \bibinfo {pages}
  {125034} (\bibinfo {year} {2013})},\ \Eprint {http://arxiv.org/abs/1302.6720}
  {arXiv:1302.6720 [astro-ph.CO]} \BibitemShut {NoStop}%
\bibitem [{\citenamefont {Mirizzi}\ \emph {et~al.}(2013)\citenamefont
  {Mirizzi}, \citenamefont {Mangano}, \citenamefont {Saviano}, \citenamefont
  {Borriello}, \citenamefont {Giunti}, \citenamefont {Miele},\ and\
  \citenamefont {Pisanti}}]{Mirizzi:2013gnd}%
  \BibitemOpen
  \bibfield  {author} {\bibinfo {author} {\bibfnamefont {A.}~\bibnamefont
  {Mirizzi}}, \bibinfo {author} {\bibfnamefont {G.}~\bibnamefont {Mangano}},
  \bibinfo {author} {\bibfnamefont {N.}~\bibnamefont {Saviano}}, \bibinfo
  {author} {\bibfnamefont {E.}~\bibnamefont {Borriello}}, \bibinfo {author}
  {\bibfnamefont {C.}~\bibnamefont {Giunti}}, \bibinfo {author} {\bibfnamefont
  {G.}~\bibnamefont {Miele}}, \ and\ \bibinfo {author} {\bibfnamefont
  {O.}~\bibnamefont {Pisanti}},\ }\href {\doibase
  10.1016/j.physletb.2013.08.015} {\bibfield  {journal} {\bibinfo  {journal}
  {Phys. Lett.}\ }\textbf {\bibinfo {volume} {B726}},\ \bibinfo {pages} {8}
  (\bibinfo {year} {2013})},\ \Eprint {http://arxiv.org/abs/1303.5368}
  {arXiv:1303.5368 [astro-ph.CO]} \BibitemShut {NoStop}%
\bibitem [{\citenamefont {Hernandez}\ \emph
  {et~al.}(2014{\natexlab{a}})\citenamefont {Hernandez}, \citenamefont
  {Kekic},\ and\ \citenamefont {Lopez-Pavon}}]{Hernandez:2013lza}%
  \BibitemOpen
  \bibfield  {author} {\bibinfo {author} {\bibfnamefont {P.}~\bibnamefont
  {Hernandez}}, \bibinfo {author} {\bibfnamefont {M.}~\bibnamefont {Kekic}}, \
  and\ \bibinfo {author} {\bibfnamefont {J.}~\bibnamefont {Lopez-Pavon}},\
  }\href {\doibase 10.1103/PhysRevD.89.073009} {\bibfield  {journal} {\bibinfo
  {journal} {Phys. Rev.}\ }\textbf {\bibinfo {volume} {D89}},\ \bibinfo {pages}
  {073009} (\bibinfo {year} {2014}{\natexlab{a}})},\ \Eprint
  {http://arxiv.org/abs/1311.2614} {arXiv:1311.2614 [hep-ph]} \BibitemShut
  {NoStop}%
\bibitem [{\citenamefont {Vincent}\ \emph {et~al.}(2015)\citenamefont
  {Vincent}, \citenamefont {Martinez}, \citenamefont {Hernández},
  \citenamefont {Lattanzi},\ and\ \citenamefont {Mena}}]{Vincent:2014rja}%
  \BibitemOpen
  \bibfield  {author} {\bibinfo {author} {\bibfnamefont {A.~C.}\ \bibnamefont
  {Vincent}}, \bibinfo {author} {\bibfnamefont {E.~F.}\ \bibnamefont
  {Martinez}}, \bibinfo {author} {\bibfnamefont {P.}~\bibnamefont
  {Hernández}}, \bibinfo {author} {\bibfnamefont {M.}~\bibnamefont
  {Lattanzi}}, \ and\ \bibinfo {author} {\bibfnamefont {O.}~\bibnamefont
  {Mena}},\ }\href {\doibase 10.1088/1475-7516/2015/04/006} {\bibfield
  {journal} {\bibinfo  {journal} {JCAP}\ }\textbf {\bibinfo {volume} {1504}},\
  \bibinfo {pages} {006} (\bibinfo {year} {2015})},\ \Eprint
  {http://arxiv.org/abs/1408.1956} {arXiv:1408.1956 [astro-ph.CO]} \BibitemShut
  {NoStop}%
\bibitem [{\citenamefont {Hernandez}\ \emph
  {et~al.}(2014{\natexlab{b}})\citenamefont {Hernandez}, \citenamefont
  {Kekic},\ and\ \citenamefont {Lopez-Pavon}}]{Hernandez:2014fha}%
  \BibitemOpen
  \bibfield  {author} {\bibinfo {author} {\bibfnamefont {P.}~\bibnamefont
  {Hernandez}}, \bibinfo {author} {\bibfnamefont {M.}~\bibnamefont {Kekic}}, \
  and\ \bibinfo {author} {\bibfnamefont {J.}~\bibnamefont {Lopez-Pavon}},\
  }\href {\doibase 10.1103/PhysRevD.90.065033} {\bibfield  {journal} {\bibinfo
  {journal} {Phys. Rev.}\ }\textbf {\bibinfo {volume} {D90}},\ \bibinfo {pages}
  {065033} (\bibinfo {year} {2014}{\natexlab{b}})},\ \Eprint
  {http://arxiv.org/abs/1406.2961} {arXiv:1406.2961 [hep-ph]} \BibitemShut
  {NoStop}%
\bibitem [{\citenamefont {Antusch}\ and\ \citenamefont
  {Fischer}(2016)}]{Antusch:2016brq}%
  \BibitemOpen
  \bibfield  {author} {\bibinfo {author} {\bibfnamefont {S.}~\bibnamefont
  {Antusch}}\ and\ \bibinfo {author} {\bibfnamefont {O.}~\bibnamefont
  {Fischer}},\ }\href@noop {} {\  (\bibinfo {year} {2016})},\ \Eprint
  {http://arxiv.org/abs/1604.00208} {arXiv:1604.00208 [hep-ph]} \BibitemShut
  {NoStop}%
\bibitem [{\citenamefont {Antusch}\ \emph {et~al.}(2016)\citenamefont
  {Antusch}, \citenamefont {Cazzato},\ and\ \citenamefont
  {Fischer}}]{Antusch:2016vyf}%
  \BibitemOpen
  \bibfield  {author} {\bibinfo {author} {\bibfnamefont {S.}~\bibnamefont
  {Antusch}}, \bibinfo {author} {\bibfnamefont {E.}~\bibnamefont {Cazzato}}, \
  and\ \bibinfo {author} {\bibfnamefont {O.}~\bibnamefont {Fischer}},\
  }\href@noop {} {\  (\bibinfo {year} {2016})},\ \Eprint
  {http://arxiv.org/abs/1604.02420} {arXiv:1604.02420 [hep-ph]} \BibitemShut
  {NoStop}%
\end{thebibliography}
\end{document}